\documentclass[12pt]{article}
\usepackage{geometry}
\usepackage{amsmath}
\usepackage{amssymb,epsfig,subfigure}
\usepackage{graphicx}
\usepackage{verbatim}
\numberwithin{equation}{section}

\newlength{\wth}
\newcommand{\startappendix}{
\setcounter{section}{0}
\renewcommand{\thesection}{\Alph{section}}}

\newcommand{\be}{\begin{equation}}
\newcommand{\ee}{\end{equation}}
\newcommand{\ba}{\begin{aligned}}
\newcommand{\ea}{\end{aligned}}

\def\m1{\left(-1\right)^{F_i}}

\makeatletter
\def\sla@#1#2#3#4#5{{%
  \setbox\z@\hbox{$\m@th#4#5$}%
  \setbox\tw@\hbox{$\m@th#4#1$}%
  \dimen4\wd\ifdim\wd\z@<\wd\tw@\tw@\else\z@\fi
  \dimen@\ht\tw@
  \advance\dimen@-\dp\tw@
  \advance\dimen@-\ht\z@
  \advance\dimen@\dp\z@
  \divide\dimen@\tw@
  \advance\dimen@-#3\ht\tw@
  \advance\dimen@-#3\dp\tw@
  \dimen@ii#2\wd\z@  \raise-\dimen@\hbox to\dimen4{%
    \hss\kern\dimen@ii\box\tw@\kern-\dimen@ii\hss}%
  \llap{\hbox to\dimen4{\hss\box\z@\hss}}}}
\def\slashed#1{%
  \expandafter\ifx\csname sla@\string#1\endcsname\relax
    {\mathpalette{\sla@/00}{#1}}%
  \else
    \csname sla@\string#1\endcsname
  \fi}
\makeatother

\begin{document}


\thispagestyle{empty}
\begin{flushright}\footnotesize
\texttt{IPPP-11-06, DCPT-11-12, EFI-11-05, KCL-MTH-11-02, PI-STRINGS-211}\\
\vspace{1.4cm}
\end{flushright}

\renewcommand{\thefootnote}{\fnsymbol{footnote}}
\setcounter{footnote}{0}

\begin{center}
{\Large\textbf{\mathversion{bold} F-theory GUTs with $U(1)$ Symmetries: \\
Generalities and Survey}\par}

\vspace{2.1cm}

\textrm{Matthew J. Dolan$^1$, Joseph Marsano$^2$, Natalia Saulina$^3$ and Sakura Sch\"afer-Nameki$^4$}

\vspace{1cm}

\textit{
$^1$ Institute for Particle Physics Phenomenology \\ University of Durham, Durham DH1 3LE, UK \\[1ex]
$^2$ Enrico Fermi Institute, University of Chicago\\ 5640 S Ellis Ave, Chicago, IL 60637, USA\\[1ex]
$^3$ Perimeter Institute for Theoretical Physics\\ 31 Caroline St N., Waterloo, Ontario N2L 2Y5, Canada \\[1ex]
$^4$ Department of Mathematics, King's College, University of London \\ The Strand, WC2R 2LS, London, England \\[1ex]
}

\texttt{m.j.dolan durham.ac.uk, marsano,  saulina, ss299  theory.caltech.edu}

\bigskip


\par\vspace{1cm}

\textbf{Abstract}\vspace{5mm}
\end{center}

\noindent
We study the structure of $SU(5)$ F-theory GUT models that engineer additional $U(1)$ symmetries.  These are highly constrained by a set of relations observed by Dudas and Palti (DP) that originate from the physics of 4D anomaly cancellation.  Using the DP relations, we describe a general tension between unification and the suppression of dimension 5 proton decay when one or more $U(1)$'s are PQ symmetries and hypercharge flux is used to break the $SU(5)$ GUT group.  We then specialize to spectral cover models, whose global completions in F-theory we know how to construct.  In that setting, we provide a technical derivation of the DP relations, construct spectral covers that yield all possible solutions to them, and provide a complete survey of spectral cover models for $SU(5)$ GUTs that exhibit two $U(1)$ symmetries.

\vspace*{\fill}

\setcounter{page}{1}
\renewcommand{\thefootnote}{\arabic{footnote}}
\setcounter{footnote}{0}

 \newpage

\tableofcontents

\section{Introduction and Summary}

Throughout the past few years, F-theory has emerged as a promising framework for engineering supersymmetric GUTs in string theory \cite{Donagi:2008ca,Beasley:2008dc,Beasley:2008kw,Donagi:2008kj}.  The main focus has been on $SU(5)$ GUTs, where 
internal fluxes make it easy to break the GUT group and remove Higgs triplets \cite{Beasley:2008kw,Donagi:2008kj}.  Proton decay and large $R$-parity violation can in principle be controlled through the introduction of symmetries \cite{Tatar:2009jk,Marsano:2009gv,Blumenhagen:2009yv}.  Most of the current literature makes use of $U(1)$ symmetries for this purpose because they seem plentiful from many points of view.  There are some subtle issues associated with them \cite{Hayashi:2010zp} but by now the task of realizing $U(1)$'s is fairly well-understood for a large class of F-theory compactifications \cite{Grimm:2010ez,Marsano:2010ix}.

Already in the early days of F-theory model building, though, it became clear that hypercharge flux and $U(1)$ symmetries are nontrivially linked.  For model-building, we would like to distribute hypercharge flux freely among the matter curves but our ability to do this is strongly restricted when $U(1)$ symmetries are present \cite{Marsano:2009gv,Marsano:2009wr}.  The nature of these restrictions can be quantified using the Dudas-Palti relations, which were first observed by Dudas and Palti in a set of spectral cover models \cite{Dudas:2010zb}{\footnote{While this paper was in preparation, the work \cite{Ludeling:2011en} appeared which studies constraints related to $R$-parity that arise in the spectral cover models of \cite{Dudas:2010zb}}.  Letting $q_I/q_a$ denote the common $U(1)$ charge of $\mathbf{10}$'s/$\mathbf{\overline{5}}$'s that localize on a matter curve $\Sigma_{\mathbf{10}}^{(i)}/\Sigma_{\mathbf{\overline{5}}}^{(a)}$, the Dudas-Palti observation can be written as{\footnote{More generally, the anomaly cancellation argument of \cite{Marsano:2010sq} leads to a generalization of \eqref{DPintro} in which each integration over $\Sigma$ is multiplied by an integer $M_{\Sigma}$ that corresponds to the rank of a vector bundle associated to matter fields on $\Sigma$.  We can always absorb this factor by redefining the curve of integration as $\Sigma\rightarrow M_{\Sigma}\Sigma$ and this will be implicit in all of our formulae.  This is motivated in large part by the observation that $M_{\Sigma}$'s different than 1 in spectral cover models arise when the matter curve in the spectral cover is $M_{\Sigma}$-fold degenerate.}}
\begin{equation}
\sum_{\mathbf{10}\text{ matter curves, }i}q_i\int_{\Sigma_{\mathbf{10}}^{(i)}}F_Y = \sum_{\mathbf{\overline{5}}\text{ matter curves, }a}q_a\int_{\Sigma_{\mathbf{\overline{5}}}^{(a)}} F_Y \,,
\label{DPintro}\end{equation}
where $F_Y$ denotes the hypercharge flux. 
These relations have a physical origin in that they reflect the inability of hypercharge flux to contribute to mixed $U(1)$ anomalies in 4-dimensions \cite{Marsano:2010sq}.  As we shall review in section \ref{subsec:anomaly}, this property of hypercharge flux seems to be a general one that applies to all $SU(5)$ F-theory GUT models regardless of how they are constructed so that \eqref{DPintro} represents a general set of constraints.
This is important because \eqref{DPintro} has strong implications for model-building.  If we demand that the massless spectrum of our model is precisely that of the MSSM, \eqref{DPintro} implies that the only flavor-blind $U(1)$ symmetry we can engineer is the unique linear combination of $U(1)_Y$ and $U(1)_{B-L}$ that commutes with $SU(5)_{\rm GUT}$ and preserves the MSSM superpotential \cite{Dudas:2010zb,Marsano:2010sq}.  This $U(1)$, which we refer to as $U(1)_{\chi}$, has two related shortcomings: the operators responsible for generating a $\mu$ term and dimension 5 proton decay are both $U(1)_{\chi}$-invariant
\begin{equation}W_{\mu}\sim \mu H_uH_d\,,\qquad W_{\text{Dim }5}\sim \frac{1}{\Lambda}Q^3L \,.
\label{mudim5}\end{equation}

As is well-known, the $\mu$ and dimension 5 proton decay problems are not completely unrelated.  After all, $W_{\mu}$ and $W_{\text{Dim 5}}$ carry opposite charges under any $U(1)$ that commutes with $SU(5)_{\rm GUT}$ and preserves the MSSM superpotential.  This means we can address both by insisting that our model exhibit a $U(1)$ symmetry with respect to which $H_u$ and $H_d$ do not carry exactly opposite charges.  Such $U(1)$'s are typically referred to as $PQ$ symmetries and have played a prominent role in both local and global studies of F-theory GUTs in the past \cite{Marsano:2008jq,Heckman:2008qt,Marsano:2009wr}.  Whenever a $U(1)_{PQ}$ symmetry is present, though, \eqref{DPintro} forces the introduction of new charged exotics into the spectrum that do not come in complete GUT multiplets.  To make progress, one must find a suitable mechanism to lift these exotics and deal with the consequences of their non-GUT nature.  

\subsection{F-theory GUTs with $U(1)_{PQ}$: Generalities}

In this paper, we begin by exploring the implications of \eqref{DPintro} for generic F-theory GUT models beyond the simple statement that $U(1)_{PQ}$ symmetries necessitate exotics.  Of interest to us is the precise nature of the exotic spectrum and how \eqref{DPintro} constrains it.  While incomplete GUT multiplets will generically spoil unification, certain combinations of them will not.  If our exotic spectrum can be made to arise in such a combination, we can reap the benefits of having one or more $U(1)_{PQ}$ symmetries without paying any of the penalties that cause a muddling of the unification picture.

Along with giving us the set of constraints \eqref{DPintro}, general anomaly arguments guarantee that the charged exotics in $U(1)_{PQ}$ models come in vector-like pairs with respect to $SU(5)_{\rm GUT}$.  This means that they can all participate in cubic couplings to $SU(5)_{\rm GUT}$ singlets $X_i$ that carry  $U(1)_{PQ}$-charge.  If the singlets $X_i$ manage to pick up nonzero expectation values, these couplings can allow the exotics to be safely lifted from the spectrum.
The simplest possible scenario is one in which a single field $X$ is sufficient to lift all of the exotics.  When this happens the exotic spectrum can be parametrized by four integers $M$, $N$, $K$, and $L$ as
\begin{equation}\begin{array}{c|c|c}\label{KLMNLabels}
SU(5)\text{ origin} & \text{Exotic Multiplet} & \text{Degeneracy}  \\ \hline
& (\mathbf{1},\mathbf{1})_{+1}\oplus (\mathbf{1},\mathbf{1})_{-1} & M+N \\
\mathbf{10}\oplus\mathbf{\overline{10}} & (\mathbf{3},\mathbf{2})_{+1/6}\oplus (\mathbf{\overline{3}},\mathbf{2})_{-1/6} & M \\
& (\mathbf{\overline{3}},\mathbf{1})_{-2/3}\oplus (\mathbf{3},\mathbf{1})_{+2/3} & M-N\\ \hline
\mathbf{\overline{5}}\oplus \mathbf{5} & (\mathbf{\overline{3}},\mathbf{1})_{+1/3} \oplus (\mathbf{3},\mathbf{1})_{-1/3} & K \\
& (\mathbf{1},\mathbf{2})_{-1/2}\oplus (\mathbf{1},\mathbf{2})_{+1/2} & K-L \\ \hline
\end{array}\end{equation}
For this parametrization to make sense, we must have
\begin{equation}M\ge |N|\qquad K\ge 0\qquad K-L\ge 0\label{exoticparam}\end{equation}
As we will demonstrate, \eqref{DPintro} gives rise to a nontrivial relation between the singlet charge $q_X$ and a particularly important combination of these parameters
\begin{equation}q_{H_u}+q_{H_d} = q_X\Delta\label{qXintro} \,, \qquad \Delta\equiv N-L \,.\end{equation}
What makes $\Delta$ important is that it precisely measures the non-universal shifts of 1-loop MSSM $\beta$ functions that arise from letting the exotics run in the loop{\footnote{Equality of the first and second lines is just the statement that $5\delta b_1-3\delta b_2 - 2\delta b_3=0$ for exotics of the type \eqref{KLMNLabels}.  This was previously noted in \cite{Marsano:2009wr}.}}
\begin{equation}\begin{split}\Delta &= \delta b_2 - \delta b_3 \\
&= \frac{1}{6}\left(5\delta b_1+3\delta b_2-8\delta b_3\right) \,.
\end{split}\label{deltshift}\end{equation}
From this, we see that whenever $q_{H_u}+q_{H_d}\ne 0$, the net exotic spectrum is guaranteed to come in a combination that splits unification.  Note that this is stronger than simply saying that the exotics come in incomplete GUT multiplets.  It could have been that the net spectrum obtained by combining exotics from all matter curves satisfied $N=L$.  Interestingly, \eqref{qXintro} tells us that this nice situation is not compatible with the presence of a PQ symmetry. 

Because it determines the $U(1)_{PQ}$ charge of $X$, the relation \eqref{qXintro} also has implications for dimension 5 proton decay.  In general, when $X$ picks up an expectation value we expect $U(1)_{PQ}$ to be broken and $W_{\text{Dim }5}$ to be regenerated with a suppression that goes like $(\langle X\rangle / \Lambda)^n$ for some $n$.  In general we must take $\langle X\rangle$ fairly close to $\Lambda\sim M_{\rm GUT}$ in order to deal with unification issues.  If the charge of $X$ is chosen correctly, though, we can hope to recover some suppression of dimension 5 proton decay by making $n$ large.
Unfortunately, the relation \eqref{qXintro} between the $U(1)$ charge of $X$, $q_X$, and the combination $\Delta$ implies that the following operator is always $U(1)$-invariant
\begin{equation}\frac{1}{\Lambda}\int\,d^2\theta\,\left(\frac{X}{\Lambda}\right)^{\Delta}Q^3L \,.
\label{dim5operatorallowed}\end{equation}
In other words, the degree $n$ of the $X/\Lambda$ suppression is the same quantity $\Delta$ that measures the distortion of unification caused by the exotics.  Because of this, any manipulation that we introduce to raise this degree will necessarily change the exotic spectrum in a way that increases the non-universality of the 1-loop $\beta$ function shifts \eqref{deltshift}.  In this way, \eqref{qXintro} codifies a {\it general tension between unification and dimension 5 proton decay} for which no obvious solution is apparent{\footnote{If $\Delta < -1$ then the operator \eqref{dim5operatorallowed} cannot be generated because $\Lambda$ appears in the numerator.  In that case, however, the operator $\Lambda^{\Delta+1}\int\,d^2\theta\, X^{-\Delta}H_uH_d$ arises and generates a $\mu$ term that is far too large.}}.  It is a simple matter to generalize this analysis to situations in which multiple singlet fields $X_i$ are needed to lift all of the exotics.  As we demonstrate, the result is again a strong tension between unification and proton decay.

To summarize, \eqref{DPintro} introduces a tension between proton decay and unification by forcing incomplete GUT-multiplets into the spectrum.  The mass of these multiplets is set by the scale at which $U(1)_{PQ}$ is broken and hence by the scale at which $W_{\text{Dim 5}}$ is regenerated.  These statements have been known for some time.  What we have seen from a more detailed analysis of \eqref{DPintro}, however, is that it obstructs the most obvious ways to reduce this tension by tying the effective $U(1)$ charges of the exotic masses to the distortion of unification that they induce.  This reasoning is quite general in that it describes a tension but not necessarily the severity of that tension.  To see whether some combination of clever model building or tuning can alleviate it will require a closer look at unification, which is already known to be somewhat spoiled in F-theory GUT models by effects due to hypercharge flux \cite{Donagi:2008kj,Blumenhagen:2008aw}.  As has been suggested elsewhere \cite{Marsano:2009wr}, this may give us some room to maneuver by allowing the effects of hypercharge flux and exotics to counterbalance one another.  We hope to say more about this issue in the future \cite{wip}.

\subsection{F-theory GUTs with $U(1)_{PQ}$: Survey}
\label{subsec:SummarySurvey}

Before we can make a detailed study of how these problems may be dealt with, though, we first need to ask a more important question.  We expect the most desirable models to be those with the smallest $\Delta=N-L$ allowed by proton decay considerations but do we even know which $\Delta$'s can actually be realized?  Beyond general restrictions obtained from \eqref{DPintro}, exactly what exotic spectra can be achieved in explicit models?  To address this, we specialize to so-called spectral cover models where the rules for model-building are well-established.  

After reviewing the basic structure and context of spectral cover models, we begin with a technical derivation of \eqref{DPintro} in that setting.  We then establish that \eqref{DPintro} represents essentially the only constraint on the distribution of hypercharge flux by constructing spectral covers for all solutions to \eqref{DPintro} that engineer 1 or 2 $U(1)$ symmetries{\footnote{Actually, there are two other well-known constraints that supplement \eqref{DPintro}.  These are $\sum_{\mathbf{10}\text{ matter curves, }i}\int_{\Sigma_{\mathbf{10}}^{(i)}}F_Y=0$ and $\sum_{\mathbf{\overline{5}}\text{ matter curves },a}\int_{\Sigma_{\mathbf{\overline{5}}}^{(a)}}F_Y=0$ and they reflect the cancellation of the $U(1)_Y^3$ and $SU(2)^2U(1)_Y$ anomalies.}}
 {\footnote{We restrict to these cases because spectral covers with 3 $U(1)$ symmetries are significantly more complicated and require a substantial amount of 'topological tuning' to avoid non-Kodaira singularities at isolated points, to say nothing for non-Kodaira singularities along holomorphic curves.  We expect that no new constraints beyond \eqref{DPintro} control the hypercharge flux there as well but we do not prove it.  Spectral cover models with more than 3 $U(1)$ symmetries cannot have a large top Yukawa coupling so should not be considered.}}.
These build upon ideas used in the multiple $U(1)$ model of \cite{Kuflik:2010dg} and represent significant generalizations of the models in \cite{Dudas:2010zb}.  Each construction is specified up to a choice of holomorphic sections that must satisfy a number of assumptions in order to avoid the development of isolated singularities of non-Kodaira type.  We clearly state these assumptions and a sample choice of sections that satisfies them for each model.  We also note that these sample choices are consistent with the local topological data required to embed the models into full F-theory compactifications based on the geometry of \cite{Marsano:2009ym}.  Sample choices aside, though, the constructions of Appendix \ref{app:DPsols} are general enough to serve as a starting point for building explicit examples of any spectral cover model with a distribution of hypercharge flux that is consistent with \eqref{DPintro}.

Once we have established that there are no constraints on hypercharge flux beyond \eqref{DPintro}, we turn to a survey of the possible exotic spectra that can arise in spectral cover models with both a $U(1)_{\chi}$ and a $U(1)_{PQ}$ symmetry.  We always insist on having an unbroken $U(1)_{\chi}$ in order to protect against dimension 4 proton decay.  The $U(1)_{PQ}$ is allowed to be broken by the expectation value of precisely one singlet field, which we require to give mass to all exotic fields.  We perform the survey by looking at the spectrum induced by the most generic distribution of hypercharge flux allowed by \eqref{DPintro}{\footnote{We do not say anything about the condition $\int_{S_{\rm GUT}}F_Y\wedge F_Y = -2$ that is needed to avoid exotic $(\mathbf{3},\mathbf{2})_{-5/6}$'s or their conjugates from propagating along $S_{\rm GUT}$.  It will be necessary to impose this as an additional constraint on any explicit model.}} combined with a completely general distribution of bulk flux.
We do not impose any constraint on the bulk or '$\gamma$'-flux.  This is motivated by past experience with $\gamma$-fluxes in which no obvious global obstructions on their distribution emerged.  Despite our confidence to the contrary, one should keep in mind that obstructions might arise when realizing some of these models in practice.

Once we have parametrized the hypercharge and bulk fluxes, we proceed to enumerate all cases in which the expectation value of one singlet field, $X$, can lift everything except the matter content of the MSSM.  In the end, we find exactly 10 `models'.  By  'model' here we mean a specific identification of MSSM and exotic matter curves.  Within any such 'model' there is a parameter space of possible flux values that leads to some variance in the net spectrum of exotics.    
It is an empirical fact, however, that the combination $\Delta=N-L$ is fixed within each of the 10 'models' that we find.  We can summarize key properties of the `models' as follows
\begin{equation}\begin{array}{c|cc|c}
\text{Model Number} & \text{Exotic Spectra} & & \text{Dim 5} \\ \hline
1,\, 2,\,9 & N-L=1 & & XQ^3L/\Lambda^2 \\
3,\, 4 & N-L=2 & K\ge M & X^2Q^3 L / \Lambda^3 \\
5,\,6,\,7,\,8 & L=2 & M=N=0 & X^{\dag\,2}Q^3L /\Lambda^4 \\
10 & N-L=1 & K-L=M & XQ^3L/\Lambda^2
\label{surveysum}\end{array}\end{equation}
The model number refers to the detailed list in appendix \ref{app:Models}.  Note that the power of $X$ responsible for generating $W_{\text{Dim }5}$ is exactly $N-L$ in each case as we expected from general reasoning.  Quite interestingly, only low values of $N-L$ seem possible.  The maximal value $\Delta=2$ may not be enough to adequately suppress proton decay but, as we have said, the additional spoilage of unification by hypercharge flux may give us enough wiggle room to allow for smaller $\langle X\rangle$. The models we construct realise $\Delta = -2,1,2$ and it is curious to note that $N-L=-1$ does not seem to appear (as we have explained earlier $\Delta =0$ is ruled out by eq.~\eqref{qXintro}).  We have no good explanation for this fact. The possibility of achieving $\Delta=-1$ is rendered more interesting by the fact that this choice appears to be the most promising from the point of view achieving unification~\cite{wip}, though it is plagued by other phenomenological issues. Let us stress that any patterns observed in \eqref{surveysum} are intrinsic to the spectral cover formalism.
It may be possible that models constructed beyond this framework \cite{Cecotti:2010bp} can realize more general $N-L$ values.

Extending the survey to models with multiple singlet fields $X_i$ and possibly also multiple $U(1)_{PQ}$ symmetries could lead to interesting new possibilities.  From a local model-building perspective, each singlet vev $\langle X_i\rangle$ gives us an extra parameter to tune.  It can increase the likelihood of finding reasonable ranges of parameters but, on the flip side, will probably require the resulting model to be even more finely tuned.

\subsection{Outline}

The rest of this paper is organized as follows.  In section \ref{sec:DP}, we review the anomaly argument of \cite{Marsano:2010sq} for the Dudas-Palti relations \eqref{DPintro} and discuss its general implications for the spectrum of exotics in F-theory GUTs with $U(1)_{PQ}$ symmetries.  We then turn to spectral cover models in section \ref{sec:sc}.  There, we review the spectral cover framework, provide a technical derivation of \eqref{DPintro} in that setting, and present in more detail the results of the survey of spectral cover models with 2 $U(1)$ symmetries.  The details of the survey, along with the construction of spectral covers that realize generic solutions to \eqref{DPintro}, are contained in the Appendices along with other supplemental calculations.


\section{Dudas-Palti Relations and General Implications}
\label{sec:DP}

In this section, we describe some general features of F-theory GUT models that engineer $U(1)$ symmetries as a means of controlling the structure of 4-dimensional physics.  We focus in particular on models that exhibit two key features
\begin{itemize}
\item Internal ``hypercharge flux" for breaking $SU(5)_{\rm GUT}\rightarrow SU(3)\times SU(2)\times U(1)_Y$

\item One or more $U(1)$ symmetries that commute with $SU(5)_{\rm GUT}$ and do not distinguish particles from different generations{\footnote{That our $U(1)$'s are not `family' symmetries is a necessary condition for realizing the flavor scenarios of \cite{Heckman:2008qa,Bouchard:2009bu,Cecotti:2009zf}.}}
\end{itemize}
At the heart of our considerations is a set of relations first noted by Dudas and Palti \cite{Dudas:2010zb} in the context of so-called 'spectral cover models'.  The applicability of these relations to F-theory GUTs in general relies on the observation of \cite{Marsano:2010sq} that they reflect the inability of ``hypercharge flux" to influence 4-dimensional mixed gauge anomalies.  After reviewing this argument, we describe the general implications of the Dudas-Palti relations in models with $U(1)_{PQ}$ symmetries.  These include the existence of charged exotics that introduce a generic tension between unification and proton longevity.  We stress that the results of this section are quite general in that they rely only on the anomaly cancellation argument of \cite{Marsano:2010sq}.  Because of this, the structure and constraints that we derive should arise in any F-theory GUT model with $U(1)$ symmetries regardless of how it might be constructed{\footnote{We are aware of only two loopholes to this.  The first is the possibility that some new mechanism can be found to allow ``hypercharge flux" to influence 4-dimensional anomalies without also generating a $U(1)_Y$ mass.  No known set of couplings in F-theory are capable of doing this and the absence of any spectral cover models that violate \eqref{DP} strongly suggests, to us anyway, that this possibility is not realized.  The second is the possibility that our description of the spectrum of $\mathbf{10}$'s and $\mathbf{\overline{5}}$'s is too limited.  We assume the structure obtained from smooth $\mathbf{10}$ and $\mathbf{\overline{5}}$ matter curves but it may be possible to engineer more exotic combinations of matter if, for instance, matter curves and/or $S_{\rm GUT}$ itself exhibit exotic singular behaviors that go beyond the types that have been considered so far.}}.
In the next section, we will see explicit realizations of this structure in our survey of spectral cover models with multiple $U(1)$ symmetries.

\subsection{Anomaly Cancellation}
\label{subsec:anomaly}

It has been known for some time that our ability to distribute ``hypercharge flux" along the matter curves of an F-theory GUT is limited in models that engineer extra $U(1)$ symmetries.  This structure can be nicely described by a set of relations first observed by Dudas and Palti \cite{Dudas:2010zb} in the context of spectral cover models, whose structure we shall review in the next section.  Letting $q_i/q_a$ denote the common $U(1)$ charge of $\mathbf{10}/\mathbf{\overline{5}}$ fields on a matter curve $\Sigma_{\mathbf{10}}^{(i)}/\Sigma_{\mathbf{\overline{5}}}^{(a)}$, the Dudas-Palti observation can be written as \cite{Dudas:2010zb}
\begin{equation}
\sum_{\mathbf{10}\text{ matter curves, }i}q_i\int_{\Sigma_{\mathbf{10}}^{(i)}}F_Y = \sum_{\mathbf{\overline{5}}\text{ matter curves, }a}q_a\int_{\Sigma_{\mathbf{\overline{5}}}^{(a)}} F_Y \,.
\label{DP}\end{equation}
It was demonstrated in \cite{Marsano:2010sq} that this simple set of relations is related to the physics of 4-dimensional anomaly cancellation.  Because of this, we expect \eqref{DP} to have general applicability to F-theory GUT models beyond the spectral cover examples that initially motivated it.

For completeness, let us review how \eqref{DP} arises from considerations of 4-dimensional anomaly cancellation.  The situation is particularly simple if we study F-theory compactifications on Calabi-Yau 4-folds that do not require the introduction of $G$-flux to satisfy the quantization condition of \cite{Witten:1996md}.  In that case, we can consider adding ``hypercharge flux" into the game and no additional bulk $G$-flux.  By construction, the ``hypercharge" flux does not induce a mass for $U(1)_Y$ or, in fact, any other $U(1)$ symmetries that we might engineer.  As a result, it cannot induce any gauge anomalies.  Of particular interest to us are mixed gauge anomalies with insertions of both MSSM and $U(1)$ currents as these anomalies only receive contributions from the charged fields that localize along matter curves.
A simple calculation reveals that \eqref{DP} is just the condition that one needs to ensure that these mixed anomalies cancel \cite{Marsano:2010sq}.

Things are slightly more tricky if we are forced to introduce a background $G$-flux to satisfy the quantization condition of \cite{Witten:1996md} because that flux on its own will induce mixed anomalies while lifting our $U(1)$'s through the St\"uckelberg mechanism.  We can argue as before, however, by noting that when we add ``hypercharge flux" to the game it will not change any of these anomalies.  The reason for this has its origin in the 4-dimensional Green-Schwarz mechanism, which cancels any gauge anomalies involving our extra $U(1)$'s that might be present before we integrate them out.  The basic ingredient of this mechanism is a 2-form axion $c_2$ that participates in 4-dimensional couplings of the form
\begin{equation}\int\,c_0\,\,F\wedge F + c_2\wedge F\label{4dGS}\end{equation}
where $F$ here can denote either a $U(1)$ field strength or any of the MSSM field strengths and $dc_0 = *dc_2$.  Because $c_2$ couples both linearly and quadratically to flux, it can propagate within tree level diagrams that can cancel nontrivial contributions to gauge anomalies that arise from triangle diagrams.  The $c_2\wedge F$ term is also important for a related reason; if it is present for a given field strength, the corresponding $U(1)$ will become massive via the St\"uckelberg mechanism.  Among our most important requirements, then, is the absence of any coupling like $c_2\wedge F_Y$ that could lift the $U(1)_Y$ gauge boson.  

In F-theory, the only way to get a 2-form axion in 4-dimensions of the type in \eqref{4dGS} is from the reduction of the RR 4-form $C_4$ which, as the potential associated to D3-branes, is nicely $SL(2,\mathbb{Z})$-invariant.  This 4-form is an honest bulk field whose coupling to flux is well-known
\begin{equation}\int_{Y_4}\,C_4\wedge G\wedge G\label{CGcoup}\end{equation}
where the integration is over our entire Calabi-Yau 4-fold, $Y_4$.  In the presence of a stack of 7-branes, like the one that gives us $SU(5)_{\rm GUT}$, we can rewrite the specific contribution to \eqref{CGcoup} that involves localized worldvolume fluxes as the familiar coupling
\begin{equation}\int_{\text{7-brane worldvolume}} C_4\wedge F\wedge F\label{branecoup}\end{equation}
When we have an internal ``hypercharge flux", we always generate the dangerous $c_2\wedge F_Y$ coupling by reducing $C_4$ along 2-forms in $Y_4$ unless $F_Y$ happens to be orthogonal to all such 2-forms.  We can ensure this orthogonality by choosing $F_Y$ to be a $(1,1)$-form that is dual to a holomorphic curve $[F_Y]$ in $S_{\rm GUT}$ that is trivial in the homology of $Y_4$ \cite{Beasley:2008kw,Donagi:2008kj}.  When $F_Y$ is of this type, though, any insertion of $F_Y$ into \eqref{branecoup} gives a vanishing result so that $C_4$ has no direct coupling at all to this flux.  This means that the coefficients of \eqref{4dGS} in the 4-dimensional theory are the same whether we introduce the ``hypercharge flux" or not.  It follows that the mixed 4-dimensional gauge anomalies also cannot change when we turn on ``hypercharge flux" and a simple calculation demonstrates that \eqref{DP} is nothing other than a mathematical statement of this condition \cite{Marsano:2010sq}.

\subsection{`Uniqueness' of $U(1)_{B-L}$}

To see why \eqref{DP} is so constraining, let us recall how ``hypercharge flux" impacts the 4-dimensional spectrum.  Charged fields in the $\mathbf{10}$ or $\mathbf{\overline{5}}$ localize along curves $\Sigma$ in the internal space and the number of chiral zero modes in the $SU(3)\times SU(2)\times U(1)_Y$ representation $R$ is determined by an index theorem
\begin{equation}n_R-n_{\overline{R}}=\int_{\Sigma}c_1(V_{\Sigma}\otimes L_Y^{Y_R})=\int_{\Sigma}\left[c_1(V_{\Sigma})+M_{\Sigma}\,c_1(L_Y^{Y_R})\right]\end{equation}
where $V_{\Sigma}$ is a bundle of rank $M_{\Sigma}$ that encodes the ``bulk" $G$-flux, $c_1(L_Y)$ is roughly the ``hypercharge flux", and $Y_R$ is the $U(1)_Y$ charge of fields in the representation $R$.  Because $V_{\Sigma}$ and $M_{\Sigma}$ are intrinsic properties of the matter curve $\Sigma$, the only way to distinguish different $SU(3)\times SU(2)\times U(1)_Y$ representations within an $SU(5)$ multiplet is through the ``hypercharge flux".  To get a model with exactly 2 Higgs doublets, $H_u$ and $H_d$, and a set of matter fields that comprise complete $SU(5)$ multiplets with common $U(1)$ charges we need a distribution of ``hypercharge flux" in which
\begin{equation}
\int_{{\Sigma}_{H_u}}F_Y = +1 \,,\qquad 
\int_{\Sigma_{H_d}}F_Y=-1 \,,\qquad \int_{\text{any other matter curve }\Sigma'}F_Y = 0
\end{equation}
and $M_{\Sigma_{H_u}}=M_{\Sigma_{H_d}}=1$.  The DP relations \eqref{DP} now tell us that any $U(1)$ symmetry consistent with this distribution of ``hypercharge flux" must satisfy $q_{\Sigma_{H_u}}-q_{\Sigma_{H_d}}=0$.  Because $q_{\Sigma_i}$ denotes the common charge of fields from the $\mathbf{\overline{5}}$ on $\Sigma_i$, we see that the charge of $H_u$ is $q_{H_u}=-q_{\Sigma_{H_u}}$ and hence that
\begin{equation}q_{H_u}+q_{H_d}=0 \,.
\label{BLcond}\end{equation}
Up to normalization, there is only one $U(1)$ symmetry that satisfies \eqref{BLcond}, commutes with $SU(5)$, and preserves the MSSM superpotential.  That symmetry is the unique linear combination of $U(1)_Y$ and $U(1)_{B-L}$ that commutes with $SU(5)$, which we denote by $U(1)_{\chi}$
\begin{equation}\begin{array}{c|cccc}
 & \mathbf{10}_M & \mathbf{\overline{5}}_M & H_u & H_d \\ \hline
U(1)_{\chi} & 1 & -3 & -2 & 2
\end{array}\end{equation}
This argument is essentially the same one used in \cite{Dudas:2010zb} to explain why none of their models had an unbroken $U(1)_{PQ}$.  As \cite{Dudas:2010zb} further emphasize, we will face two important phenomenological problems in general if $U(1)_{\chi}$ represents the only control that we have over the theory.  These are, respectively, the $\mu$ problem and dimension 5 proton decay, which are associated with the operators
\begin{equation}\mu\int\,d^2\theta\, H_uH_d\qquad\text{ and }\qquad \frac{1}{\Lambda}\int\,d^2\theta\,Q^3L \,.
\end{equation}
These operators carry opposite charges under any $U(1)$ symmetry that preserves the MSSM superpotential so their fates are related.  In the absence of a symmetry to indicate otherwise, we expect $\mu$ to be large and $\Lambda^{-1}Q^3L$ to be generated with $\Lambda\sim M_{\rm GUT}$ by massive Kaluza-Klein modes at the GUT scale without further suppression.

\subsection{$U(1)_{PQ}$ Symmetries and Exotics}
\label{subsec:PQ}

Because $H_uH_d$ and $Q^3L$ carry opposite charges under any $U(1)$ symmetry that preserves the MSSM superpotential, we can try to take care of both problems simultaneously by introducing one or more new $U(1)$'s that satisfy
\begin{equation}q_{H_u}+q_{H_d}\ne 0\end{equation}
Symmetries of this type, which we refer to as $PQ$ symmetries, have played a prominent role in past studies of F-theory models \cite{Marsano:2008jq, Heckman:2008qt,Marsano:2009wr}.  From the DP relations \eqref{DP}, though, we know that the presence of such a symmetry will not allow a distribution of ``hypercharge flux" whose spectrum of non-GUT fields consists of exactly one pair of Higgs doublets.  If we insist on engineering all MSSM matter fields as complete GUT multiplets with common $U(1)$ charges, we will be forced to introduce new non-GUT multiplets in addition to the Higgs doublets.  These new fields represent charged exotics that must be dealt with in some way.

Anomaly considerations allow us to make an important observation about the structure of the new non-GUT fields.  In particular, the cancellation of $U(1)_Y^3$ anomalies in the presence of ``hypercharge flux" leads to the standard relations
\begin{equation}\sum_{\mathbf{10}\text{ matter curves, }i}\int_{\Sigma_{\mathbf{10}}^{(i)}}F_Y=0\,,\qquad 
\sum_{\mathbf{\overline{5}}\text{ matter curves, }a}\int_{\Sigma_{\mathbf{\overline{5}}}^{(a)}}F_Y = 0\end{equation}
which forces the non-GUT exotics to come in vector-like pairs with respect to $SU(3)\times SU(2)\times U(1)_Y$.  Though they will not be vector-like with respect to our additional $U(1)$ symmetries, in general, this fact allows them to participate in cubic couplings with MSSM singlets that carry appropriate $U(1)$ charges
\begin{equation}W\supset X_i f_{\text{Exotic,}i}\overline{f}_{\text{Exotic,}i}\label{cubicmass}\end{equation}
If enough singlets $X_i$ pick up nonzero expectation values then the exotics can be safely lifted from the spectrum.

Even though we can remove the exotics in this way, we must be aware of two potential phenomenological problems that can arise \cite{Dudas:2010zb}.  The first is that the expectation values required to lift them may lead to a stronger breaking of $U(1)$ symmetries than we can allow.  These $U(1)$'s are expected to become massive from the St\"uckelberg mechanism, which always allows for the possibility that $U(1)$-violating couplings are generated by nonperturbative effects.  One typically assumes that these violations are acceptably small.  Expectation values of scalar fields, on the other hand, can lead to larger violations depending on the physics that drives them.

The second problem is that our charged exotics, even when lifted from the zero mode spectrum, can still make significant contributions to the 1-loop $\beta$ functions for the MSSM gauge couplings that spoil unification depending on how massive they become.  On general grounds, we expect a tension between proton decay, which should favor depressed $U(1)_{PQ}$-breaking with small exotic masses, and unification, which should favor large exotic masses.

\subsubsection{The Exotic Spectrum}

We could in principle solve both of these problems if given enough control over the exotic spectrum.  Even though the exotics on each individual matter curve will give non-universal shifts of the 1-loop $\beta$ functions, it could be that the net shift of all exotics happens to be universal.  To see that this is not obviously far-fetched, consider the generic spectrum on a $\mathbf{10}$ or $\mathbf{\overline{5}}$ matter curve.  Using index theory, the zero mode chiralities are seen to follow a simple pattern \cite{Donagi:2008ca,Beasley:2008dc}
\begin{equation}
\mathbf{10}\text{ matter curve }\Sigma_{\mathbf{10}}^{(i)}\quad \leftrightarrow \quad\begin{array}{ll} n_{(\mathbf{1},\mathbf{1})_{+1}}-n_{(\mathbf{1},\mathbf{1})_{-1}} &= m_i+n_i \\ 
n_{(\mathbf{3},\mathbf{2})_{+1/6}}-n_{(\mathbf{\overline{3}},\mathbf{2})_{-1/6}} &= m_i \\ 
n_{(\mathbf{\overline{3}},\mathbf{1})_{-2/3}}-n_{\mathbf{3},\mathbf{1})_{+2/3}}&= m_i-n_i \\ 
\end{array}\label{10param}\end{equation}

\begin{equation}\mathbf{\overline{5}}\text{ matter curve }\Sigma_{\mathbf{\overline{5}}}^{(a)}\quad \leftrightarrow\quad \begin{array}{ll}n_{(\mathbf{\overline{3}},\mathbf{1})_{+1/3}}-n_{(\mathbf{3},\mathbf{1})_{-1/3}} &= k_a \\
n_{(\mathbf{1},\mathbf{2})_{-1/2}}-n_{(\mathbf{1},\mathbf{2})_{+1/2}} &= k_a-\ell_a \\ 
\end{array}\label{5barparam}\end{equation}
This parametrization is useful because the $n_i$ and $\ell_a$ correspond roughly to the net ``hypercharge flux" threading the matter curves $\Sigma_{\mathbf{10}}^{(i)}$ and $\Sigma_{\mathbf{\overline{5}}}^{(a)}$, which are directly constrained by \eqref{DP}.  By a standard calculation, one can see that the combined spectrum from one $\mathbf{10}$ matter curve and one $\mathbf{\overline{5}}$ matter curve gives a universal shift to the 1-loop MSSM $\beta$ functions provided $n_i=\ell_a$.  When $n_i,\ell_a\ne 0$, the spectrum is comprised of incomplete GUT multiplets but nevertheless the special choice $n_i=\ell_a$ eliminates any (1-loop) distortion of unification that they might have caused, provided of course that their masses are nearly degenerate.

It is therefore not obvious at all that the non-GUT exotics we get from introducing a $U(1)_{PQ}$ symmetry have to cause a problem for unification.  We might have to introduce a small tuning to make sure their masses do not differ by very much but, even then, we could hope to make this automatic without excessive tuning by ensuring that one singlet vev is sufficient to lift everything.

The question now is whether we have enough control over the exotic spectrum to make a scenario like this work.  Because the Dudas-Palti relations \eqref{DP} restrict our ability to distribute ``hypercharge flux", they will have something to say about the feasibility of this approach.  To study their effect, we must first specify a convenient parametrization for the exotics and their masses.  We assume at the outset that all exotics are lifted so we can start by identifying a subset of all cubic terms of the form \eqref{cubicmass} in which each exotic field couples to exactly one singlet.  Let us call these the `initial masses' and any remaining couplings the `mixings':  
\begin{equation}W_{\text{masses}}\sim W_{\text{`initial' mass}}+W_{\text{mixings}}\label{genmass}\end{equation}
This identification allows us to break the exotics into groups according to the singlet that provides their `initial' mass.  Such a separation is already useful but it will become more so if we can use the singlet vevs in $W_{\text{`initial' mass}}$ as an order of magnitude estimate for the actual exotic masses.
This would be fine if there were no mixing terms but, in the presence of the additional couplings, $W_{\text{mixings}}$, this is not always possible.  To illustrate this simple point and how we deal with it, consider a collection of 2 vector-like pairs of exotics, $f_1/\bar{f}_1$ and $f_2/\bar{f}_2$ with couplings
\begin{equation}W\sim X f_1\bar{f}_2 + Y (f_1\bar{f}_1+f_2\bar{f}_2)\end{equation}
In this case, we would have
\begin{equation}W_{\text{`initial' mass}}\sim Y(f_1\bar{f}_1+f_2\bar{f}_2)\qquad W_{\text{mixings}}\sim Xf_1\bar{f}_2 \,.
\end{equation}
Based on $W_{\text{`initial' mass}}$ alone, we would like to say that $\langle Y\rangle$ determines all exotic masses.  Even in the presence of $W_{\text{mixings}}$, this is a fine assumption in the absence of tuning provided $\langle X\rangle$ doesn't get too large.  If we have $\langle X\rangle \gg \langle Y\rangle$, however, we get a seesaw in the mass matrix and $\langle Y\rangle$ alone doesn't accurately capture any of the exotic masses.  We can nevertheless obtain an effective form for the exotic masses that involves no mixings by diagonalizing the mass matrix to leading order in $\langle Y\rangle / \langle X\rangle$.  The leading order mass eigenstates remain $f_1/\bar{f}_1$ and $f_2/\bar{f}_2$ but we find effective masses
\begin{equation}W_{\text{eff}}\sim X f_1\bar{f}_2 + \frac{Y^2}{X}f_2\bar{f}_1 + \ldots \,.
\end{equation}
If we like, we can define a new quantity $Z=\frac{Y^2}{X}$ and write a set of mass couplings of the form \eqref{genmass} that includes only `initial' masses and no mixing terms.

More generally, the identification of `initial' masses with actual exotic masses is reasonable, at least as an order of magnitude estimate, provided there are no large hierarchies between singlet vevs appearing in $W_{\text{`initial' mass}}$ and $W_{\text{mixing}}$.  In the presence of such hierarchies, though, we can always diagonalize the mass matrix to leading order, in which case we obtain an effective set of mass couplings of the form \eqref{genmass} with no mixings and such that all `masses' are rational functions of our initial singlet fields{\footnote{The hierarchical separation is needed to ensure that the masses are rational functions of the initial singlets.}}.

In what follows, we will therefore always assume a mass term structure of the form \eqref{genmass} with $W_{\text{mixings}}=0$ where the singlets $X_i$ may be rational functions of fundamental singlets rather than fundamental singlets themselves.  Once we have done this, we can associate to each $X_i$ the collection of vector-like pairs of exotics that it lifts.  A convenient parametrization for the fields of that collection is the following
\begin{equation}\begin{split}
n_{(\mathbf{1},\mathbf{1})_{+1}}+n_{(\mathbf{1},\mathbf{1})_{-1}} &= M_i+P_i \\
n_{(\mathbf{3},\mathbf{2})_{-1/6}}+n_{(\mathbf{\overline{3}},\mathbf{2})_{+1/6}} &= M_i \\
n_{(\mathbf{\overline{3}},\mathbf{1})_{-2/3}}+n_{(\mathbf{3},\mathbf{1})_{+2/3}} &= M_i-N_i \\
n_{(\mathbf{\overline{3}},\mathbf{1})_{+1/3}}+n_{(\mathbf{3},\mathbf{1})_{-1/3}} &= K_i \\
n_{(\mathbf{1},\mathbf{2})_{-1/2}} + n_{(\mathbf{1},\mathbf{2})_{+1/2}} &= K_i-L_i \,.
\end{split}\label{exoticparametrization}\end{equation}
In other words, we say that $X_i$ gives an `initial' mass to $K_i$ triplet pairs of exotics, $K_i-L_i$ doublet pairs, and so on.  One might naively think that we should set $P_i=N_i$ based on \eqref{10param} but, as explained in Appendix \ref{app:exotics}, this need not be the case.

With this parametrization, we can study the implications of the DP relations \eqref{DP} on the exotic spectrum and the singlets that can remove them.  As we show in Appendix \ref{app:exotics}, the result actually takes a fairly simple form
\begin{equation}q_{H_u}+q_{H_d} = \sum_{\text{Singlets }X_i}q_{X_i}(N_i-L_i) = \sum_{\text{Singlets }X_i}q_{X_i}(P_i-L_i)\label{DPimplications}\,.\end{equation}
Implicit in this equation is the identity
\begin{equation}\sum_{\text{Singlets }X_i}q_{X_i}(N_i-P_i)=0\,,\end{equation}
whose origin is the fact that the exotic spectrum on each $\mathbf{10}$ matter curve is controlled by two integers rather than three \eqref{10param}.

\subsubsection{Proton Decay and Unification}

Using the parametrization \eqref{exoticparametrization} and the constraint \eqref{DPimplications}, we can now make the general tension between unification and proton decay quite precise.  We start by computing the 1-loop $\beta$ function shifts that are induced by the collection of exotics whose mass is set by the singlet $X_i$
\begin{equation}\begin{split}\delta b_{1,i} &= 3M_i + K_i + \frac{1}{5}\left(6P_i - 8N_i-3L_i\right) \\
\delta b_{2,i} &= 3M_i + K_i - L_i \\
\delta b_{3,i} &= 3M_i-N_i+K_i \,.
\end{split}\end{equation}
Because we have 3 $\beta$ functions there is, in general, a 2-parameter family of distortions to unification.  The above results demonstrate that $N_i-L_i$ and $P_i-L_i$ correspond to one particularly simple choice of basis for this family
\begin{equation}N_i-L_i = \delta b_{2,i}-\delta b_{3,i}\equiv \Delta_i^{(N)}\qquad P_i-L_i = \frac{1}{6}\left(5\delta b_{1,i} + 3\delta b_{2,i} - 8\delta b_{3,i}\right)\equiv \Delta_i^{(P)}  \,.
\label{betadistsNLP}\end{equation}
Already from \eqref{DPimplications} and \eqref{betadistsNLP} we can see that models with a $U(1)_{PQ}$ can never realize our dream of engineering a set of exotics that are lifted by just one singlet field, $X$, and induce a universal shift of the 1-loop $\beta$ functions.  In the case of one singlet, \eqref{DPimplications} and \eqref{betadistsNLP} simplify to
\begin{equation}q_{H_u}+q_{H_d} = q_X (N-L)\,,\qquad N=P\label{DPimplicationsoneX}\end{equation}
\begin{equation}\Delta = N-L=P-L=\delta b_{2}-\delta b_{3} = \frac{1}{6}\left(5\delta b_{1} + 3\delta b_{2} - 8\delta b_{3}\right) \,.\end{equation}
Because $q_{H_u}+q_{H_d}\ne 0$ and $q_X\ne 0$, we see that $N\ne L$ and hence that a distortion of unification is always introduced.  Further, a second consequence of \eqref{DPimplicationsoneX} is that the following operator is always gauge invariant
\begin{equation}\frac{1}{\Lambda}\int\,d^2\theta\,\left(\frac{X}{\Lambda}\right)^{\Delta}\,Q^3L\label{dim5oneX}\,.\end{equation}
While we might want to minimize the effect on unification by taking $\langle X\rangle$ to sit near the unification scale $\Lambda\sim M_{KK}\sim M_{\rm GUT}$, we see that this effectively eliminates any hope we might have for suppressing dimension 5 proton decay.  What we have found is an intrinsic tension between unification and proton decay that seems unavoidable.

Let us now return to the possibility that multiple singlets $X_i$ play a role in lifting the exotics.  For this, it is useful to introduce separate notation for the the sum over $\Delta_i^{(N)}$'s or $\Delta_i^{(P)}$'s
\begin{equation}\begin{split}\Delta^{(N)} &= \sum_i\Delta_i^{(N)} \\
\Delta^{(P)} &= \sum_i \Delta_i^{(P)}  \\ 
\end{split}\end{equation}
Just as \eqref{DPimplicationsoneX} guaranteed the gauge invariance of \eqref{dim5oneX} in the case of one singlet $X$, the more general condition \eqref{DPimplications} implies that the following operators are always invariant under all $U(1)$ symmetries
\begin{equation}{1\over \Lambda^{\Delta^{(N)}}}\int\,d^4\theta\,
\prod_i \left[X_i^{\dagger \Delta_i^{(N)}}\right]\,H_uH_d,\qquad \frac{1}{\Lambda^{\Delta^{(N)}+1}}\int\,d^2\theta\,\left[\prod_i X_i^{\Delta_i^{(N)}}\right]\,Q^3L\label{allowedoperators}\end{equation}
and similar for $N\leftrightarrow P$.  If we want to achieve an exotic spectrum that sits at a common mass and leads to universal shifts of the 1-loop MSSM $\beta$ functions, then we must have $\Delta^{(N)}=\Delta^{(P)}=0$.  In this case, however, any suppression of dimension 5 proton decay operators comes from ratios of the singlet vevs that determine the exotic masses in the first place.  Introducing hierarchies to achieve a suppression willÊ
necessarily split the exotic masses into groups that each have $\Delta_i^{(N)}\ne 0$.  Unification will be spoiled here because each such collection induces $\beta$ function distortions that are individually non-universal and the energy ranges over which they run are hierarchically separated.

In the end, then, we find a general tension between unification and proton decay.  That is not to say that the issue cannot be circumvented through a combination of clever model-building and tuning.  The most obvious solution, however, would have been to engineer non-GUT multiplets that combine to yield universal $\beta$ function shifts.  Sadly, it seems that this cannot be done without allowing unsuppressed dimension 5 proton decay.


\section{Spectral Cover and Dudas-Palti Relations}
\label{sec:sc}

So far, we have encountered a number of general constraints on the structure of F-theory GUTs with $U(1)$ symmetries.  It remains to be seen, however, whether additional surprises might be waiting for us when we start to  look at concrete models.  Several examples have been constructed in recent years and, in all cases, they can be successfully studied with spectral cover techniques.  In the remainder of this paper, we perform a systematic study of such spectral cover models with a particular focus on the allowed spectrum of exotics.  We start by reviewing the conventional methods for building global F-theory GUTs in the current literature and then provide a direct technical derivation of the Dudas-Palti relations \eqref{DP} in that setting.  We then turn to a general survey of spectral cover models that exhibit one or two $U(1)$ symmetries and the most generic distributions of hypercharge flux  that are consistent with Dudas-Palti.  In Appendix \ref{app:DPsols}, we provide `in principle' constructions for all such models that can be promoted to complete local and global models by making a suitable choices of sections and using the procedure outlined below to embed them into honest F-theory compactifications.

\subsection{Calabi-Yau 4-folds and Higgs Bundles}

A complete F-theory GUT model is specified by the geometry of an elliptically fibered Calabi-Yau 4-fold $Y_4$ along with a set of fluxes that controls the chiral spectrum.  Though several different constructions have been achieved in the literature \cite{Andreas:2009uf,Marsano:2009ym,Collinucci:2008zs,Collinucci:2009uh,Blumenhagen:2009up,Marsano:2009gv,Blumenhagen:2009yv,Marsano:2009wr,Grimm:2009yu,Cvetic:2010rq,Chen:2010tp,Chen:2010ts,Chung:2010bn,Chen:2010tg,Knapp:2011wk}, all of them follow roughly the same basic strategy.  We can start with a 3-fold $B_3$ that will serve as the IIB compactification geometry and identify a holomorphic divisor $S_{\rm GUT}$ inside $B_3$ that will support the charged degrees of freedom of our $SU(5)$ GUT.  Letting $z$ denote the holomorphic section on $B_3$ whose vanishing defines $S_{\rm GUT}$, we can then build our Calabi-Yau 4-fold $Y_4$ by specifying a `Tate model' for the elliptic fibration
\begin{equation}y^2 = x^3 + a_0z^5 + a_2 xz^3 + a_3 yz^2 + a_4 x^2z + a_5 xy\label{TateModel}\,,\end{equation}
where $x$ and $y$ are sections of ${\cal{O}}(-2K_{B_3})$ and ${\cal{O}}(-3K_{B_3})$, respectively, and the $a_m$ are sections of ${\cal{O}}((m-6)K_{B_3}+(m-5)S_{\rm GUT})$.  By construction, \eqref{TateModel} exhibits the $SU(5)$ singularity that we need along $S_{\rm GUT}$ to realize our $SU(5)_{\rm GUT}$ gauge group.  Charged matter in the $\mathbf{10}$ and $\mathbf{\overline{5}}$ representations can then be found along 'matter curves' of $S_{\rm GUT}$ where the singularity type enhances to $SO(10)$ or $SU(6)$
\begin{equation}SO(10):\quad z=a_5=0\qquad  \quad SU(6):\quad z= a_0 a_5^2  -a_2 a_3 a_5 +a_3^2 a_4 =0\label{mattcurvesF}\,.\end{equation}
  Using insight from Heterotic duality, a set of $G$-fluxes for engineering chiral matter was recently identified in \cite{Marsano:2010ix} and a formalism for determining the chiral spectrum that they generate was presented.  With this toolbox, we can determine the full matter content and symmetry structure of the model directly in F-theory from the choice of sections $a_m$ and the collection of $G$-fluxes that are present.  When the $a_m$ are generic, we find a single $\mathbf{10}$ matter curve, a single $\mathbf{\overline{5}}$ matter curve, and no symmetry other than $SU(5)_{\rm GUT}$ to control the physics.

Since the charged degrees of freedom localize near $S_{\rm GUT}$, it should not be necessary to talk about the full compactification geometry to describe them.  Rather, their low energy physics can be captured by an 8-dimensional gauge theory that propagates along $\mathbb{R}^{3,1}\times S_{\rm GUT}$ and this, in turn, depends only on the geometry (and fluxes) at small $z$.  We can explicitly present the local geometry there by restricting each of the sections $a_m$ on $B_3$ to corresponding sections $b_m$ on $S_{\rm GUT}$
\begin{equation}a_m|_{S_{\rm GUT}} = b_m\end{equation}
and writing
\begin{equation}y^2 = x^3 + b_0 z^5 + b_2xz^3 + b_3yz^2 + b_4x^2z+b_5xy\label{localmodel} \,.\end{equation}
The $b_m$'s make their appearance in the local gauge theory description of the physics as parameters that control a Higgs bundle \cite{Hayashi:2009ge,Donagi:2009ra}.  A heuristic way to see this is by noting that the geometry \eqref{localmodel} appears to arise from an $E_8$ singularity at generic points that is unfolded by the $b_m$'s to $SU(5)_{\rm GUT}$.  This unfolding has a natural worldvolume interpretation: we start with an $E_8$ gauge theory on $\mathbb{R}^{3,1}\times S_{\rm GUT}$ and the $b_m$'s (along with the fluxes) specify a nontrivial configuration for the adjoint scalar $\phi$ (and gauge connection $A$) that breaks $E_8\rightarrow SU(5)_{\rm GUT}$.  The resulting Higgs bundle takes values in the $SU(5)_{\perp}$ commutant of $SU(5)_{\rm GUT}$ inside $E_8$ and the $b_m$'s are nothing more than the Casimirs of $\phi$
\begin{equation}b_m\sim b_0\text{tr}\phi^m\,.\end{equation}
In other words, the $b_m$'s tell us about the spectral data, or eigenvalues, of $\phi$ as we move along $S_{\rm GUT}$.  When the spectral data is generic, it is actually sufficient to determine the entire $\phi$ configuration up to gauge equivalence \cite{Cecotti:2010bp}.

An important check on this whole story is that the gauge theory description reproduces the matter curves, symmetries, and chiral spectrum that we expect from a direct analysis of the F-theory compactification.  To see how $\phi$ influences the spectrum, we first recall that all matter descends from the adjoint of $E_8$, whose decomposition under $E_8\rightarrow SU(5)_{\rm GUT}$ can be succinctly written in terms of $SU(5)_{\rm GUT}\times SU(5)_{\perp}$ representations as
\begin{equation}\mathbf{248}\rightarrow (\mathbf{24},\mathbf{1})\oplus (\mathbf{1},\mathbf{24})\oplus (\mathbf{10},\mathbf{5})\oplus (\mathbf{\overline{5}},\mathbf{10})\label{248decomp}\,.\end{equation}

The nontrivial configuration of $\phi$ generates masses for the bifundamental fields of this reduction that vary as we move along $S_{\rm GUT}$.  The wave functions of bifundamental fields localize where their `$S_{\rm GUT}$-dependent' masses vanish and it is this behavior that leads to the structure of matter curves that one directly observes in the F-theory picture.
To understand this in detail, 
it is often helpful to think in terms of the eigenvalues of $\phi$, which can be diagonalized in a generic coordinate patch
\begin{equation}\phi\sim_{\text{patch}} \begin{pmatrix}t_1 & 0 & 0 & 0 & 0 \\ 0 & t_2 & 0 & 0 & 0 \\ 0 & 0 & t_3 & 0 & 0 \\ 0 & 0 & 0 & t_4 & 0 \\ 0 & 0 & 0 & 0 & t_5\end{pmatrix}\qquad \sum_{i=1}^5t_i=0\end{equation}
If $\phi$ were diagonal everywhere, such an expectation value would induce a mass proportional to $t_i$ for each of the five $\mathbf{10}$'s, $\mathbf{10}_{t_i}$, of $SU(5)_{\rm GUT}$ that descend from the $\mathbf{248}$ in \eqref{248decomp}.  The eigenvalues $t_i$, as nontrivial holomorphic sections, vanish along curves of $S_{\rm GUT}$ in general, and it is along these curves that the $\mathbf{10}_{t_i}$ wave functions localize.  One gets a similar story for the ten $\mathbf{\overline{5}}_{t_i+t_j}$ fields in \eqref{248decomp}, which can be labeled by the combinations $t_i+t_j$ with $i\ne j$ that correspond to the weights of a $\mathbf{10}$ of $SU(5)_{\perp}$.  In total, then, we get a naive picture that includes five $\mathbf{10}_{t_i}$ matter curves and ten $\mathbf{\overline{5}}_{t_i+t_j}$ matter curves that house matter fields with the corresponding $SU(5)_{\perp}$ weights.

In general, the story is not this simple because nothing requires the eigenvalues themselves to be globally defined along $S_{\rm GUT}$; only gauge invariant objects, like symmetric polynomials $s_n(t_i)$, need have this property.  Individual eigenvalues can be interchanged with one another through monodromies as we move throughout $S_{\rm GUT}$ and this is precisely what happens when the $b_m$ are generic holomorphic sections.  The effect of this monodromy action is crucially important because it impacts the structure of localized bifundamental fields as well as the presence or absence of $U(1)$ symmetries that might otherwise survive the breaking $E_8\rightarrow SU(5)_{\rm GUT}$.

When the $b_m$'s are completely generic, the Higgs bundle exhibits a maximal monodromy group which, for $SU(5)_{\perp}$, is the symmetric group on 5 objects.  This identifies all $\mathbf{10}_{t_i}$ fields, all $\mathbf{\overline{5}}_{t_i+t_j}$ fields, and removes all $U(1)$ factors from $SU(5)_{\perp}$ that could have remained from the underlying $E_8$.  What we are left with is one $\mathbf{10}$ matter curve, one $\mathbf{\overline{5}}$ matter curve, and nothing other than $SU(5)_{\rm GUT}$ to constrain the physics in general.  This is exactly what we expected from the geometry \eqref{TateModel} for a generic choice of sections and, indeed, the explicit description of matter curves from the Higgs bundle picture
\begin{equation}0=b_5\sim \prod_i t_i\qquad 0 = b_0 b_5^2  -b_2 b_3 b_5 +b_3^2 b_4\sim \prod_{i<j}(t_i+t_j)\label{mattcurvesH}\end{equation}
matches the singularity enhancements of \eqref{mattcurvesF}.

In what follows, we will often be interested in Higgs bundles with a monodromy group that is nontrivial so it will be helpful to have a way to directly visualize the monodromic structure.  One useful object for this is the spectral cover, which is a 5-sheeted cover of $S_{\rm GUT}$ inside the total space of the canonical bundle{\footnote{The canonical bundle arises here because $\phi$ itself is a section of this bundle.}}
$(K_{S_{\rm GUT}}\rightarrow S_{\rm GUT})$
\begin{equation}{\cal{C}}:\quad b_0s^5 + b_2s^3 + b_3 s^2 + b_4s + b_5=0\label{speccover}\,.\end{equation}
The five sheets of ${\cal{C}}$ roughly correspond to local eigenvalues $t_i$ and monodromies are manifested by the fact that ${\cal{C}}$ is smooth and irreducible for generic $b_m$, with branch cuts connecting all of its sheets.  The spectral cover is of course useful for much more than visualizing monodromies as we will review in a bit.

\subsection{$U(1)$ Symmetries and Spectral Cover Models}

We now turn to the task of engineering geometries with extra $U(1)$'s that couple to the charged degrees of freedom of our GUT.
Actually obtaining $U(1)$ symmetries of this type can be quite subtle in F-theory \cite{Hayashi:2010zp} but recent progress \cite{Grimm:2010ez,Marsano:2010ix} has suggested one way to proceed when the geometry is built as a `Tate model' \eqref{TateModel}.   In the language of \cite{Marsano:2010ix}, we consider an object in $Y_4$ referred to as the 'spectral divisor'{\footnote{More specifically, what we mean by $\hat{\cal{C}}_{SD}$ is the proper transform of \eqref{specdiv} when the singularities of $Y_4$ are resolved.}}
\begin{equation}\hat{\cal{C}}_{SD}:\quad a_0z^5 + a_2xz^3 + a_3yz^2 + a_4x^2z + a_5xy=0\label{specdiv}\end{equation}
which reduces to the spectral cover of the Higgs bundle in a suitable local limit \cite{Marsano:2010ix}.  To see this, note that the section
\begin{equation}t=\frac{y}{x}\end{equation}
is meromorphic on $Y_4$ but restricts to a holomorphic section on $\hat{\cal{C}}_{SD}$.  In terms of $t$, we can write \eqref{specdiv} as
\begin{equation}a_0z^5 + a_2t^2z^3 + a_3t^3z^2+a_4t^4z+a_5t^5 =0\,.\end{equation}
We recover \eqref{speccover} by sending $z,t\rightarrow 0$ while holding $s=z/t$ fixed.  In the same way that \eqref{speccover} tells us something about the monodromic structure of the Higgs bundle, the behavior of \eqref{specdiv} near $z=0$ captures the monodromic structure of the local geometry \eqref{localmodel} when viewed as an ALE fibration over $S_{\rm GUT}$.  Unlike \eqref{speccover}, however, \eqref{specdiv} is a global object that tells us, in a sense, how this local structure fits into the full geometry of $Y_4$.
To get an extra $U(1)$ symmetry, the prescription of \cite{Grimm:2010ez} and \cite{Marsano:2010ix} is to choose the $a_m$ so that $\hat{\cal{C}}_{SD}$ splits into multiple components{\footnote{The $U(1)$'s are most easily seen in the M-theory language, where they come from reduction of $C_3$ along suitable $(1,1)$-forms in $Y_4$.  The $(1,1)$-forms for our 'new' $U(1)$'s are specified by suitable ('traceless') combinations of the various components of $\hat{\cal{C}}_{SD}$.}}.  When we do this, it is easy to verify that the matter curves \eqref{mattcurvesH} also split into factors and that this splitting can be directly attributed to a reduction of the monodromy group of the local geometry \eqref{localmodel}.

When we choose the $a_m$ in this way, the spectral cover of the gauge theory description \eqref{speccover} also splits into multiple components.  This immediately signals a reduction in the Higgs bundle monodromy group because the eigenvalues associated to sheets of one component do not mix with those of the others.  This also gives us some intuition for the appearance of extra $U(1)$ symmetries from the gauge theory perspective because a nongeneric monodromy group will not project out all of the $U(1)$ factors in $SU(5)_{\perp}\subset E_8$.
Unfortunately, this special choice of $a_m$'s also introduces a slight ambiguity into the story that was recently emphasized by the authors of \cite{Cecotti:2010bp}.  While a generic choice of spectral data $b_m=a_m|_{S_{\rm GUT}}$ uniquely specifies $\phi$ up to gauge equivalence, this is no longer true when the $b_m$ are sufficiently nongeneric that ${\cal{C}}$ becomes singular, as it does when \eqref{speccover} factors.  Given such a collection of $b_m$'s one must therefore take more care in identifying the field configuration for $\phi$ that accurately captures the physics near $S_{\rm GUT}$.

One particularly natural possibility that requires us to make an additional assumption but requires no new data is to use each component of ${\cal{C}}$ to specify a separate Higgs bundle of smaller rank.  If ${\cal{C}}={\cal{C}}^{(2)}{\cal{C}}^{(3)}$ splits into quadratic and cubic factors, for instance, they uniquely determine $U(2)$ and $U(3)$ bundles, respectively.  Provided we choose the spectral data so that the coefficient $b_1\sim b_0\text{tr}\phi$ in the full product ${\cal{C}}={\cal{C}}^{(2)}{\cal{C}}^{(3)}$ vanishes, these two bundles can be combined to yield an $SU(5)_{\perp}$ bundle.  Such bundles are said to be ``block reconstructible" and allow a naive extension of the eigenvalue-based analysis that we used to study the structure of generic $SU(5)_{\perp}$ bundles above.  We will abuse language in this paper and refer to such bundles as ``spectral cover bundles" and the resulting models as ``spectral cover models" even though one could in principle write a spectral cover for a non-Abelian Higgs bundle that does not satisfy the ``block reconstructible" property{\footnote{We feel this language is justified because Higgs bundles that are not ``block reconstructible" will require additional data, in addition to that contained in the spectral cover, to specify and describe.}}.  Essentially all Higgs bundles that have been studied in the F-theory literature are of this type.  The reason is that they appear to be the right Higgs bundles for describing the Calabi-Yau 4-folds that we know how to construct as Tate models \eqref{TateModel} with factored spectral divisor \eqref{specdiv}.  The dictionary relating spectral data to local geometric moduli can be used to compare the structure of matter curves and $U(1)$ symmetries that we obtain from the local Higgs bundleÊ 
description and the 4-fold geometry \eqref{TateModel}.  Even the chiral spectrum can be seen to match using the formalism of \cite{Marsano:2010ix}.

Throughout the rest of this paper, we will focus on the local description based on geometries of the type \eqref{localmodel} or, equivalently, the 8-dimensional $E_8$ gauge theory on $\mathbb{R}^{3,1}\times S_{\rm GUT}$ with Higgs bundle specified by a spectral cover \eqref{speccover} and taken to be block reconstructible whenever \eqref{speccover} factors.  When we talk about models with extra $U(1)$ symmetries in this way, we make an intrinsic assumption that our local model is embeddable into a global one in which the global object \eqref{specdiv} factors.  This does not always have to be the case; as emphasized in \cite{Hayashi:2010zp,Grimm:2010ez,Marsano:2010ix}, our local model might indicate the presence of $U(1)$ symmetries that are not actually present in the global completion.  This reflects the fact that local models only capture the physics of $SU(5)_{\rm GUT}$-charged degrees of freedom and do not know about the dynamics of GUT-singlet fields, which can break our $U(1)$ symmetries by attaining large expectation values.

\subsection{Technical Matters and Matter Curves}

We now turn to some technical details for how to work with ``spectral cover bundles" with a particular focus on the structure of matter curves.  We refer the reader to \cite{Marsano:2009gv} for a more complete discussion.

In spectral cover models, the effect of the monodromy group is to effectively quotient the spectrum by removing the distinction between fields whose $SU(5)_{\perp}$ weights lie in the same orbit.  As a result, we get one $\mathbf{10}_{t_i}$ field for each distinct orbit of the $t_i$'s under monodromy with a similar story for $\mathbf{\overline{5}}_{t_i+t_j}$'s.  This means that we get one $\mathbf{10}^{(a)}$ for each component, ${\cal{C}}^{(a)}$, of ${\cal{C}}$ while we get one $\mathbf{\overline{5}}^{(ab)}$ for each pair of components, ${\cal{C}}^{(a)}$ and ${\cal{C}}^{(b)}$, of ${\cal{C}}$ where we allow for the case $a=b$ when ${\cal{C}}^{(a)}$ has multiple sheets.  We emphasize this relation between matter curves and components of of the spectral cover because the wave functions of charged fields are properly described not on curves inside $S$ but rather by suitable lifts of those curves to ${\cal{C}}$.
In what follows, we will use the notation $\Sigma$ for matter curves inside $S$ and $\tilde{\Sigma}$ for the corresponding lifts to ${\cal{C}}$.  

The curves $\tilde{\Sigma}$ are most easily described as topological classes in the total space of the canonical bundle over $S$, $(K_S\rightarrow S)$, since this is the ambient space in which the spectral cover is defined.  Following Donagi and Wijnholt \cite{Donagi:2009ra}, one can perform computations by compactifying this space provided that care is taken to remove any spurious contributions that arise at $\infty$.  This compactified space, $X=\mathbb{P}({\cal{O}}\oplus K_S)$, takes the form of a $\mathbb{P}^1$-bundle over $S$ and comes equipped with two sections, $\sigma$ and $\sigma_{\infty}$, that satisfy $\sigma\cdot \sigma_{\infty}=0$.  We will use $\pi$ to denote the projection map $\pi:X\rightarrow S$.  That our original space was $(K_S\rightarrow S)$ is encoded in the fact that $\sigma_{\infty}=\sigma+\pi^*c_1$ where $c_1$ is the conventional shorthand for $c_1(S)$.

In this language, the full spectral cover ${\cal{C}}$ \eqref{speccover} is a divisor inside $X$ with topological class
\begin{equation}{\cal{C}}=5\sigma+\pi^*\eta\,,\end{equation}
where $\eta$ is a divisor class in $S$ that we are free to choose.  In mapping the spectral data to algebraic deformations of a local Calabi-Yau 4-fold $Y_4$, the class $\eta$ becomes identified with the combination $6c_1-t$ where $-t$ is shorthand for the normal bundle to $S$ inside the base $B_3$ of $Y_4$.

For a generic spectral cover ${\cal{C}}$, we expect $\mathbf{10}$ fields to localize whenver any of the eigenvalues of $\phi$ vanish.  Viewing the sheets as local eigenvalues, this means that the $\mathbf{10}$ matter curve should arise when ${\cal{C}}$ meets the section $\sigma$
\begin{equation}\tilde{\Sigma}_{\mathbf{10}}={\cal{C}}\cdot \sigma\qquad \Sigma_{\mathbf{10}} = \eta-5c_1\,.\end{equation}
The result for $\Sigma_{\mathbf{10}}$ agrees with the $\mathbf{10}$ matter curve in the local Calabi-Yau 4-fold specified by the same spectral data \cite{Donagi:2009ra, Hayashi:2008ba, Marsano:2009gv}  under the identification $\eta=6c_1-t$.  Similarly, the $\mathbf{\overline{5}}$ matter curve corresponds to a locus where $t_i+t_j=0$ for some pair of eigenvalues $t_i$ and $t_j$ with $i\ne j$.  We can get this by considering the intersection ${\cal{C}}\cap \tau {\cal{C}}$ where $\tau$ is the involution that sends the holomorphic section $V$ associated to $\sigma_{\infty}$ from $V\rightarrow -V$.  This effectively multiplies all eigenvalues by $-1$ so that
\begin{equation}{\cal{C}}\cap \tau {\cal{C}}\supset \left(t_i=0\text{ locus}\right) + \left(t_i+t_j=0\text{ locus}\right) + \left(t_i,t_j\rightarrow\infty\text{ locus}\right)\,.\end{equation}
The $\mathbf{\overline{5}}$ matter curve can therefore be determined by removing the $\mathbf{10}$ matter curve and component at $\infty$ from ${\cal{C}}\cap \tau {\cal{C}}$.  The result is that
\begin{equation}\tilde{\Sigma}_{\mathbf{\overline{5}}}=2\sigma\cdot \pi^*(8c_1-3t)+\pi^*()\cdot \pi^*()\qquad \Sigma_{\mathbf{\overline{5}}}=8c_1-3t\,.\end{equation}

When ${\cal{C}}$ splits into factors, the determination of matter curves is a straightforward generalization of the above procedure.  For instance, when ${\cal{C}}={\cal{C}}^{(3)}{\cal{C}}^{(2)}$ for cubic and quadratic components ${\cal{C}}^{(3)}$ and ${\cal{C}}^{(2)}$, the matter curve for $\mathbf{10}$'s associated with orbits of the first three eigenvalues $\{t_1,t_2,t_3\}$ is given by
\begin{equation}\tilde{\Sigma}_{\mathbf{10}}^{(3)}={\cal{C}}^{(3)}\cap \sigma\,.\end{equation}
The determination of $\mathbf{\overline{5}}$ matter curves can be a bit tricker but is nevertheless completely straightforward.  In all cases, the matter curves in $S$ can be matched to singularity enhancements of a local Calabi-Yau 4-fold whose algebraic complex structure deformations are specified by the same spectral data as the spectral cover Higgs bundle.


\subsection{Dudas-Palti Relations}

While $U(1)$ symmetries are important for phenomenology, F-theory GUTs typically make use of another crucial ingredient.  To break $SU(5)_{\rm GUT}$ down to the Standard Model gauge group $SU(3)\times SU(2)\times U(1)_Y$, it is conventional to introduce a nontrivial internal flux $F_Y$ along the $U(1)_Y$ direction.  This mechanism is particularly useful because the hypercharge flux can be used to lift any leptoquarks that might survive from $SU(5)_{\rm GUT}$ as well as the triplet partners of the Higgs doublets $H_u$ and $H_d$.  In order to avoid lifting the $U(1)_Y$ gauge boson, however, this flux has to satisfy a particularly special condition: it must be dual to a curve inside $S$ that is trivial in the homology of $B_3$ \cite{Beasley:2008kw, Donagi:2008kj}.  This condition has important implications for how hypercharge flux can be distributed among the matter curves.  For instance, we immediately see that $F_Y\cdot \Sigma=0$ for any curve $\Sigma$ in $S$ that descends from a nontrivial divisor class inside $B_3$.  Because $c_1$ and $t$ are both curves of this type, we have
\begin{equation}F_Y\cdot c_1 = F_Y\cdot t = F_Y\cdot \eta=0\label{FYc1t}\end{equation}
and hence
\begin{equation}\begin{split}\sum_{\mathbf{10}\text{ matter curves, }i}\int_{\Sigma_{\mathbf{10}^{(i)}}}F_Y &= (c_1-t)\cdot F_Y = 0\\
\sum_{\mathbf{\overline{5}}\text{ matter curves, }a}\int_{\Sigma_{\mathbf{\overline{5}}^{(a)}}}F_Y&=(8c_1-3t)\cdot F_Y = 0\end{split}\label{FYortho}\,.\end{equation}
It is well-known that these relations, along with the statement
\begin{equation}3\sum_{\mathbf{10}\text{ matter curves, }i}\Sigma_{\mathbf{10}}^{(i)}-\sum_{\mathbf{\overline{5}}\text{ matter curves, }a}\Sigma_{\mathbf{\overline{5}}}^{(a)} + 5c_1=0\label{mattcurverel}\end{equation}
combine to ensure that the resulting 4-dimensional theory does not exhibit any MSSM gauge anomalies.

In the course of building spectral cover models with $U(1)$ symmetries, additional constraints on the distribution of $U(1)_Y$ flux arose.  In early work \cite{Marsano:2009gv}  it was noted that models with a spectral cover that splits into quartic and linear factors must have $U(1)_Y$ flux on at least one $\mathbf{10}$ curve whenver $U(1)_Y$ flux threads any of the $\mathbf{\overline{5}}$ matter curves{\footnote{It was erroneously claimed in \cite{Marsano:2009gv} that this property generalizes to all models with factored spectral covers.  As we will explicitly see later, this is not the case.
}}.  

More recently, Dudas and Palti \cite{Dudas:2010zb} developed a framework for describing a large collection of spectral cover models with multiple $U(1)$'s and observed an interesting pattern.  Phrasing it in a language that does not explicitly depend on spectral cover, we can write the Dudas-Palti observation as
\begin{equation}\sum_{\mathbf{10}\text{ matter curves },i}q_i\int_{\Sigma_{\mathbf{10}}^{(i)}}F_Y = \sum_{\mathbf{\overline{5}}\text{ matter curves },a}q_a\int_{\Sigma_{\mathbf{\overline{5}}^{(a)}}}F_Y\label{DPP}\,,\end{equation}
where $q_i/q_a$ is the common charge of all fields on the corresponding matter curve under one of the $U(1)$ symmetries that has been engineered.  Recently, it has been shown that this relation, like \eqref{FYortho} and \eqref{mattcurverel}, is a consequence of 4-dimensional anomaly cancellation.  In the following, however, we would like to give an explicit technical derivation of \eqref{DP} for spectral cover models.


\subsection{Deriving the Dudas-Palti Relations for Spectral Cover Models}

To proceed, consider a spectral cover with $N_{\text{comp}}$ components so that $(N_{\text{comp}}-1)>0$ $U(1)$ factors are retained from the underlying $E_8$ symmetry.  
Let us introduce some notation for the $N_{\text{comp}}$ components
\begin{equation}{\cal{C}}=\sum_a {\cal{C}}^{(a)}\end{equation}
and write their topological classes in $X$ as
\begin{equation}[{\cal{C}}^{(a)}] = n_a\sigma + \pi^*(\xi_a)\end{equation}
with
\begin{equation}\sum_{\text{components, }a} n_a = 5\qquad \sum_{\text{components, }a}\xi_a = \eta\,.\end{equation}
In general, each component ${\cal{C}}^{(a)}$ gives rise to a distinct $\mathbf{10}$ matter curve from
\begin{equation}\hat{\Sigma}_{\mathbf{10}}^{(a)} = \sigma\cdot {\cal{C}}^{(a)} = \sigma\cdot \pi^*(\Sigma_{\mathbf{10}}^{(a)})\end{equation}
associated to the fields $\mathbf{10}^{(a)}$.  Explicit computation shows that, for our spectral cover
\begin{equation}\Sigma_{\mathbf{10}}^{(a)}=\xi_a-n_ac_1\label{10matt}\,.\end{equation}
Now, let us consider the $\mathbf{\overline{5}}$ matter curves involving ${\cal{C}}^{(a)}$.  We get one matter curve, $\hat{\Sigma}_{\mathbf{\overline{5}}}^{(aa)}$ from ${\cal{C}}^{(a)}\cap \tau {\cal{C}}^{(a)}\,\,$ associated to the fields $\mathbf{\overline{5}}^{(aa)}$.  We also get additional matter curves, $\hat{\Sigma}_{\mathbf{\overline{5}}}^{(a)}$, from ${\cal{C}}^{(a)}\cap \tau {\cal{C}}^{(i)}$ for $i\ne a$ associated to the fields $\mathbf{\overline{5}}^{(ai)}$.  Both sheets of $\hat{\Sigma}_{\mathbf{\overline{5}}}^{(aa)}$ sit inside ${\cal{C}}^{(a)}$ so its topological class will take the form
\begin{equation}\hat{\Sigma}_{\mathbf{\overline{5}}}^{(aa)} = 2\sigma\cdot \pi^*(\Sigma_{\mathbf{\overline{5}}}^{(aa)}) + \pi^*(\ast)\cdot \pi^*(\ast)\,.\end{equation}
On the other hand, only one sheet of $\hat{\Sigma}_{\mathbf{\overline{5}}}^{(ai)}$ sits inside ${\cal{C}}^{(a)}$ (the other sits inside ${\cal{C}}^{(i)}$).  The topological class of that sheet is
\begin{equation}\hat{\Sigma}_{\mathbf{\overline{5}}}^{(ai)} = \sigma\cdot\pi^*(\Sigma_{\mathbf{\overline{5}}}^{(ai)})+\pi^*(\ast)\cdot \pi^*(\ast)\,.\end{equation}
With this information, let us now consider what results from the topological intersection{\footnote{We emphasize that this is a topological intersection.  Strictly speaking, $\mathbf{\overline{5}}$ curves sit inside ${\cal{C}}^{(a)}\cap \tau{\cal{C}}^{(b)}$ where $\tau$ is a $\mathbb{Z}_2$ involution on the ambient space.  Since ${\cal{C}}^{(b)}$ and $\tau {\cal{C}}^{(b)}$ are in the same topological class, we do not worry about this for topological intersections.}} of ${\cal{C}}^{(a)}$ with the full spectral cover ${\cal{C}}$.  In general, we have
\begin{equation}{\cal{C}}^{(a)}\cdot {\cal{C}} = \left(\mathbf{10}\text{ matter curve in }{\cal{C}}^{(a)}\right)+\left(\mathbf{\overline{5}}\text{ matter curves in }{\cal{C}}^{(a)}\right)+\left(\text{component at }\infty\text{ in }{\cal{C}}^{(a)}\right)\,.\end{equation}
That we get a copy of the $\mathbf{10}$ matter curve in ${\cal{C}}^{(a)}$ reflects the fact that this curve sits in both ${\cal{C}}^{(a)}$ and ${\cal{C}}$.  The component at $\infty$, on the other hand, represents the part of the spectral cover that lies along $\infty$ in the compactification $X$ of the canonical bundle $(K_S\rightarrow S)$.  As we said before, this component must be removed when extracting matter curves as described, for instance, in \cite{Marsano:2009gv}.  The full component at $\infty$ for ${\cal{C}}\cdot {\cal{C}}$ is simply $3\sigma_{\infty}\cdot \pi^*\eta$ and the part of this that sits inside ${\cal{C}}^{(a)}\cdot {\cal{C}}$ is given by $3\sigma_{\infty}\cdot \pi^*\xi_a${\footnote{A review of these computations can be found in Appendix \ref{app:spectech}.}}.  With this, we expect on general grounds that
\begin{equation}{\cal{C}}^{(a)}\cdot {\cal{C}} = \sigma\cdot \left[\Sigma_{\mathbf{10}}^{(a)} + 2\Sigma_{\mathbf{\overline{5}}}^{(aa)} + \sum_{i\ne a}\Sigma_{\mathbf{\overline{5}}}^{(ai)} + 3\pi^*\xi_a\right]+\pi^*(\ast)+\pi^*(\ast)\label{CCgen}\,.\end{equation}
Explicit computation, however, yields
\begin{equation}\begin{split}{\cal{C}}^{(a)}\cdot {\cal{C}} &= (n_a\sigma + \pi^*\xi_a)\cdot (5\sigma+\pi^*\eta) \\
&= \sigma\cdot (n_a\pi^*\eta + 5\pi^*\xi_a) + \pi^*(\ast)\cdot \pi^(\ast)
\end{split}\label{CCcomp}\,.\end{equation}
Comparing \eqref{CCgen} to \eqref{CCcomp} and making use of \eqref{10matt}, we see that our matter curves must satisfy the relation
\begin{equation}\Sigma_{\mathbf{10}}^{(a)} = 2\Sigma_{\mathbf{\overline{5}}}^{(aa)} + \sum_{i\ne a}\Sigma_{\mathbf{\overline{5}}}^{(ai)} -n_a(\eta+2c_1)\,.\end{equation}
Because $F_Y$ is orthogonal to both $\eta$ and $c_1$ \eqref{FYortho}, this implies that
\begin{equation}F_Y\cdot \Sigma_{\mathbf{10}}^{(a)} = F_Y\cdot \left(2\Sigma_{\mathbf{\overline{5}}}^{(aa)} + \sum_{i=1}^a \Sigma_{\mathbf{\overline{5}}}^{(ai)}\right)\label{DPspec}\,.\end{equation}

This is a mathematical statement of the observation by Dudas and Palti of how hypercharge flux was distributed on the matter curves of their models \cite{Dudas:2010zb}.  We get one equation for each component ${\cal{C}}^{(a)}$ of the spectral cover with each side representing a sum over matter curves weighted by the $SU(5)_{\perp}$ weight of that component.  Recalling that each $U(1)$ is given by a traceless linear combination of these weights, though, it is natural to ask what we get by taking the corresponding linear combinations of \eqref{DPspec}.  Doing this, we get
\begin{equation}\sum_a q_a F_Y\cdot \Sigma_{\mathbf{10}}^{(a)} = \sum_i q_i F_Y\cdot \Sigma_{\mathbf{\overline{5}}}^{(i)}\label{DPapp}\,,\end{equation}
where $q_a$ ($q_i$) denotes the $U(1)$ charge of $\mathbf{10}^{(a)}$ ($\mathbf{\overline{5}}^{(i)}$) fields and the sum over $i$ now runs over all $\mathbf{\overline{5}}$ matter curves.  This set of $N_{\text{comp}}-1$ equations is precisely the version of the DP relations that we quoted in \eqref{DP}.  Because \eqref{DPspec} consisted of $N_{\text{comp}}$ equations, \eqref{DPapp} seems to be missing one relation.  This is obtained by simply summing both sides over all components and leads to
\begin{equation}F_Y\cdot \sum_a \Sigma_{\mathbf{10}}^{(a)} = F_Y\cdot \sum_i \Sigma_{\mathbf{\overline{5}}}^{(i)}\,,\end{equation}
which is a trivial equation; both sides vanish identically from \eqref{FYortho} because $\sum_a\Sigma_{\mathbf{10}}^{(a)}=c_1-t$ and $\sum_i\Sigma_{\mathbf{\overline{5}}}^{(i)}=8c_1-3t$. 


\subsection{Survey}

Our general analysis of the consequences of the DP relations in Section \ref{sec:DP} showed that generically it will be difficult to build models that are both consistent with unification and sufficiently suppress proton decay.
In this section we give a survey of solutions to the DP relations, and analyze them with respect to
\begin{itemize}
\item Unification
\item Proton Decay
\item $\mu$-Term
\end{itemize}
The detailed analysis is provided in Appendix \ref{app:Survey} and a summary of all models follows in the next subsection.
The survey is comprehensive for all models with two $U(1)$ symmetries that can arise from spectral cover constructions and has been summarized in the Introduction. In particular, this restricts the integers $N_i$ and $P_i$ in \eqref{exoticparametrization} to be equal. We constrain ourselves to models with at least two unbroken $U(1)$s as one will be broken subsequently by the vev of the charged singlet, and we require at least another $U(1)$ that is not affected by the singlet vev and can protect against proton decay operators. Note that the $U(1)$ symmetries are realized in terms of the spectral cover $\mathcal{C}$  by requiring a factorization into $n+1$ factors: 
\begin{equation}
U(1)^{n}  \qquad \hbox{requires} \qquad \mathcal{C} = \prod_{i=1}^{n+1} \mathcal{C}^{(i)} \,.
\end{equation}
Furthermore, when realizing these models, the only constraints on fluxes that we impose are the Dudas-Palti relations, which will restrict the hypercharge flux $F_Y$. There may be further restrictions on other fluxes, once they are realized in a full-fledged global model, however, we do not impose such restrictions. In this sense, the class of models in the survey may get even further restricted.

We presented the summary table already in Section \ref{subsec:SummarySurvey}. The salient features of the models were as follows.  Labeling the exotic spectrum as in Table \ref{KLMNLabels} by the integers $K, L, M, N$, we show in Appendix \ref{app:Survey} that the only choices are as follows:
\begin{equation}\begin{array}{c|cc|c}
\text{Models} & \text{Exotic Spectra} & & \text{Dim 5} \\ \hline
1,\, 2,\,9 & N-L=1 & & XQ^3L/\Lambda^2 \\
3,\, 4 & N-L=2 & K\ge M & X^2Q^3 L / \Lambda^3 \\
5,\,6,\,7,\,8 & L=2 & M=N=0 & X^{\dag\,2}Q^3L /\Lambda^4 \\
10 & N-L=1 & K-L=M & XQ^3L/\Lambda^2
\end{array}\end{equation}
In the appendices \ref{app:Models} and  \ref{app:Survey} we provide a detailed description of each of these models, including the $U(1)$ charge assignments as well as the detailed exotic content.
The last column in this table summarizes which dimension 5 proton decay operators are still present in these models, where $\Lambda$ is the high-scale relevant for these models.

The survey is performed systematically by first picking one of the two spectral cover factorizations, ${\bf 2+2+1}$ or ${\bf 3+1+1}$.  For each factorization, there are several embeddings of matter and Higgs curves into the spectral cover, and for each such matter curve assignment, there are flux and hypercharge flux choices.
We systematically analyze all such models, by first fixing
 a singlet with specific charges $\lambda_i - \lambda_j$, and then surveying which models allow for all exotics to be lifted by veving this singlet. For each such singlet and matter curve assignment we then present
 the most general flux and hypercharge flux assignment that is a solution to the DP relations \eqref{DP} and furthermore gives rise to a three-generation model.
 
Note that in models 5,6,7 and 8 the singlet's vev will give rise to a large $\mu$-term due to
the coupling
$$ {1\over \Lambda}\int d^2\theta X^2H_uH_d,$$
and so these models are clearly not good for particle phenomenology.
Meanwhile, models 1,2,9 and 10 likely have too large dimension 5 operators that are 
dangerous for proton decay. This leaves only models 3 and 4 deserving of further investigation. Namely, computing the dynamical
vev of $X$ will answer the question if dimension 5 operators in  models 3 and 4
are sufficiently suppressed.


\section*{Acknowledgements}

We thank W.~Taylor and T.~Watari for interesting discussions.  J.M. is also grateful to C.~Cordova and J.~Heckman for very helpful correspondence about T-branes.  We would like to thank the organizers of the MPI Munich workshop 'GUTs and Strings' for providing a stimulating research environment where this collaboration began.  JM and NS are grateful to The Ohio State University and the organizers of the String Vacuum Project Fall Meeting for hospitality during the course of this work.  NS would also like to thank the Enrico Fermi Institute and the University of Chicago theory group for hospitality.  SSN thanks the Caltech theory group, the Center for Theoretical Physics at MIT, and the Mathematical Institute in Oxford for hospitality. MJD thanks the Peierls Centre for Theoretical Physics in Oxford for hospitality. The work of JM is supported by DOE grant DE-FG02-90ER-40560 and NSF grant PHY-0855039.



\startappendix

\newpage

\section{List of Models}
\label{app:Models}

In this Appendix, we provide a list of the spectral cover models from our survey that could realize a $U(1)_{\chi}$ symmetry, a $U(1)_{PQ}$ symmetry, and lift all charged fields except for precisely the matter content of the MSSM by giving an expectation value to exactly one $SU(5)_{\rm GUT}$ singlet field.  The models are of two basic types depending on the factorization structure of the spectral cover.  In all cases, we have three factors and we label them as
\begin{equation}{\cal{C}} = {\cal{C}}^{(a)}{\cal{C}}^{(d)}{\cal{C}}^{(e)}\end{equation}
For the two factorization structures, these labels correspond to factors of the following degrees
\begin{equation}\begin{array}{c|ccc}\text{Factorization Structure} & \text{Deg of }{\cal{C}}^{(a)} & \text{Deg of }{\cal{C}}^{(d)} & \text{Deg of }{\cal{C}}^{(e)} \\ \hline
2+2+1 & 2 & 2 & 1 \\
3+1+1 & 3 & 1 & 1
\end{array}\end{equation}
The description of each model begins by identifying the singlet $X$ whose expectation value will lift the non-MSSM exotics.  More specifically we list the $SU(5)_{\perp}$ weights.  We label $\mathbf{10}$s according to their corresponding component so that $\mathbf{10}^{(x)}$ localizes on the matter curve ${\cal{C}}^{(x)}\cdot \sigma$ and $\mathbf{\overline{5}}^{(xy)}$ localizes on a matter curve contained in ${\cal{C}}^{(x)}\cap \tau {\cal{C}}^{(y)}$.  The $SU(5)_{\perp}$ weight of $\mathbf{10}^{(x)}$ is denoted $\lambda_x$ while that of $\mathbf{\overline{5}}^{(xy)}$ is denoted $\lambda_x+\lambda_y$.  Singlet weights are of the form $\lambda_i-\lambda_j$ for $i,j$ labeling some components of ${\cal{C}}$.  Note that singlets with weights $\lambda_i-\lambda_j$ and $\lambda_j-\lambda_i$ are conjugates of one another.  We will only assume an expectation value to one singlet field; its conjugate will not obtain a nonzero expectation value.   

For each matter curve, we will provide the number of units of bulk $G$-flux, labeled $G$, and hypercharge flux, labeled $F_Y$, and then list the net chirality of the different zero modes that localize there.  The fluxes, and hence the spectra, will typically be parametrized by a number of integers whose ranges are specified explicitly.  We next identify the $U(1)$ subgroups of $SU(5)_{\perp}$ that are preserved in a notation that is hopefully clear from the context.  We write explicitly the $U(1)$ charges of all fields and present the operator that can generate dimension 5 proton decay when the singlet $X$ picks up a nonzero expectation value.  Finally, we summarize the exotic spectrum using the $M,N,K,L$ parametrization of \eqref{KLMNLabels}, which we repeat here for clarity

\begin{equation}\begin{array}{c|c|c}\label{KLMNLabelsapp}
SU(5)\text{ origin} & \text{Exotic Multiplet} & \text{Degeneracy}  \\ \hline
& (\mathbf{1},\mathbf{1})_{+1}\oplus (\mathbf{1},\mathbf{1})_{-1} & M+N \\
\mathbf{10}\oplus\mathbf{\overline{10}} & (\mathbf{3},\mathbf{2})_{+1/6}\oplus (\mathbf{\overline{3}},\mathbf{2})_{-1/6} & M \\
& (\mathbf{\overline{3}},\mathbf{1})_{-2/3}\oplus (\mathbf{3},\mathbf{1})_{+2/3} & M-N\\ \hline
\mathbf{\overline{5}}\oplus \mathbf{5} & (\mathbf{\overline{3}},\mathbf{1})_{+1/3} \oplus (\mathbf{3},\mathbf{1})_{-1/3} & K \\
& (\mathbf{1},\mathbf{2})_{-1/2}\oplus (\mathbf{1},\mathbf{2})_{+1/2} & K-L \\ \hline
\mathbf{5}_H & (\mathbf{\overline{3}},\mathbf{1})_{1/3} & 0 \\
& (\mathbf{1},\mathbf{2})_{-1/2} & -1 \\
\hline
\mathbf{\overline{5}}_H & (\mathbf{\overline{3}},\mathbf{1})_{+1/3} & 0\\
& (\mathbf{1},\mathbf{2})_{-1/2} & 1 \end{array}\end{equation}
For this parametrization to make sense, we must have
\begin{equation}M\ge |N|\qquad K\ge 0\qquad K-L\ge 0\label{exoticparamapp}\end{equation}

We now move on to the list of 'models'.

\subsection{Model 1}

This is a 2+2+1 model with quadratic components ${\cal{C}}^{(a)}$ and ${\cal{C}}^{(d)}$ and linear component ${\cal{C}}^{(e)}$.

\begin{equation}\text{Singlet weight is }\lambda_d-\lambda_a\end{equation}

\begin{equation}\begin{array}{c|c|c|c|c|c}
\text{Matter Curve} & G & F_Y & (1,1)_{+1} & (3,2)_{+1/6} & (\overline{3},1)_{-2/3} \\ \hline
\mathbf{10}^{(a)}\leftrightarrow\mathbf{10}_M & 3+\tilde{G} & P+1 & 3+\tilde{G}+(P+1) & 3+\tilde{G} & 3+\tilde{G}-(P+1)\\
\mathbf{10}^{(d)} & -\tilde{G} & -(P+1) & -\tilde{G}-(P+1) & -\tilde{G} & -\tilde{G}+(P+1) \\
\mathbf{10}^{(e)} & 0 & 0 & 0 & 0 & 0
\end{array}\end{equation}

\begin{equation}\begin{array}{c|c|c|c|c}
\text{Matter Curve} & G & F_Y & (\overline{3},1)_{+1/3} & (1,2)_{-1/2} \\ \hline
\mathbf{\overline{5}}^{(aa)}\leftrightarrow\mathbf{5}_H & 0 & 1 & 0 & -1\\
\mathbf{\overline{5}}^{(ad)}\leftrightarrow \mathbf{\overline{5}}_H & G_{ad} & P-1 & G_{ad} & (G_{ad}-P)+1 \\
\mathbf{\overline{5}}^{(dd)} & -G_{ad} & -P & -G_{ad} & -(G_{ad}-P) \\
\mathbf{\overline{5}}^{(ae)} & 0 & 0 & 0 & 0 \\
\mathbf{\overline{5}}^{(de)}\leftrightarrow\mathbf{\overline{5}}_M & 3 & 0 & 3 & 3
\end{array}\end{equation}
where
\begin{equation}G_{ad}\ge0\qquad (G_{ad}-P)\ge 0\qquad \tilde{G}\ge |P+1|\end{equation}

\begin{equation}U(1)_1\sim \begin{pmatrix} 1 & 0 & 0 & 0 & 0 \\ 0 & 1 & 0 & 0 & 0 \\ 0 & 0 & 1 & 0 & 0 \\ 0 & 0 & 0 & 1 & 0 \\ 0 & 0 & 0 & 0 & -4\end{pmatrix}\qquad U(1)_2\sim \begin{pmatrix}1 & 0 & 0 & 0 & 0 \\ 0 & 1 & 0 & 0 & 0 \\ 0 & 0 & -1 & 0 & 0 \\ 0 & 0 & 0 & -1 & 0 \\ 0 & 0 & 0 & 0 & 0\end{pmatrix}\end{equation}

\begin{equation}\begin{array}{c|ccccc|c}
\text{Field} & \mathbf{10}_M & \mathbf{\overline{5}}_M & H_u & H_d & X & Q^3L \\ \hline
U(1)_1 & 1 & -3 & -2 & 2 & 0 & 0 \\
U(1)_2 & 1 & -1 & -2 & 0 & -2 & 2
\end{array}\end{equation}

\begin{equation}\frac{1}{\Lambda^2}XQ^3L\text{ is allowed}\end{equation}

\hrule

\begin{center}\textbf{Exotic Spectrum}\end{center}
\begin{equation}\begin{array}{c|c|c|c}
M & N & K & L \\ \hline
\tilde{G} & P+1 & G_{ad} & P
\end{array}\end{equation}

\newpage

\subsection{Model 2}

This is a 2+2+1 model with quadratic components ${\cal{C}}^{(a)}$ and ${\cal{C}}^{(d)}$ and linear component ${\cal{C}}^{(e)}$.

\begin{equation}\text{Singlet weight is }\lambda_d-\lambda_a\end{equation}

\begin{equation}\begin{array}{c|c|c|c|c|c}
\text{M.C.} & G & F_Y & (1,1)_{+1} & (3,2)_{+1/6} & (\overline{3},1)_{-2/3} \\ \hline
\mathbf{10}^{(a)}\leftrightarrow\mathbf{10}_M & 3+\tilde{G} & P+G_{aa}+1 & 3+\tilde{G}+(P+G_{aa}+1) & 3+\tilde{G} & 3+\tilde{G}-(P+G_{aa}+1) \\
\mathbf{10}^{(d)} & -\tilde{G} & -(P+G_{aa}+1) & -\tilde{G}-(P+G_{aa}+1) & -\tilde{G} & -\tilde{G}+(P+G_{aa}+1) \\
\mathbf{10}^{(e)} & 0 & 0 & 0 & 0 & 0
\end{array}\end{equation}

\begin{equation}\begin{array}{c|c|c|c|c}
\text{Matter Curve} & G & F_Y & (\overline{3},1)_{+1/3} & (1,2)_{-1/2} \\ \hline
\mathbf{\overline{5}}^{(aa)}\leftrightarrow\mathbf{5}_H & G_{aa} & 1+G_{aa} & G_{aa} & -1 \\
\mathbf{\overline{5}}^{(ad)}\leftrightarrow\mathbf{\overline{5}}_H & -G_{aa} & P-1-G_{aa} & -G_{aa} & 1-P \\
\mathbf{\overline{5}}^{(dd)} & 0 & -P & 0 & P \\
\mathbf{\overline{5}}^{(ae)} & 0 & 0 & 0 & 0 \\
\mathbf{\overline{5}}^{(de)}\leftrightarrow\mathbf{\overline{5}}_M & 3 & 0 & 3 & 3
\end{array}\end{equation}
where
\begin{equation}G_{aa}\ge 0\qquad (-P)\ge 0\qquad \tilde{G}\ge|P+G_{aa}+1|\end{equation}

\begin{equation}U(1)_1\sim \begin{pmatrix} 1 & 0 & 0 & 0 & 0 \\ 0 & 1 & 0 & 0 & 0 \\ 0 & 0 & 1 & 0 & 0 \\ 0 & 0 & 0 & 1 & 0 \\ 0 & 0 & 0 & 0 & -4\end{pmatrix}\qquad U(1)_2\sim \begin{pmatrix}1 & 0 & 0 & 0 & 0 \\ 0 & 1 & 0 & 0 & 0 \\ 0 & 0 & -1 & 0 & 0 \\ 0 & 0 & 0 & -1 & 0 \\ 0 & 0 & 0 & 0 & 0\end{pmatrix}\end{equation}

\begin{equation}\begin{array}{c|ccccc|c}
\text{Field} & \mathbf{10}_M & \mathbf{\overline{5}}_M & H_u & H_d & X & Q^3L \\ \hline
U(1)_1 & 1 & -3 & -2 & 2 & 0 & 0 \\
U(1)_2 & 1 & -1 & -2 & 0 & -2 & 2
\end{array}\end{equation}

\begin{equation}\frac{1}{\Lambda^2}XQ^3L\text{ is allowed}\end{equation}

\hrule

\begin{center}\textbf{Exotic Spectrum}\end{center}
\begin{equation}\begin{array}{c|c|c|c}
M & N & K & L \\ \hline
\tilde{G} & P+G_{aa}+1 & G_{aa} & G_{aa}+P
\end{array}\end{equation}

\newpage

\subsection{Model 3}

This is a 2+2+1 model with quadratic components ${\cal{C}}^{(a)}$ and ${\cal{C}}^{(d)}$ and linear component ${\cal{C}}^{(e)}$.

\begin{equation}\text{Singlet weight is }\lambda_d-\lambda_a\end{equation}

\begin{equation}\begin{array}{c|c|c|c|c|c}
\text{M.C.} & G & F_Y & (1,1)_{+1} & (3,2)_{+1/6} & (\overline{3},1)_{-2/3} \\ \hline
\mathbf{10}^{(a)}\leftrightarrow\mathbf{10}_M & 3+\tilde{G} & P+1 & 3+\tilde{G} +(P+1) & 3+\tilde{G} & 3+\tilde{G}-(P+1) \\
\mathbf{10}^{(d)} & -\tilde{G} & -(P+1) & -\tilde{G}-(P+1) & -\tilde{G} & -\tilde{G}+(P+1)  \\
\mathbf{10}^{(e)} & 0 & 0 & 0 & 0 & 0
\end{array}\end{equation}

\begin{equation}\begin{array}{c|c|c|c|c}
\text{Matter Curve} & G & F_Y & (\overline{3},1)_{+1/3} & (1,2)_{-1/2} \\ \hline
\mathbf{\overline{5}}^{(aa)}\leftrightarrow\mathbf{5}_H & 0 & 1 & 0 & -1 \\
\mathbf{\overline{5}}^{(ad)} & G_{ad} & G_{ad} & G_{ad} & 0 \\
\mathbf{\overline{5}}^{(dd)}\leftrightarrow\mathbf{\overline{5}}_H & -G_{ad} & -G_{ad}-1 & -G_{ad} & 1 \\
\mathbf{\overline{5}}^{(ae)}\leftrightarrow\mathbf{\overline{5}}_M & 3+\hat{G} & P-G_{ad}-1 & 3+\hat{G} & 3+\hat{G}+G_{ad}+1-P \\
\mathbf{\overline{5}}^{(de)} & -\hat{G} & -P+G_{ad}+1 & -\hat{G} & -\hat{G}-G_{ad}-1+P 
\end{array}\end{equation}
where
\begin{equation}G_{ad}\ge 0\qquad \hat{G}\ge 0\qquad \hat{G}+G_{ad}+1-P\ge 0\qquad \tilde{G}\ge |P+1|\end{equation}

\begin{equation}U(1)_1\sim \begin{pmatrix} 1 & 0 & 0 & 0 & 0 \\ 0 & 1 & 0 & 0 & 0 \\ 0 & 0 & 1 & 0 & 0 \\ 0 & 0 & 0 & 1 & 0 \\ 0 & 0 & 0 & 0 & -4\end{pmatrix}\qquad U(1)_2\sim \begin{pmatrix}1 & 0 & 0 & 0 & 0 \\ 0 & 1 & 0 & 0 & 0 \\ 0 & 0 & -1 & 0 & 0 \\ 0 & 0 & 0 & -1 & 0 \\ 0 & 0 & 0 & 0 & 0\end{pmatrix}\end{equation}

\begin{equation}\begin{array}{c|ccccc|c}
\text{Field} & \mathbf{10}_M & \mathbf{\overline{5}}_M & H_u & H_d & X & Q^3L \\ \hline
U(1)_1 & 1 & -3 & -2 & 2 & 0 & 0 \\
U(1)_2 & 1 & 1 & -2 & -2 & -2 & 4
\end{array}\end{equation}

\begin{equation}\frac{1}{\Lambda^3}X^2Q^3L\text{ is allowed}\end{equation}

\hrule

\begin{center}\textbf{Exotic Spectrum}\end{center}
\begin{equation}\begin{array}{c|c|c|c}
M & N & K & L \\ \hline
\tilde{G} & P+1 & G_{ad}+\hat{G} & P-1
\end{array}\end{equation}

\newpage

\subsection{Model 4}

This is a 2+2+1 model with quadratic components ${\cal{C}}^{(a)}$ and ${\cal{C}}^{(d)}$ and linear component ${\cal{C}}^{(e)}$.

\begin{equation}\text{Singlet weight is }\lambda_d-\lambda_a\end{equation}

\begin{equation}\begin{array}{c|c|c|c|c|c}
\text{M.C.} & G & F_Y & (1,1)_{+1} & (3,2)_{+1/6} & (\overline{3},1)_{-2/3} \\ \hline
\mathbf{10}^{(a)}\leftrightarrow\mathbf{10}_M & 3+\tilde{G} & G_{aa}+P+1 & 3+\tilde{G}+(G_{aa}+P+1) & 3+\tilde{G} & 3+\tilde{G}-(G_{aa}+P+1) \\
\mathbf{10}^{(d)} & -\tilde{G} & -(G_{aa}+P+1)& -\tilde{G}-(G_{aa}+P+1)& -\tilde{G} & -\tilde{G}+(G_{aa}+P+1  \\
\mathbf{10}^{(e)} & 0 & 0 & 0 & 0 & 0
\end{array}\end{equation}

\begin{equation}\begin{array}{c|c|c|c|c}
\text{Matter Curve} & G & F_Y & (\overline{3},1)_{+1/3} & (1,2)_{-1/2} \\ \hline
\mathbf{\overline{5}}^{(aa)}\leftrightarrow\mathbf{5}_H & G_{aa} & G_{aa}+1 & G_{aa} & -1 \\
\mathbf{\overline{5}}^{(ad)} & -G_{aa} & -G_{aa} & -G_{aa} & 0 \\
\mathbf{\overline{5}}^{(dd)}\leftrightarrow\mathbf{\overline{5}}_H & 0 & -1 & 0 & 1\\
\mathbf{\overline{5}}^{(ae)}\leftrightarrow\mathbf{\overline{5}}_M & 3+\hat{G} & P-1 & 3+\hat{G} & 3+\hat{G}-(P-1)\\
\mathbf{\overline{5}}^{(de)} & -\hat{G} & -(P-1) & -\hat{G} & -\hat{G}+(P-1) \\
\end{array}\end{equation}
where
\begin{equation}G_{aa}\ge 0\qquad \hat{G}\ge 0 \qquad \hat{G}-(P-1)\ge 0\qquad \tilde{G}\ge |G_{aa}+P+1|\end{equation}

\begin{equation}U(1)_1\sim \begin{pmatrix} 1 & 0 & 0 & 0 & 0 \\ 0 & 1 & 0 & 0 & 0 \\ 0 & 0 & 1 & 0 & 0 \\ 0 & 0 & 0 & 1 & 0 \\ 0 & 0 & 0 & 0 & -4\end{pmatrix}\qquad U(1)_2\sim \begin{pmatrix}1 & 0 & 0 & 0 & 0 \\ 0 & 1 & 0 & 0 & 0 \\ 0 & 0 & -1 & 0 & 0 \\ 0 & 0 & 0 & -1 & 0 \\ 0 & 0 & 0 & 0 & 0\end{pmatrix}\end{equation}

\begin{equation}\begin{array}{c|ccccc|c}
\text{Field} & \mathbf{10}_M & \mathbf{\overline{5}}_M & H_u & H_d & X & Q^3L \\ \hline
U(1)_1 & 1 & -3 & -2 & 2 & 0 & 0 \\
U(1)_2 & 1 & 1 & -2 & -2 & -2 & 4
\end{array}\end{equation}

\begin{equation}\frac{1}{\Lambda^3}X^2Q^3L\text{ is allowed}\end{equation}

\hrule

\begin{center}\textbf{Exotic Spectrum}\end{center}
\begin{equation}\begin{array}{c|c|c|c}
M & N & K & L \\ \hline
\tilde{G} & G_{aa}+P+1 & G_{aa}+\hat{G} & G_{aa}+P-1
\end{array}\end{equation}

\newpage

\subsection{Model 5}

This is a 2+2+1 model with quadratic components ${\cal{C}}^{(a)}$ and ${\cal{C}}^{(d)}$ and linear component ${\cal{C}}^{(e)}$.

\begin{equation}\text{Singlet weight is }\lambda_a-\lambda_d\end{equation}

\begin{equation}\begin{array}{c|c|c|c|c|c}
\text{M.C.} & G & F_Y & (1,1)_{+1} & (3,2)_{+1/6} & (\overline{3},1)_{-2/3} \\ \hline
\mathbf{10}^{(a)} & 3 & 0 & 3 & 3 & 3 \\
\mathbf{10}^{(d)} & 0 & 0 & 0 & 0 & 0 \\
\mathbf{10}^{(e)} & 0 & 0 & 0 & 0 & 0
\end{array}\end{equation}

\begin{equation}\begin{array}{c|c|c|c|c}
\text{Matter Curve} & G & F_Y & (\overline{3},1)_{+1/3} & (1,2)_{-1/2} \\ \hline
\mathbf{\overline{5}}^{(aa)} & -G_{ad} & -1 & -G_{ad} & -G_{ad}+1 \\
\mathbf{\overline{5}}^{(ad)} & G_{ad} & 2 & G_{ad} & G_{ad}-2 \\
\mathbf{\overline{5}}^{(dd)} & 0 & -1 & 0 & 1 \\
\mathbf{\overline{5}}^{(ae)} & 3 & 0 & 3 & 3 \\
\mathbf{\overline{5}}^{(de)} & 0 & 0 & 0 & 0 \\
\end{array}\end{equation}
where
\begin{equation}G_{ad}\ge 2\end{equation}

\begin{equation}U(1)_1\sim \begin{pmatrix} 1 & 0 & 0 & 0 & 0 \\ 0 & 1 & 0 & 0 & 0 \\ 0 & 0 & 1 & 0 & 0 \\ 0 & 0 & 0 & 1 & 0 \\ 0 & 0 & 0 & 0 & -4\end{pmatrix}\qquad U(1)_2\sim \begin{pmatrix}1 & 0 & 0 & 0 & 0 \\ 0 & 1 & 0 & 0 & 0 \\ 0 & 0 & -1 & 0 & 0 \\ 0 & 0 & 0 & -1 & 0 \\ 0 & 0 & 0 & 0 & 0\end{pmatrix}\end{equation}

\begin{equation}\begin{array}{c|ccccc|c}
\text{Field} & \mathbf{10}_M & \mathbf{\overline{5}}_M & H_u & H_d & X & Q^3L \\ \hline
U(1)_1 & 1 & -3 & -2 & 2 & 0 & 0 \\
U(1)_2 & 1 & 1 & -2 & -2 & 2 & 4
\end{array}\end{equation}

\begin{equation}\frac{1}{\Lambda^4}X^{\dag\,2}Q^3L\text{ is allowed}\end{equation}

\hrule

\begin{center}\textbf{Exotic Spectrum}\end{center}
\begin{equation}\begin{array}{c|c|c|c}
M & N & K & L \\ \hline
0 & 0 & G_{ad} & 2
\end{array}\end{equation}

\newpage

\subsection{Model 6}

This is a 2+2+1 model with quadratic components ${\cal{C}}^{(a)}$ and ${\cal{C}}^{(d)}$ and linear component ${\cal{C}}^{(e)}$.

\begin{equation}\text{Singlet weight is }\lambda_a-\lambda_d\end{equation}

\begin{equation}\begin{array}{c|c|c|c|c|c}
\text{M.C.} & G & F_Y & (1,1)_{+1} & (3,2)_{+1/6} & (\overline{3},1)_{-2/3} \\ \hline
\mathbf{10}^{(a)} & 3 & 0 & 3 & 3 & 3 \\
\mathbf{10}^{(d)} & 0 & 0 & 0 & 0 & 0 \\
\mathbf{10}^{(e)} & 0 & 0 & 0 & 0 & 0
\end{array}\end{equation}

\begin{equation}\begin{array}{c|c|c|c|c}
\text{Matter Curve} & G & F_Y & (\overline{3},1)_{+1/3} & (1,2)_{-1/2} \\ \hline
\mathbf{\overline{5}}^{(aa)} & -G_{ad} & -G_{ad}+1 & -G_{ad} & -1 \\
\mathbf{\overline{5}}^{(ad)} & G_{ad} & 2G_{ad}-2 & G_{ad} & 2-G_{ad} \\
\mathbf{\overline{5}}^{(dd)} & 0 & -G_{ad}+1 & 0 & G_{ad}-1 \\
\mathbf{\overline{5}}^{(ae)} & 3 & 0 & 3 & 3 \\
\mathbf{\overline{5}}^{(de)} & 0 & 0 & 0 & 0 \\
\end{array}\end{equation}
where
\begin{equation}G_{ad}\ge 2\end{equation}

\begin{equation}U(1)_1\sim \begin{pmatrix} 1 & 0 & 0 & 0 & 0 \\ 0 & 1 & 0 & 0 & 0 \\ 0 & 0 & 1 & 0 & 0 \\ 0 & 0 & 0 & 1 & 0 \\ 0 & 0 & 0 & 0 & -4\end{pmatrix}\qquad U(1)_2\sim \begin{pmatrix}1 & 0 & 0 & 0 & 0 \\ 0 & 1 & 0 & 0 & 0 \\ 0 & 0 & -1 & 0 & 0 \\ 0 & 0 & 0 & -1 & 0 \\ 0 & 0 & 0 & 0 & 0\end{pmatrix}\end{equation}

\begin{equation}\begin{array}{c|ccccc|c}
\text{Field} & \mathbf{10}_M & \mathbf{\overline{5}}_M & H_u & H_d & X & Q^3L \\ \hline
U(1)_1 & 1 & -3 & -2 & 2 & 0 & 0 \\
U(1)_2 & 1 & 1 & -2 & -2 & 2 & 4
\end{array}\end{equation}

\begin{equation}\frac{1}{\Lambda^4}X^{\dag\,2}Q^3L\text{ is allowed}\end{equation}

\hrule

\begin{center}\textbf{Exotic Spectrum}\end{center}
\begin{equation}\begin{array}{c|c|c|c}
M & N & K & L \\ \hline
0 & 0 & G_{ad} & 2
\end{array}\end{equation}

\newpage

\subsection{Model 7}

This is a 2+2+1 model with quadratic components ${\cal{C}}^{(a)}$ and ${\cal{C}}^{(d)}$ and linear component ${\cal{C}}^{(e)}$.

\begin{equation}\text{Singlet weight is }\lambda_a-\lambda_d\end{equation}

\begin{equation}\begin{array}{c|c|c|c|c|c}
\text{M.C.} & G & F_Y & (1,1)_{+1} & (3,2)_{+1/6} & (\overline{3},1)_{-2/3} \\ \hline
\mathbf{10}^{(a)} & 3 & 0 & 3 & 3 & 3 \\
\mathbf{10}^{(d)} & 0 & 0 & 0 & 0 & 0 \\
\mathbf{10}^{(e)} & 0 & 0 & 0 & 0 & 0
\end{array}\end{equation}

\begin{equation}\begin{array}{c|c|c|c|c}
\text{Matter Curve} & G & F_Y & (\overline{3},1)_{+1/3} & (1,2)_{-1/2} \\ \hline
\mathbf{\overline{5}}^{(aa)} & 0 & G_{dd}-1 & 0 & 1-G_{dd} \\
\mathbf{\overline{5}}^{(ad)} & -G_{dd} & 2(1-G_{dd}) & -G_{dd} & -2+G_{dd} \\
\mathbf{\overline{5}}^{(dd)} & G_{dd} & G_{dd}-1 & G_{dd} & 1 \\
\mathbf{\overline{5}}^{(ae)} & 3 & 0 & 3 & 3 \\
\mathbf{\overline{5}}^{(de)} & 0 & 0 & 0 & 0 \\
\end{array}\end{equation}
where
\begin{equation}G_{dd}\ge 2\end{equation}

\begin{equation}U(1)_1\sim \begin{pmatrix} 1 & 0 & 0 & 0 & 0 \\ 0 & 1 & 0 & 0 & 0 \\ 0 & 0 & 1 & 0 & 0 \\ 0 & 0 & 0 & 1 & 0 \\ 0 & 0 & 0 & 0 & -4\end{pmatrix}\qquad U(1)_2\sim \begin{pmatrix}1 & 0 & 0 & 0 & 0 \\ 0 & 1 & 0 & 0 & 0 \\ 0 & 0 & -1 & 0 & 0 \\ 0 & 0 & 0 & -1 & 0 \\ 0 & 0 & 0 & 0 & 0\end{pmatrix}\end{equation}

\begin{equation}\begin{array}{c|ccccc|c}
\text{Field} & \mathbf{10}_M & \mathbf{\overline{5}}_M & H_u & H_d & X & Q^3L \\ \hline
U(1)_1 & 1 & -3 & -2 & 2 & 0 & 0 \\
U(1)_2 & 1 & 1 & -2 & -2 & 2 & 4
\end{array}\end{equation}

\begin{equation}\frac{1}{\Lambda^4}X^{\dag\,2}Q^3L\text{ is allowed}\end{equation}

\hrule

\begin{center}\textbf{Exotic Spectrum}\end{center}
\begin{equation}\begin{array}{c|c|c|c}
M & N & K & L \\ \hline
0 & 0 & G_{dd} & 2
\end{array}\end{equation}

\newpage

\subsection{Model 8}

This is a 2+2+1 model with quadratic components ${\cal{C}}^{(a)}$ and ${\cal{C}}^{(d)}$ and linear component ${\cal{C}}^{(e)}$.

\begin{equation}\text{Singlet weight is }\lambda_a-\lambda_d\end{equation}

\begin{equation}\begin{array}{c|c|c|c|c|c}
\text{M.C.} & G & F_Y & (1,1)_{+1} & (3,2)_{+1/6} & (\overline{3},1)_{-2/3} \\ \hline
\mathbf{10}^{(a)} & 3 & 0 & 3 & 3 & 3 \\
\mathbf{10}^{(d)} & 0 & 0 & 0 & 0 & 0 \\
\mathbf{10}^{(e)} & 0 & 0 & 0 & 0 & 0
\end{array}\end{equation}

\begin{equation}\begin{array}{c|c|c|c|c}
\text{Matter Curve} & G & F_Y & (\overline{3},1)_{+1/3} & (1,2)_{-1/2} \\ \hline
\mathbf{\overline{5}}^{(aa)} & 0 & 1 & 0 & -1 \\
\mathbf{\overline{5}}^{(ad)} & -G_{dd} & -2 & -G_{dd} & -G_{dd}+2 \\
\mathbf{\overline{5}}^{(dd)} & G_{dd} & 1 & G_{dd} & G_{dd}-1 \\
\mathbf{\overline{5}}^{(ae)} & 3 & 0 & 3 & 3 \\
\mathbf{\overline{5}}^{(de)} & 0 & 0 & 0 & 0 \\
\end{array}\end{equation}
where
\begin{equation}G_{dd}\ge 2\end{equation}

\begin{equation}U(1)_1\sim \begin{pmatrix} 1 & 0 & 0 & 0 & 0 \\ 0 & 1 & 0 & 0 & 0 \\ 0 & 0 & 1 & 0 & 0 \\ 0 & 0 & 0 & 1 & 0 \\ 0 & 0 & 0 & 0 & -4\end{pmatrix}\qquad U(1)_2\sim \begin{pmatrix}1 & 0 & 0 & 0 & 0 \\ 0 & 1 & 0 & 0 & 0 \\ 0 & 0 & -1 & 0 & 0 \\ 0 & 0 & 0 & -1 & 0 \\ 0 & 0 & 0 & 0 & 0\end{pmatrix}\end{equation}

\begin{equation}\begin{array}{c|ccccc|c}
\text{Field} & \mathbf{10}_M & \mathbf{\overline{5}}_M & H_u & H_d & X & Q^3L \\ \hline
U(1)_1 & 1 & -3 & -2 & 2 & 0 & 0 \\
U(1)_2 & 1 & 1 & -2 & -2 & 2 & 4
\end{array}\end{equation}

\begin{equation}\frac{1}{\Lambda^4}X^{\dag\,2}Q^3L\text{ is allowed}\end{equation}

\hrule

\begin{center}\textbf{Exotic Spectrum}\end{center}
\begin{equation}\begin{array}{c|c|c|c}
M & N & K & L \\ \hline
0 & 0 & G_{dd} & 2
\end{array}\end{equation}

\newpage

\subsection{Model 9}

This is a 2+2+1 model with quadratic components ${\cal{C}}^{(a)}$ and ${\cal{C}}^{(d)}$ and linear component ${\cal{C}}^{(e)}$.

\begin{equation}\text{Singlet weight is }\lambda_e-\lambda_a\end{equation}

\begin{equation}\begin{array}{c|c|c|c|c|c}
\text{M.C.} & G & F_Y & (1,1)_{+1} & (3,2)_{+1/6} & (\overline{3},1)_{-2/3} \\ \hline
\mathbf{10}^{(a)}\leftrightarrow\mathbf{10}_M & 3+\tilde{G} & G_{aa}+P+1 & 3+\tilde{G}+(G_{aa}+P+1) & 3+\tilde{G} & 3+\tilde{G} - (G_{aa}+P+1) \\
\mathbf{10}^{(d)} & 0 & 0 & 0 & 0 & 0 \\
\mathbf{10}^{(e)} & -\tilde{G} & -(G_{aa}+P+1) & -\tilde{G}-(G_{aa}+P+1) & -\tilde{G} & -\tilde{G}+(G_{aa}+P+1) \\
\end{array}\end{equation}

\begin{equation}\begin{array}{c|c|c|c|c}
\text{Matter Curve} & G & F_Y & (\overline{3},1)_{+1/3} & (1,2)_{-1/2} \\ \hline
\mathbf{\overline{5}}^{(aa)}\leftrightarrow\mathbf{5}_H & G_{aa} & G_{aa}+1 & G_{aa} & -1 \\
\mathbf{\overline{5}}^{(ad)} & G_{ad} & P & G_{ad} & G_{ad}-P \\
\mathbf{\overline{5}}^{(dd)}\leftrightarrow\mathbf{\overline{5}}_M & 3 & 0 & 3 & 3 \\
\mathbf{\overline{5}}^{(ae)}\leftrightarrow\mathbf{\overline{5}}_H & -G_{aa} & -G_{aa}-1 & -G_{aa} & 1 \\
\mathbf{\overline{5}}^{(de)} & -G_{ad} & -P & -G_{ad} & -(G_{ad}-P) \\
\end{array}\end{equation}
where
\begin{equation}G_{aa}\ge 0\qquad G_{ad}\ge 0\qquad (G_{ad}-P)\ge 0\qquad \tilde{G}\ge |G_{aa}+P+1|\end{equation}

\begin{equation}U(1)_1\sim \begin{pmatrix}-2 & 0 & 0 & 0 & 0 \\ 0 & -2 & 0 & 0 & 0 \\ 0 & 0 & 3 & 0 & 0 \\ 0 & 0 & 0 & 3 & 0 \\ 0 & 0 & 0 & 0 & -2\end{pmatrix}\qquad U(1)_2\sim \begin{pmatrix}1 & 0 & 0 & 0 & 0 \\ 0 & 1 & 0 & 0 & 0 \\ 0 & 0 & 1 & 0 & 0 \\ 0 & 0 & 0 & 1 & 0 \\ 0 & 0 & 0 & 0 & -4\end{pmatrix}\end{equation}

\begin{equation}\begin{array}{c|ccccc|c}
\text{Field} & \mathbf{10}_M & \mathbf{\overline{5}}_M & H_u & H_d & X & Q^3L \\ \hline
U(1)_1 & -2 & 6 & 4 & -4 & 0 & 0 \\
U(1)_2 & 1 & 2 & -2 & -3 & -5 &5
\end{array}\end{equation}

\begin{equation}\frac{1}{\Lambda^2}XQ^3L\text{ is allowed}\end{equation}

\hrule

\begin{center}\textbf{Exotic Spectrum}\end{center}
\begin{equation}\begin{array}{c|c|c|c}
M & N & K & L \\ \hline
\tilde{G} & G_{aa}+P+1 & G_{aa}+G_{ad} & G_{aa}+P
\end{array}\end{equation}

\newpage

\subsection{Model 10}

This is a 3+1+1 model with cubic component ${\cal{C}}^{(a)}$ and linear components ${\cal{C}}^{(d)}$ and ${\cal{C}}^{(e)}$.

\begin{equation}\text{Singlet weight is }\lambda_d-\lambda_a\end{equation}

\begin{equation}\begin{array}{c|c|c|c|c|c}
\text{M.C.} & G & F_Y & (1,1)_{+1} & (3,2)_{+1/6} & (\overline{3},1)_{-2/3} \\ \hline
\mathbf{10}^{(a)}\leftrightarrow\mathbf{10}_M & 3+\tilde{G} & G_{aa}+1 & 3+\tilde{G}+(G_{aa}+1) & 3+\tilde{G} & 3+\tilde{G}-(G_{aa}+1)\\
\mathbf{10}^{(d)} & -\tilde{G} &-(G_{aa}+1) & -\tilde{G}-(G_{aa}+1) & -\tilde{G} & -\tilde{G}+(G_{aa}+1) \\
\mathbf{10}^{(e)} & 0 & 0 & 0 & 0 & 0 \\
\end{array}\end{equation}

\begin{equation}\begin{array}{c|c|c|c|c}
\text{Matter Curve} & G & F_Y & (\overline{3},1)_{+1/3} & (1,2)_{-1/2} \\ \hline
\mathbf{\overline{5}}^{(aa)}\leftrightarrow\mathbf{5}_H & G_{aa} & G_{aa}+1 & G_{aa} & -1 \\
\mathbf{\overline{5}}^{(ad)}\leftrightarrow\mathbf{\overline{5}}_H & -G_{aa} & -G_{aa}-1 & -G_{aa} & 1 \\
\mathbf{\overline{5}}^{(ae)}\leftrightarrow\mathbf{\overline{5}}_M & 3+\hat{G} & 0 & 3+\hat{G} & 3+\hat{G} \\
\mathbf{\overline{5}}^{(de)} & -\hat{G} & 0 & -\hat{G} & -\hat{G} 
\end{array}\end{equation}
where
\begin{equation}G_{aa}\ge 0\qquad \hat{G}\ge 0\qquad \tilde{G}\ge |G_{aa}+1|\end{equation}

\begin{equation}U(1)_1\sim \begin{pmatrix} 1 & 0 & 0 & 0 & 0 \\ 0 & 1 & 0 & 0 & 0 \\ 0 & 0 & 1 & 0 & 0 \\ 0 & 0 & 0 & 1 & 0 \\ 0 & 0 & 0 & 0 & -4\end{pmatrix}\qquad U(1)_2\sim \begin{pmatrix}1 & 0 & 0 & 0 & 0 \\ 0 & 1 & 0 & 0 & 0 \\ 0 & 0 & 1 & 0 & 0 \\ 0 & 0 & 0 & -3 & 0 \\ 0 & 0 & 0 & 0 & 0\end{pmatrix}\end{equation}

\begin{equation}\begin{array}{c|ccccc|c}
\text{Field} & \mathbf{10}_M & \mathbf{\overline{5}}_M & H_u & H_d & X & Q^3L \\ \hline
U(1)_1 & 1 & -3 & -2 & 2 & 0 & 0 \\
U(1)_2 & 1 & 1 & -2 & -2 & -4 & 4
\end{array}\end{equation}

\begin{equation}\frac{1}{\Lambda^2}XQ^3L\text{ is allowed}\end{equation}

\hrule

\begin{center}\textbf{Exotic Spectrum}\end{center}
\begin{equation}\begin{array}{c|c|c|c}
M & N & K & L \\ \hline
\tilde{G} & G_{aa}+1 & G_{aa}+\hat{G} & G_{aa}
\end{array}\end{equation}



\section{Exotic Spectra}
\label{app:exotics}

In this appendix, we use the Dudas-Palti relations \eqref{DP} to derive the constraint \eqref{DPimplications} on the exotic spectrum.  What makes this somewhat tricky is that we are not interested in net chiralities of exotics but rather the number of vector-like pairs that we get by summing the exotic spectra from different matter curves.

Consider, for instance, the set of exotics that localize on a $\mathbf{\overline{5}}$ matter curve, $\Sigma_{\mathbf{\overline{5}}}^{(0)}$.  The chiral spectrum is determined by two integers, $k_0$ and $\ell_0$, that encode the bulk flux and ``hypercharge flux" as in \eqref{10param} and \eqref{5barparam}.  Because $k_i$ and $\ell_j$ carry information about chirality, though, we cannot determine the net spectrum by simply summing them over all $\mathbf{\overline{5}}$ matter curves.  This sum will yield the net chirality of exotics, which vanishes, rather than the net number of doublet or triplet pairs.  To get information about the net spectrum itself, we need to compute weighted sums
\begin{equation}\sum_{\mathbf{\overline{5}}\text{ matter curves, }i}q_i \left(n_{(\mathbf{\overline{3}},\mathbf{1})_{+1/3}}-n_{(\mathbf{3},\mathbf{1})_{-1/3}}\right)\quad\text{and}\quad \sum_{\mathbf{\overline{5}}\text{ matter curves, }i}q_i \left(n_{(\mathbf{1},\mathbf{2})_{-1/2}}-n_{(\mathbf{1},\mathbf{2})_{+1/2}}\right)\,,\end{equation}
where $q_i$ is the $U(1)$ charge associated to the $\mathbf{\overline{5}}$ fields on a given $\mathbf{\overline{5}}$ matter curve, $\Sigma_{\mathbf{\overline{5}},i}$.  The $U(1)$ charge of doublets and triplets from the $\mathbf{\overline{5}}$ is $q_a$ while the $U(1)$ charge of doublets and triplets from the $\mathbf{5}$ is $-q_a$.  This means that the above computes the net number of doublets and triplets weighted not by their chirality but rather by their $U(1)$ charge
\begin{equation}\begin{split}\sum_{\mathbf{\overline{5}}\text{ matter curves, }i}q_i \left(n_{(\mathbf{\overline{3}},\mathbf{1})_{+1/3}}-n_{(\mathbf{3},\mathbf{1})_{-1/3}}\right) &= \sum_{\substack{\text{Triplets of}\\ \text{charge }Q_p}}Q_p n_{\text{trips},p} \\
\sum_{\mathbf{\overline{5}}\text{ matter curves, }i}q_i \left(n_{(\mathbf{1},\mathbf{2})_{-1/2}}-n_{(\mathbf{1},\mathbf{2})_{+1/2}}\right) &= \sum_{\substack{\text{Doublets of}\\ \text{charge }Q_p}}Q_p n_{\text{doubs},p}
 \end{split}\label{doubtripnos}\end{equation}
 where $n_{\text{doubs},p}$ and $n_{\text{trips},p}$ count the number of doublets or triplets irrespective of their chiralities.  Let us emphasize this by writing
 \begin{equation}n_{\text{doubs},p}\ge 0\qquad n_{\text{trips},p}\ge 0\,.\end{equation}
Now, using \eqref{5barparam} we recognize the difference of the right-hand sides in \eqref{doubtripnos} as nothing other than the right-hand-side of the Dudas-Palti relations \eqref{DP}
\begin{equation}\begin{split}\sum_{\substack{\text{Triplet} \\ \text{charges }Q_p}}Q_p n_{\text{trips},p} - \sum_{\substack{\text{Doublet} \\ \text{charges }Q_q}}Q_q n_{\text{doubs},q} &= \sum_{\mathbf{\overline{5}}\text{ matter curves, }i}q_i\left[\left(n_{(\mathbf{\overline{3}},\mathbf{1})_{+1/3}}-n_{(\mathbf{3},\mathbf{1})_{-1/3}}\right)-\left(n_{(\mathbf{1},\mathbf{2})_{-1/2}}-n_{(\mathbf{1},\mathbf{2})_{+1/2}}\right)\right] \\
&= \sum_{\mathbf{\overline{5}}\text{ matter curves, }i}q_i\int_{\Sigma_{\mathbf{\overline{5}}}^{(i)}}F_Y
\end{split}\end{equation}
We can use similar reasoning to relate the left-hand-side of the Dudas-Palti relations \eqref{DP} to the weighted spectra of exotics that arise from $\mathbf{10}$ matter curves.  To write the result, we use $n_{(1,1),p}$ to denote the net number of $(\mathbf{1},\mathbf{1})_{+1}$'s and $(\mathbf{1},\mathbf{1})_{-1}$'s of charge $Q_p$ and $n_{(3,2),p}$ and $n_{(3,1),p}$ for the similar numbers of $(\mathbf{\overline{3}},\mathbf{2})_{+1/6}$'s/$(\mathbf{3},\mathbf{2})_{-1/6}$'s and $(\mathbf{\overline{3}},\mathbf{1})_{-2/3}$/$(\mathbf{3},\mathbf{1})_{+2/3}$'s, respectively.  With this notation, we find two relations    
\begin{equation}\begin{split}
\sum_{\substack{(\mathbf{1},\mathbf{1})\text{ or cc} \\ \text{charges }Q_p}}Q_p n_{(1,1),p}- \sum_{\substack{(\mathbf{3},\mathbf{2})\text{ or cc} \\ \text{charges, }Q_q}}Q_qn_{(3,2),p} &= \sum_{\mathbf{10}\text{ matter curves, }a}q_a\left[\left(n_{(\mathbf{1},\mathbf{1})_{+1}}-n_{(\mathbf{1},\mathbf{1})_{-1}}\right) - \left(n_{(\mathbf{3},\mathbf{2})_{+1/6}}-n_{(\mathbf{\overline{3}},\mathbf{2})_{-1/6}}\right)\right] \\
&= \sum_{\mathbf{10}\text{ matter curves},a}q_a\int_{\Sigma_{\mathbf{10}}^{(a)}}F_Y\end{split}\label{10exotics1}\end{equation}
\begin{equation}\begin{split}
\sum_{\substack{(\mathbf{3},\mathbf{2})\text{ or cc} \\ \text{charges, }Q_p}}Q_pn_{(3,2),p} - \sum_{\substack{(\mathbf{\overline{3}},\mathbf{1})\text{ or cc} \\ \text{charges, }Q_q}}Q_qn_{(3,1),q}
&= \sum_{\mathbf{10}\text{ matter curves},a}q_a\left[\left(n_{(\mathbf{3},\mathbf{2})_{+1/6}}-n_{(\mathbf{\overline{3}},\mathbf{2})_{-1/6}}\right) - \left(n_{(\mathbf{\overline{3}},\mathbf{1})_{-2/3}}-n_{(\mathbf{3},\mathbf{1})_{+2/3}}\right)\right] \\
 &= \sum_{\mathbf{10}\text{ matter curves},a}q_a\int_{\Sigma_{\mathbf{10}}^{(a)}} F_Y
\end{split}\label{10exotics2}\end{equation}
Note that the right-hand-sides of both \eqref{10exotics1} and \eqref{10exotics2} are equivalent because the spectrum on each $\mathbf{10}$ matter curve in \eqref{10param} is controlled by two integers rather than three.  In the end, we conclude that the Dudas-Palti relations have the following direct effect on the non-GUT part of the spectrum
\begin{equation}\begin{split}\sum_{\substack{\text{Triplet} \\ \text{charges }Q_p}}Q_p n_{\text{trips},p} - \sum_{\substack{\text{Doublet} \\ \text{charges }Q_q}}Q_q n_{\text{doubs},q} &= \sum_{\substack{(\mathbf{1},\mathbf{1})\text{ or cc} \\ \text{charges }Q_p}}Q_p n_{(1,1),p}- \sum_{\substack{(\mathbf{3},\mathbf{2})\text{ or cc} \\ \text{charges, }Q_q}}Q_qn_{(3,2),p} \\
&= \sum_{\substack{(\mathbf{3},\mathbf{2})\text{ or cc} \\ \text{charges, }Q_p}}Q_pn_{(3,2),p} - \sum_{\substack{(\mathbf{\overline{3}},\mathbf{1})\text{ or cc} \\ \text{charges, }Q_q}}Q_qn_{(3,1),q}
\end{split}\label{prelimDPimp}\end{equation}
Equation \eqref{DPimplications} is nothing more than a suggestive way of writing these relations.  To derive it, let us assume as in section \ref{subsec:PQ} that that the exotics get their masses from a cubic superpotential of the form
\begin{equation}W_0\sim \sum_i X_i \left[\sum_j f_{\text{exotic},ij}\overline{f}_{\text{exotic},ij}\right]\label{initialmasses}\end{equation}
in which it is assumed that each exotic multiplet appears exactly once.  As discussed in section \ref{subsec:PQ}, this superpotential must be a subset of the full superpotential since we assume all exotics are lifted from the spectrum.  Restricting to the couplings in $W_0$ allows us to unambiguously split the exotics into groups according to the singlets that they couple to{\footnote{Of course there may not be a unique way to choose $W_0$ in the end.  We assume in this Appendix that a specific choice has been made.}}.  For each singlet $X_i$ in \eqref{initialmasses}, then, we parametrize the set of vector-like exotics that it couples to in \eqref{initialmasses} as in \eqref{exoticparametrization}
\begin{equation}\begin{split}
n_{(\mathbf{1},\mathbf{1})_{+1}}+n_{(\mathbf{1},\mathbf{1})_{-1}} &= M_i+P_i \\
n_{(\mathbf{3},\mathbf{2})_{-1/6}}+n_{(\mathbf{\overline{3}},\mathbf{2})_{+1/6}} &= M_i \\
n_{(\mathbf{\overline{3}},\mathbf{1})_{-2/3}}+n_{(\mathbf{3},\mathbf{1})_{+2/3}} &= M_i-N_i \\
n_{(\mathbf{\overline{3}},\mathbf{1})_{+1/3}}+n_{(\mathbf{3},\mathbf{1})_{-1/3}} &= K_i \\
n_{(\mathbf{1},\mathbf{2})_{-1/2}} + n_{(\mathbf{1},\mathbf{2})_{+1/2}} &= K_i-L_i
\end{split}\label{exoticparametrizationapp}\end{equation}
We can now rewrite \eqref{prelimDPimp} using the fact that the $U(1)$ charges of any two fields that couple to $X_i$ must sum to minus the charge of $X_i$, $-q_{X_i}$.  Remembering to add in the contribution from $H_u$ and $H_d$, which are not lifted by assumption, we find that \eqref{prelimDPimp} becomes
\begin{equation}\begin{split}-q_{H_u}-q_{H_d} - \sum_{\text{Singlets, }i}q_{X_i}L_i &= -\sum_{\text{Singlets, }i}q_{X_i}P_i \\
&= -\sum_{\text{Singlets, }i}q_{X_i}N_i
\end{split}\end{equation}
which is nothing other than \eqref{DPimplications}.
 

\section{Engineering Solutions to the Dudas-Palti Relations}
\label{app:DPsols}

Once we have arrived at the condition \eqref{DPapp}, it is natural to ask if it represents the only constraint on the distribution of hypercharge flux in spectral cover models.  There are, of course, two additional conditions that we encountered earlier but repeat again here for clarity
\begin{equation}F_Y\cdot \sum_a \Sigma_{\mathbf{10}}^{(a)} = F_Y\cdot \sum_i\Sigma_{\mathbf{\overline{5}}}^{(i)}\label{MSSManomalycond}\end{equation}
While \eqref{DPapp} reflects the cancellation of mixed MSSM and $U(1)$ anomalies \cite{Marsano:2010sq}, \eqref{MSSManomalycond} reflects the cancellation of pure MSSM anomalies that involve $U(1)_Y$.  We would like to claim that \eqref{DPapp} and \eqref{MSSManomalycond}, which can both be understood as a consequence of 4-dimensional anomaly cancellation, represent the only nontrivial restrictions on the distribution of hypercharge flux in spectral cover models.  More specifically, we will present ``in principle" constructions of spectral cover models with up to 2 $U(1)$ symmetries that realize the most generic hypercharge flux distributions allowed by \eqref{DPapp} and \eqref{MSSManomalycond}.  Such constructions are only models ``in principle" because we will be forced to introduce a variety of new sections; one must always check that these sections actually exist for a given choice of $S_{\rm GUT}$ and normal bundle.  In some cases, we will also have to make some assumptions about the vanishing locus of pairs of sections in order to avoid singularities that are not of Kodaira type.  Further restrictions that extend \eqref{DPapp} and \eqref{MSSManomalycond} may arise if we make special requirements of $S_{\rm GUT}$, for instance if we insist that it be a del Pezzo surface, or the normal bundle of $S_{\rm GUT}$ inside $B_3$.  We view these as ``model-building" restrictions, however, as opposed to physical obstructions like \eqref{DPapp} and \eqref{MSSManomalycond}.

We now proceed to construct models with one and two $U(1)$ symmetries.  The models with two $U(1)$ symmetries represent generalizations of the constructions in \cite{Dudas:2010zb}.

\subsection{Models with one $U(1)$ Symmetry}

We begin by considering models that engineer a single $U(1)$ symmetry.  Many explicit models of this type have been constructed in the literature in both the local and global settings \cite{Blumenhagen:2009up,Marsano:2009gv,Blumenhagen:2009yv,Marsano:2009wr,Grimm:2009yu,Cvetic:2010rq,Chen:2010ts,Chung:2010bn,Knapp:2011wk}.  The spectral cover must factor into two components in order to realize a single $U(1)$ symmetry.  These are two possibilities for how this occurs and we treat each one in turn.

\subsubsection{4+1 Factorization}

We begin with a factorization into a quartic and linear piece
\begin{equation}{\cal{C}}_{4+1}={\cal{C}}^{(4)}{\cal{C}}^{(1)} = \left(a_4V^4+a_3V^3U+a_2V^2U^2+a_1VU^3+a_0U^4\right)\left(e_1V+e_0U\right)\,.\end{equation}
In general, there will be two $\mathbf{10}$ matter curves and two $\mathbf{\overline{5}}$ matter curves in such a model.  The most general distribution of hypercharge flux consistent with \eqref{DPapp} and \eqref{MSSManomalycond} is easy to determine and is described below

\begin{equation}\begin{array}{c|c|c}\text{Matter Curve} & \text{Origin} & F_Y \\ \hline
\mathbf{10}^{(4)} & {\cal{C}}^{(4)} & -N \\
\mathbf{10}^{(1)} & {\cal{C}}^{(1)} & N \\
\mathbf{\overline{5}}^{(44)} & {\cal{C}}^{(4)}-{\cal{C}}^{(4)} & -N \\
\mathbf{\overline{5}}^{(41)} & {\cal{C}}^{(4)}-{\cal{C}}^{(1)} & N
\end{array}\label{41hyperdist}\end{equation}

We now describe a spectral cover construction capable of yielding precisely this configuration of hypercharge flux.  In addition to choosing the sections $a_m$ and $e_n$, we also have one choice of bundle that can be thought of as choosing the topological class of say ${\cal{C}}^{(1)}$.  More specifically, we take the $a_m$ and $e_n$ to be sections of the bundles described below
\begin{equation}\begin{array}{c|c}\text{Section} & \text{Bundle} \\ \hline
a_m & \eta-(m+1)c_1-\xi \\
e_n & (1-n)c_1+\xi
\end{array}\end{equation}
As we saw in \eqref{FYc1t}, the hypercharge flux is orthogonal to both $\eta$ and $c_1$ but it need not be orthogonal to $\xi$.  In principle, then, we can have
\begin{equation}F_Y\cdot \xi = N\end{equation}
for any integer $N$.  It is in this way that we can distribute hypercharge flux along the matter curves.  To actually construct a model, however, we must first solve the traceless condition $b_1=0$ which, in this case, amounts to
\begin{equation}a_1e_0+a_0e_1=0\,.\end{equation}
To avoid a non-Kodaira type singularity, we are actually forced to assume that $e_1$ and $e_0$ do not simultaneously vanish anywhere
\begin{equation}[e_0]\cdot [e_1]=0\end{equation}
so $e_1$ must divide $a_1$ and, correspondingly, $e_0$ must divide $a_0$.  We are therefore forced to introduce a new section $\alpha$ in terms of which 
\begin{equation}a_0 = \alpha e_0\qquad a_1 = -\alpha e_1\,.\end{equation}
We see that $\alpha$ must be a section of $\eta-2c_1-2\xi$.

With this, we now turn to the structure of our four matter curves
\begin{equation}\begin{array}{c|c|c|c|c}
\text{Matter Curve} & \text{Origin} & \text{Equation} & \text{Class} & F_Y \\ \hline
\mathbf{10}^{(4)} & {\cal{C}}^{(4)} & a_4 & \eta-5c_1-\xi & -N \\
\mathbf{10}^{(1)} & {\cal{C}}^{(1)} & e_1 & \xi & N \\
\mathbf{\overline{5}}^{(44)} & {\cal{C}}^{(4)}-{\cal{C}}^{(4)} & a_3^2e_0 + a_2a_3e_1+a_4e_1^2\alpha & 2\eta-7c_1-\xi & -N \\
\mathbf{\overline{5}}^{(41)} & {\cal{C}}^{(4)}-{\cal{C}}^{(1)} & a_4e_0^2+a_3e_0e_1+a_2e_1^2 & \eta-3c_1+\xi & N
\end{array}\end{equation}
This gives an explicit realization of the hypercharge flux distribution in \eqref{41hyperdist}.  To make everything completely explicit, we now list all of the sections that are needed to build the model along with the corresponding bundles
\begin{equation}\begin{array}{c|c}\text{Section} & \text{Bundle} \\ \hline
a_m & \eta -(m+1)c_1-\xi \\
e_n & (1-n)c_1+\xi \\
\alpha & \eta-2c_1-2\xi
\end{array}\label{41sections}\end{equation}
where $m$ runs from 0 to 4 and $n$ from 0 to 1.  To obtain an actual model, we must specify a complex surface $S_{\rm GUT}$ along with bundles $\eta$ and $\xi$ subject to the assumption that holomorphic sections in \eqref{41sections} all exist.  By this, we mean that the bundles appearing in \eqref{41sections} must all admit honest holomorphic sections.  To avoid non-Kodaira type singularities, we must also require that
\begin{equation}[e_0]\cdot [e_1]=0\end{equation}
or equivalently
\begin{equation}c_1\dot (c_1+\xi)=0\end{equation}

\subsubsection{3+2 Factorization}
\label{app:subsubsec:32}

We now consider a factorization of ${\cal{C}}$ into cubic and quadratic pieces
\begin{equation}{\cal{C}}_{3+2}={\cal{C}}^{(3)}{\cal{C}}^{(2)} = \left(a_3V^3+a_2V^2U+a_1VU^2+a_0U^3\right)\left(e_2V^2+e_1VU+e_0U^2\right)\,.\end{equation}
Here, we expect to find two $\mathbf{10}$ matter curves and three $\mathbf{\overline{5}}$ matter curves.  The most general distribution of hypercharge flux consistent with \eqref{DPapp} and \eqref{MSSManomalycond} is given by
\begin{equation}\begin{array}{c|c|c}\text{Matter Curve} & \text{Origin} & F_Y \\ \hline
\mathbf{10}^{(3)} & {\cal{C}}^{(3)} & -M-N \\
\mathbf{10}^{(2)} & {\cal{C}}^{(2)} & M+N \\
\mathbf{\overline{5}}^{(33)} & {\cal{C}}^{(3)}-{\cal{C}}^{(3)} & -M \\
\mathbf{\overline{5}}^{(32)} & {\cal{C}}^{(3)}-{\cal{C}}^{(2)} & M-N \\
\mathbf{\overline{5}}^{(22)} & {\cal{C}}^{(2)}-{\cal{C}}^{(2)} & N
\end{array}\label{32hyperdist}\end{equation}
At first glance, it seems difficult to realize a two-parameter family of hypercharge flux distributions because it seems that we only have the freedom to introduce one new bundle, $\xi$, which determines the relative class between the factors ${\cal{C}}^{(3)}$ and ${\cal{C}}^{(2)}$.  A closer glance at the traceless condition, however, will suggest additional freedom.  This condition takes a similar form to the previous example
\begin{equation}a_1e_0+a_0e_1=0\,,\end{equation}
but here it is no longer necessary to require that $e_0$ and $e_1$ have no common zeroes.  In particular, this means that $e_1$ need not divide $a_1$.  We are free to take $e_1$ to be a product of the form $e_1=AB$ where $A$ divides $e_0$ and $B$ divides $a_1$, leading to the solution
\begin{equation}\begin{split}e_0 &= \tilde{e}_0 A \\
e_1 &= AB \\
a_0 &= -\tilde{a}_1 \tilde{e}_0 \\
a_1 &= \tilde{a}_1 B
\end{split}\end{equation}
In choosing a bundle for $A$ we have introduced a new parameter into the game.  Let us make everything explicit by writing all sections and their corresponding bundles
\begin{equation}\begin{array}{c|c}\text{Section} & \text{Bundle} \\ \hline
a_3 & \eta-5c_1-\xi_A-\xi_B \\
a_2 & \eta-4c_1-\xi_A-\xi_B \\
\tilde{a}_1 & \eta-3c_1-\xi_A-2\xi_B \\
e_2 & \xi_A+\xi_B \\
\tilde{e}_0 &  c_1+\xi_B \\
A & c_1+\xi_A \\
B & \xi_B \\
\end{array}\label{32sections}\end{equation}
The structure of our five matter curves is now easily determined
\begin{equation}\begin{array}{c|c|c|c|c}
\text{Matter Curve} & \text{Origin} & \text{Equation} & \text{Class} & F_Y \\ \hline
\mathbf{10}^{(3)} & {\cal{C}}^{(3)} & a_3 & \eta-5c_1-\xi_A-\xi_B & -M-N \\
\mathbf{10}^{(2)} & {\cal{C}}^{(2)} & e_2 & \xi_A+\xi_B & M+N \\
\mathbf{\overline{5}}^{(33)} & {\cal{C}}^{(3)} - {\cal{C}}^{(3)} & a_2B + a_3\tilde{e}_0 & \eta-4c_1-\xi_A & -M\\
\mathbf{\overline{5}}^{(32)} & {\cal{C}}^{(3)} - {\cal{C}}^{(2)} & a_3A^3(a_2B+a_3\tilde{e}_0) + A^2e_2(a_2^2+\tilde{a}_1a_3B) & \\
& & +2\tilde{a}_1a_2Ae_2^2+\tilde{a}_1^2e_2^3 & 2\eta-6c_1+\xi_A-\xi_B & M-N\\
\mathbf{\overline{5}}^{(22)} & {\cal{C}}^{(2)} - {\cal{C}}^{(2)} & B & \xi_B & N
\end{array}\end{equation}
where we defined
\begin{equation}M=F_Y\cdot \xi_A\qquad N=F_Y\cdot \xi_B\,.\end{equation}
The entire 2-parameter family of solutions to \eqref{DPapp} and \eqref{MSSManomalycond} has therefore emerged.  Note that this family includes solutions with $M+N=0$ that have no hypercharge flux on any $\mathbf{10}$ matter curves despite having hypercharge flux threading some $\mathbf{\overline{5}}$ matter curves.  This possibility was missed in \cite{Marsano:2009gv} but arises here for a relatively simple reason.  In \cite{Marsano:2009gv}, it was assumed that only the relative class of the spectral cover components could affect the structure of matter curves.  Here, we see another way to adjust the relative classes of matter curves.  Nominally, the $\mathbf{\overline{5}}^{(22)}$ matter curve would be given by $e_1=0$ but, when $e_0$ and $e_1$ have $A$ as a common factor, the $A=0$ part of that curve moves off to $\infty$.  A corresponding factor then moves in 'from $\infty$' to contribute to $\mathbf{\overline{5}}^{(32)}${\footnote{In other words, different ways of solving the traceless condition can allow us to adjust how the component at $\infty$ of ${\cal{C}}^{(2)}\cdot \tau({\cal{C}}^{(2)}+{\cal{C}}^{(3)})$ is distributed between ${\cal{C}}^{(2)}\cdot \tau{\cal{C}}^{(2)}$ and ${\cal{C}}^{(2)}\cap \tau {\cal{C}}^{(3)}$.  Note that the net 'component at $\infty$' for ${\cal{C}}^{(2)}\cap\tau ({\cal{C}}^{(2)}+{\cal{C}}^{(3)})$ cannot be similarly adjusted.}}.  Through this type of phenomenon the $\mathbf{\overline{5}}$ matter curves are able to depend on combinations of $\xi_A$ and $\xi_B$ that are different from the sum $\xi_A+\xi_B$ that enters into the $\mathbf{10}$ matter curves.

As usual, to build explicit models we will need to choose $S_{\rm GUT}$, $\eta$, $\xi_A$, and $\xi_B$ subject to some assumptions.  In this case, all of the holomorphic sections in \eqref{32sections} must exist and, further, we must have that
\begin{equation}\xi_A\cdot (\xi_A+\xi_B)=0\end{equation}
in order to avoid the non-Kodaira type singularities that arise wherever $A=e_2=0$.

\subsection{Models with two $U(1)$ Symmetries}

We now turn to models that exhibit 2 $U(1)$ symmetries.  We again have two possibilities for the factorization structure that we consider in turn.

\subsubsection{2+2+1}

We begin with a factorization into two quadratic pieces and a linear piece
\begin{equation}{\cal{C}}_{2+2+1} = {\cal{C}}^{(2)}{\cal{C}}^{(2\prime)}{\cal{C}}^{(1)} = \left(a_2V^2+a_1VU+a_0U^2\right)\left(d_2V^2+d_1VU+d_0U^2\right)\left(e_1V+e_0U\right)\,.\end{equation}
Here, we expect to find 3 $\mathbf{10}$ matter curves and 5 $\mathbf{\overline{5}}$ matter curves.  The most general distribution of hypercharge flux consistent with \eqref{DPapp} and \eqref{MSSManomalycond} is given by
\begin{equation}\begin{array}{c|c|c}\text{Matter Curve} & \text{Origin} & F_Y \\ \hline
\mathbf{10}^{(2)} & {\cal{C}}^{(2)} & -M-N \\
\mathbf{10}^{(2\prime)} & {\cal{C}}^{(2\prime)} & M\\
\mathbf{10}^{(1)} & {\cal{C}}^{(1)} & N\\
\mathbf{\overline{5}}^{(22)} & {\cal{C}}^{(2)}-{\cal{C}}^{(2)} & -M-N-P\\
\mathbf{\overline{5}}^{(2\prime\, 2\prime)} & {\cal{C}}^{(2\prime)}-{\cal{C}}^{(2\prime)} & M-Q\\
\mathbf{\overline{5}}^{(22\prime)} & {\cal{C}}^{(2)}-{\cal{C}}^{(2\prime} & P+Q\\
\mathbf{\overline{5}}^{(21)} & {\cal{C}}^{(2)}-{\cal{C}}^{(1)} & M+N+P-Q\\
\mathbf{\overline{5}}^{(2\prime\,1)} & {\cal{C}}^{(2\prime)}-{\cal{C}}^{(1)} & -M-P+Q
\end{array}\label{221hyperdist}\end{equation}
For models with multiple $U(1)$ symmetries, the traceless condition becomes significantly more complicated.  Here, it takes the form
\begin{equation}b_1=a_1d_0e_0+d_1a_0e_0+e_1a_0d_0=0\label{b1221}\,.\end{equation}

One simple way to get a solution is to set
\begin{equation}d_0 = a_0 e_0 \qquad d_1 = -a_1e_0 - a_0e_1\,.\end{equation}
We don't have much freedom to choose new bundles with this, though, so we take advantage of the scaling properties of \eqref{b1221} to construct from this another solution
\begin{equation}\begin{split}a_0 &= \nu\tilde{a}_0 \\
d_0 &= \nu\tilde{a}_0\tilde{e}_0 \\
e_0 &= \nu\tilde{e}_0 \\
d_1 &= -a_1\tilde{e}_0 - e_1\tilde{a}_0
\end{split}\label{221ansatz}\end{equation}
It is easy to see that we will have two additional choices of bundle with this solution.  To get a few more, let us use the trick that we learned from the 3+2 factorization in section \ref{app:subsubsec:32}.  There, we learned that we can adjust $\mathbf{\overline{5}}$ matter curves by splitting them into components and using our solution to the traceless condition to move various components to and from $\infty$.  In this case, the matter curves $\mathbf{\overline{5}}^{(22)}$ and $\mathbf{\overline{5}}^{(2\prime\,2\prime)}$ are given by the equations $a_1=0$ and $d_1=0$, respectively.  We can move a component off to $\infty$ by allowing $a_1$ and $a_0$ (respectively $d_1$ and $d_0$) to have a common factor.  This will introduce a topological tuning because this factor cannot have any simultaneous zeroes with $a_2$ ($d_2$) but this is something we will have to live with.   From the scaling of \eqref{221ansatz}, we can start with any solution and obtain a new one by simultaneously rescaling both $a_1$ and $a_0$ (respectively $d_1$ and $d_0$).  We therefore arrive at the following solution
\begin{equation}\begin{split}a_0 &= \nu\tilde{a}_0 \delta_a \\
a_1 &= \tilde{a}_1\delta_a \\
d_0 &= \nu\tilde{a}_0\tilde{e}_0 \delta_d \\
d_1 &= -\delta_d\left(\tilde{a}_1\tilde{e}_0 + e_1\tilde{a}_0\right) \\
e_0 &= \nu\tilde{e}_0
\end{split}\end{equation}
It is now a simple matter to write out the sections 
\begin{equation}\begin{array}{c|c}
\text{Section} & \text{Bundle} \\ \hline
a_2 & \eta-5c_1-\xi_1-\xi_2 \\
\tilde{a}_1 & \eta-4c_1-\xi_1-\xi_2-\Delta_a \\
\tilde{a}_0 & c_1-\Delta_d+\xi_1-\xi_2 \\
\delta_a & \Delta_a \\
d_2 & \xi_1 \\
\delta_d & \Delta_d \\
e_1 & \xi_2 \\
\tilde{e}_0 & 5c_1+2\xi_1+\xi_2+\Delta_a-\Delta_d-\eta \\
\nu & \eta-4c_1-2\xi_1+\Delta_d -\Delta_a
\end{array}\label{221sections}\end{equation}
and work out the matter curves
\begin{equation}\begin{array}{c|c|c|c|c}
\text{Matter Curve} & \text{Origin} & \text{Equation} & \text{Class} & F_Y \\ \hline
\mathbf{10}^{(2)} & {\cal{C}}^{(2)} & a_2 & \eta-5c_1-\xi_1-\xi_2 & -M-N \\
\mathbf{10}^{(2\prime)} & {\cal{C}}^{(2\prime)} & d_2 & \xi_1 & M\\
\mathbf{10}^{(1)} & {\cal{C}}^{(1)} & e_1 & \xi_2 & N\\
\mathbf{\overline{5}}^{(22)} & {\cal{C}}^{(2)}-{\cal{C}}^{(2)}& \tilde{a}_1 & \eta-4c_1-\xi_1-\xi_2-\Delta_a & -M-N-P \\ 
\mathbf{\overline{5}}^{(2\prime\,2\prime)} & {\cal{C}}^{(2\prime)}-{\cal{C}}^{(2\prime)} & (\tilde{a}_1\tilde{e}_0 + e_1\tilde{a}_0) & c_1-\Delta_d+\xi_1 & M-Q \\
\mathbf{\overline{5}}^{(2\,2\prime)} & {\cal{C}}^{(2)}-{\cal{C}}^{(2\prime)} & \delta_a\delta_de_1(d_2\tilde{a}_1\delta_a-\tilde{a}_1a_2\delta_d\tilde{e}_0-\tilde{a}_0a_2\delta_de_1) & & \\
& & -(d_2\delta_a-a_2\delta_d\tilde{e}_0)^2\nu & \eta-4c_1+\Delta_a+\Delta_d & P+Q \\
\mathbf{\overline{5}}^{(21)} & {\cal{C}}^{(2)}-{\cal{C}}^{(1)} & a_2\tilde{e}_0^2\nu + \delta_ae_1(\tilde{a}_1\tilde{e}_0+\tilde{a}_0e_1) & c_1+\xi_1+\xi_2+\Delta_a-\Delta_d & M+N+P-Q\\
\mathbf{\overline{5}}^{(2\prime\,1)} & {\cal{C}}^{(2\prime)}-{\cal{C}}^{(1)} & d_2\nu-\tilde{a}_1\delta_de_1 & \eta-4c_1-\xi_1-\Delta_a+\Delta_d & -M-P+Q
\end{array}\end{equation}
where
\begin{equation}\begin{split}M &= F_Y\cdot\xi_1 \\
N &= F_Y\cdot \xi_2 \\
P &= F_Y\cdot \Delta_a \\
Q &= F_Y\cdot \Delta_d
\end{split}\end{equation}
We have therefore managed to realize the entire 4-parameter family of solutions to \eqref{DPapp} and \eqref{MSSManomalycond} described in \eqref{221hyperdist}.  As usual, however, any explicit model must satisfy a few assumptions.  In addition to requiring all of the holomorphic sections in \eqref{221sections} to exist, we also need to ensure that
\begin{equation}[\nu]\cdot [e_1]=0\qquad [\tilde{e}_0]\cdot [e_1]=0\qquad [a_2]\cdot [\delta_a]=0\qquad [d_2]\cdot [\delta_d]=0\end{equation}
in order to avoid non-Kodaira singularities.  Equivalently, we need
\begin{equation}\begin{split}0 &= (\eta-4c_1-2\xi_1+\Delta_d-\Delta_a)\cdot \xi_2\\
&= (5c_1+2\xi_1+\xi_2+\Delta_a-\Delta_d-\eta)\cdot \xi_2\\
&= (\eta-5c_1-\xi_1-\xi_2)\cdot \Delta_a\\
&= \xi_1\cdot\Delta_2=0\end{split}\end{equation}

Let us give an example of {\bf 2+2+1} factorization based on the 3-fold constructed in \cite{Marsano:2009ym}. In this case $S_{\rm GUT}=dP_2$ and $t=h.$
We can choose 
\be
\xi_1=h-e_1,\qquad \xi_2=e_1 \,.
\ee
Then, we obtain solutions
\be
\ba
\Delta_a &=m_a(h-e_1-e_2)+n_ae_1+(m_a-n_a)e_2\cr
\Delta_d &=(m_a-n_a)(h-e_1-e_2)+p_de_2
\ea
\ee
Requiring that all bundles  admit holomorhic sections we find
constraints:
\be 
\ba
&0\le m_a\le 4,\quad 0\le n_a\le 2,\quad 0 \le m_a-n_a\le 2\cr
&0\le p_d\le 3,\quad p_d\le 1+m_a-n_a,\quad m_a-n_a\le 1+p_d
\ea
\ee
It is easy to find solutions of these constraints, for example:
\be
p_d=0, \quad m_a=n_a+1,\quad n_a=0,1,2\,.
\ee


\subsubsection{3+1+1}

Finally, let us turn to a factorization with a cubic piece and two linear pieces
\begin{equation}{\cal{C}}_{3+1+1}={\cal{C}}^{(3)}{\cal{C}}^{(1)}{\cal{C}}^{(1\prime)} = \left(a_3V^3+a_2V^2U+a_1VU^2+a_0U^3\right)\left(d_1V+d_0U\right)\left(e_1V+e_0U\right)\,.
\end{equation}
We expect to find 3 $\mathbf{10}$ matter curves and 4 $\mathbf{\overline{5}}$ matter curves.  The most general distribution of hypercharge flux consistent with \eqref{DPapp} and \eqref{MSSManomalycond} is given by
\begin{equation}\begin{array}{c|c|c}\text{Matter Curve} & \text{Origin} & F_Y \\ \hline
\mathbf{10}^{(3)} & {\cal{C}}^{(3)} & -M-N \\
\mathbf{10}^{(1)} & {\cal{C}}^{(1)} & M \\
\mathbf{10}^{(1\prime)} & {\cal{C}}^{(1\prime)} & N \\
\mathbf{\overline{5}}^{(33)} & {\cal{C}}^{(3)}-{\cal{C}}^{(3)} & -2(M+N)-P \\
\mathbf{\overline{5}}^{(31)} & {\cal{C}}^{(3)}-{\cal{C}}^{(1)} & 2M+N+P \\
\mathbf{\overline{5}}^{(31\prime)} & {\cal{C}}^{(3)}-{\cal{C}}^{(1\prime)} & M+2N+P \\
\mathbf{\overline{5}}^{(11\prime)} & {\cal{C}}^{(1)}-{\cal{C}}^{(1\prime)} & -M-N-P
\end{array}\label{311hyperdist}\end{equation}
The traceless constraint takes essentially the same form as the 2+2+1 example
\begin{equation}b_1 = a_1d_0e_0 + d_1a_0e_0 + e_1a_0d_0=0\,,\end{equation}
and we can start with a similar solution
\begin{equation}a_0 = d_0 e_0 \qquad a_1 = -d_1e_0-d_0e_1\,.\end{equation}
We can now follow a similar procedure to the 2+2+1 example to construct more general solutions from this.  In so doing, we have to be a bit careful because $d_1$ and $d_0$ cannot have simultaneous zeroes and similar for $e_1$ and $e_0$.  To proceed, we first scale $a_0$, $d_0$, and $e_0$ to get a solution of the form
\begin{equation}\begin{split}d_0 &= \nu\tilde{d}_0 \\
e_0 &= \nu\tilde{e}_0 \\
a_0 &= \tilde{d}_0\tilde{e}_0 \nu \\
a_1 &= -(d_1\tilde{e}_0 + \tilde{d}_0 e_1)
\end{split}\end{equation}
Now, we scale $a_0$ and $a_1$ to get the further solution
\begin{equation}\begin{split}d_0 &= \nu\tilde{d}_0 \\
e_0 &= \nu\tilde{e}_0 \\
a_0 &= \delta_a\tilde{d}_0\tilde{e}_0\nu \\
a_1 &= -\delta_a(d_1\tilde{e}_0 + \tilde{d}_0 e_1)
\end{split}\end{equation}
It is easy to write out the sections
\begin{equation}\begin{array}{c|c}\text{Section} & \text{Bundle} \\ \hline
a_3 & \eta-5c_1-\xi_1-\xi_2 \\
a_2 & \eta-4c_1-\xi_1-\xi_2 \\
d_1 & \xi_1 \\
\tilde{d}_0 & \eta-3c_1-\xi_1-2\xi_2-\Delta_a \\
e_1 & \xi_2 \\
\tilde{e}_0 & \eta-3c_1-2\xi_1-\xi_2-\Delta_a \\
\nu & 4c_1+2\xi_1+2\xi_2+\Delta_a-\eta \\
\delta_a & \Delta_a
\end{array}\label{311sections}\end{equation}
and work out the matter curves
\begin{equation}\begin{array}{c|c|c|c|c}
\text{Matter Curve} & \text{Origin} & \text{Equation} & \text{Class} & F_Y \\ \hline
\mathbf{10}^{(3)} & {\cal{C}}^{(3)} & a_3 & \eta-5c_1-\xi_1-\xi_2 & -M-N \\
\mathbf{10}^{(1)} & {\cal{C}}^{(1)} & d_1 & \xi_1 & M \\
\mathbf{10}^{(1\prime)} & {\cal{C}}^{(1\prime)} & e_1 & \xi_2 & N \\
\mathbf{\overline{5}}^{(33)} & {\cal{C}}^{(3)}-{\cal{C}}^{(3)} & a_2d_1\tilde{e}_0 + a_2\tilde{d}_0 e_1 + a_3\tilde{d}_0\tilde{e}_0\nu & 2\eta-7c_1-2\xi_1-2\xi_2-\Delta_a & -2(M+N)-P \\
\mathbf{\overline{5}}^{(31)} & {\cal{C}}^{(3)}-{\cal{C}}^{(1)} & d_1^2\delta_a e_1-a_2d_1\nu-a_3\tilde{d}_0\nu^2 & 2\xi_1 + \xi_2 +\Delta_a & 2M+N+P \\
\mathbf{\overline{5}}^{(31\prime)} & {\cal{C}}^{(3)}-{\cal{C}}^{(1\prime)} & d_1\delta_a e_1^2 - \nu (a_2e_1+a_3\tilde{e}_0\nu) & \xi_1+2\xi_2 +\Delta_a & M+2N+P \\
\mathbf{\overline{5}}^{(1\,1\prime)} & {\cal{C}}^{(1)}-{\cal{C}}^{(1\prime)} & d_1\tilde{e}_0 + \tilde{d}_0 e_1 & \eta-3c_1-\xi_1-\xi_2-\Delta_a & -M-N-P
\end{array}\end{equation}
where
\begin{equation}\begin{split}M &= F_Y\cdot\xi_1 \\
N &= F_Y\cdot \xi_2 \\
P &= F_Y\cdot \Delta_a
\end{split}\end{equation}
This realizes the entire 3-parameter family of solutions to \eqref{DPapp} and \eqref{MSSManomalycond} described in \eqref{311hyperdist}.  Again, any explicit model must satisfy a few assumptions.  In addition to requiring all of the holomorphic sections in \eqref{311sections} to exist, we also need to ensure that
\begin{equation}[\nu]\cdot [d_1] = [\nu]\cdot [e_1] = [\tilde{d}_0]\cdot [d_1] = [\tilde{e}_0] \cdot [e_1] = 0\end{equation}
in order to avoid non-Kodaira singularities.  Equivalently, we need
\begin{equation}\begin{split} 0 &= (4c_1+2\xi_1+2\xi_2+\Delta_a-\eta)\cdot \xi_1 \\
&= (4c_1+2\xi_1+2\xi_2+\Delta_a-\eta)\cdot \xi_2 \\
&= (\eta-3c_1-\xi_1-2\xi_2-\Delta_a)\cdot \xi_1 \\
&= (\eta-3c_1-2\xi_1-\xi_2-\Delta_a)\cdot \xi_2
\end{split}\end{equation}

Let us give an example of the {\bf 3+1+1}  factorization based on the
3-fold constructed in \cite{Marsano:2009ym}. In this case $S_{\rm GUT}=dP_2$ and $t=h$.
We can choose 
\be
\xi_1=h-e_1-e_2,\qquad \xi_2=e_1\ ,.
\ee
Then,
\be
\Delta_a=(n_a+2)(h-e_1-e_2)+n_ae_1+3e_2 \,,
\ee
is a solution.
Requiring that all bundles in eq 3.38 admit holomorhic sections we find: $n_a=1,2,3$.

\section{Some Spectral Cover Technicalities}
\label{app:spectech}

In this appendix, we describe how to compute the 'components at $\infty$' that must be removed from ${\cal{C}}^{(a)}\cdot \tau{\cal{C}}$ when computing matter curves for ${\cal{C}}^{(a)}$ a component of the full spectral cover ${\cal{C}}$.  This is all well-known and we include this appendix only for illustration.  Let us start by recalling the nature of the net component at $\infty$.  We can write generic 5-sheeted spectral cover as
\begin{equation}{\cal{C}} = b_5V^5 + b_4V^4U+b_3V^3U^2+b_2V^2U^3+b_0U^5\end{equation}
The locus ${\cal{C}}\cap \tau {\cal{C}}$ is described by the equations
\begin{equation}\begin{split}0 &= U\left(b_0U^4+b_2U^2V^2+b_4V^4\right) \\
0 &= V^3\left(b_3U^2+b_5V^2\right)
\end{split}\end{equation}
The $\mathbf{10}$ matter curve is the solution
\begin{equation}\Sigma_{\mathbf{10}}:\quad U=b_5=0\end{equation}
The $\mathbf{\overline{5}}$ matter curve is given by
\begin{equation}\Sigma_{\mathbf{\overline{5}}}:\quad b_0U^4+b_2U^2V^2+b_4V^4=b_3U^2+b_5V^2=0\end{equation}
which is easily seen to be a 2-sheeted cover of the curve $b_0b_5^2-b_2b_3b_5+b_3^2b_4=0$ inside $S_{\rm GUT}$.  What remains is the 'component at $\infty$' that must be subtracted off, namely
\begin{equation}3\times\left[V = b_0=0\right]\end{equation}
where we have explicitly indicated the multiplicity of 3 with which this component appears.  The homological class of this component is simply{\footnote{Note that we could, in principle, artifically move some parts of $\Sigma_{\mathbf{\overline{5}}}$ off to infinity by allowing $b_0$ and $b_3$ to have common factors.  There are several reasons to expect that this does not change our identification of the part of ${\cal{C}}\cap \tau {\cal{C}}$ that corresponds to $\Sigma_{\mathbf{\overline{5}}}$.  In particular, even though we artifically moved some component of $\Sigma_{\mathbf{\overline{5}}}$ off to $\infty$, that component should still be considered part of the $\mathbf{\overline{5}}$ matter curve.  One can see this by looking at the singularity structure of the local Calabi-Yau geometry that the $b_m$ specify as the equation for the $SU(6)$ enhancement locus does not change when $b_3$ and $b_0$ have a common factor.  Alternatively, we can look directly at the antisymmetric spectral cover, ${\cal{C}}_{\Lambda^2E}$, in terms of which the $\mathbf{\overline{5}}$ matter curve is just ${\cal{C}}_{\Lambda^2E}\cdot \sigma$ \cite{Donagi:2009ra}.  Nothing special happens when $b_3$ and $b_0$ carry a common factor other than the splitting of this matter curve into a few components.  In what follows we will just assume generic $b_3$ and $b_0$.  Similar concerns enter when discussing the so-called 'components at $\infty$' associated to ${\cal{C}}^{(a)}\cap\tau {\cal{C}}$ below.}}
\begin{equation}3\sigma_{\infty}\cdot \pi^*\eta\end{equation}
Let us now turn to ${\cal{C}}^{(a)}\cap \tau {\cal{C}}$ for components ${\cal{C}}^{(n)}$ of varying degree $n$.  In general, we can write such a component as
\begin{equation}d_n V^n + \ldots + d_0 U^n\end{equation}
for some sections $d_m$.  The net homological class of ${\cal{C}}^{(n)}$ is
\begin{equation}{\cal{C}}^{(n)} = n\sigma + \pi^*\xi_n\end{equation}
where $d_0$ is a section of the bundle $\xi_n$ on $S_{\rm GUT}$.  What we want to see explicitly is that the 'component at $\infty$' of ${\cal{C}}^{(n)}\cap \tau {\cal{C}}$ that must be removed in the course of computing matter curves is simply
\begin{equation}3\sigma_{\infty}\cdot \pi^*\xi_n\end{equation}
Since the full spectral cover ${\cal{C}}$ is in the class
\begin{equation}{\cal{C}} = 5\sigma + \pi^*\eta\end{equation}
we know that
\begin{equation}\sum_{\text{components}}\xi_n = \eta\label{sumxis}\end{equation}

\subsection{Linear and Quartic Components}

Let us now start by considering a linear component ${\cal{C}}^{(1)}$ so that
\begin{equation}\begin{split}{\cal{C}} &= (d_1V + d_0 U)\left(a_4V^4+a_3V^3U+a_2V^2U^2+a_1VU^3+a_0U^4\right)\\
{\cal{C}}^{(1)} &= d_1V+d_0U\label{14app}\end{split}\end{equation}
The traceless condition implies that
\begin{equation}a_1d_0+a_0d_1=0\end{equation}
Note that the quartic piece could factor further.  Whether or not this actually happens, though, will not have any effect on the analysis that follows.  We can study ${\cal{C}}^{(1)}\cap \tau{\cal{C}}$ by starting with $\tau {\cal{C}}$ and repeatedly utilizing the relations $d_0U=-d_1V$ and $a_1d_0+a_0d_1=0$.  Doing this, we find an explicit equation for ${\cal{C}}^{(1)}\cap \tau {\cal{C}}$ that takes the form
\begin{equation}\begin{split}0 &= d_0U+d_1V \\
&= d_1V^3\left(a_2U^2-a_3UV+a_4V^2\right)
\end{split}\end{equation}
From this, we see that the 'component at $\infty$' is simply
\begin{equation}3\times \left[V=d_0=0\right]\end{equation}
which is in the class
\begin{equation}3\sigma_{\infty}\cdot \pi^*\xi_1\end{equation}
We can also immediately read off the component at $\infty$ of ${\cal{C}}^{(4)}\cdot \tau{\cal{C}}$ since we can just take ${\cal{C}}^{(4)}$ to be the quartic factor in \eqref{14app}.  With our above analysis, it is clear that it is in the class
\begin{equation}3\sigma_{\infty}\cdot \pi^*\left(\eta-\xi_1\right) = 3\sigma_{\infty}\cdot \pi^*\xi_4\label{4compinf}\end{equation}
because $\xi_1+\xi_4=\eta$ from \eqref{sumxis}.  We can also write directly the equations for ${\cal{C}}^{(4)}\cap \tau {\cal{C}}$ as
\begin{equation}\begin{split}0 &= a_0U^4+a_1U^3V+a_2U^2V^2+a_3UV^3+a_4V^4 \\
&= V^3\left(a_3d_0U^2+a_2d_1U^2+a_4d_1V^2\right)
\end{split}\end{equation}
from which \eqref{4compinf} also follows.

\subsection{Quadratic and Cubic Components}

Let us now turn to the cases of quadratic and cubic factors ${\cal{C}}^{(2)}$ and ${\cal{C}}^{(3)}$.  To study a quadratic factor ${\cal{C}}^{(2)}$ we write
\begin{equation}\begin{split}{\cal{C}} &= (a_2V^2+a_1VU+a_0U^2)(e_3V^3+e_2V^2U+e_1VU^2+e_0U^3)\\
{\cal{C}}^{(2)}&=a_2V^2+a_1VU+a_0U^2\end{split}\end{equation}
where traceless tells us that
\begin{equation}a_1e_0+a_0e_1=0\label{c2traceless}\end{equation}
The cubic piece could factor further but this will not affect the analysis that follows for ${\cal{C}}^{(2)}$.  We can now write simple equations for ${\cal{C}}^{(2)}\cap \tau {\cal{C}}$ by starting with $\tau{\cal{C}}$ and repeatedly using the defining equation for ${\cal{C}}$ as well as the traceless condition \eqref{c2traceless} to further simplify things.  In the end, we find
\begin{equation}\begin{split}0 &= a_0U^2+a_1UV+a_2V^2 \\
&= V^3\left(-(2a_2e_1+2a_1e_2+a_0e_3)U^2+a_1e_3UV-a_2e_3V^2\right)
\end{split}\end{equation}
The part that must be subtracted when determining matter curves is precisely
\begin{equation}3\times \left[V=a_0=0\right]\end{equation}
which is in the class
\begin{equation}3\sigma_{\infty}\cdot \pi^*\xi_2\end{equation}

Now, it is a simple matter to compute ${\cal{C}}^{(3)}\cdot \tau {\cal{C}}$.  The reasoning above does not depend on whether ${\cal{C}}^{(2)}$ factors further so we know that the component at $\infty$ that must be removed from ${\cal{C}}^{(3)}\cdot \tau {\cal{C}}$ is just
\begin{equation}3\sigma_{\infty}\cdot \pi^*(\eta-\xi_2) = 3\sigma_{\infty}\cdot \pi^*\xi_3\label{3compinf}\end{equation}
We can also write directly the equations for ${\cal{C}}^{(4)}\cap \tau {\cal{C}}$ as
\begin{equation}\begin{split}0 &= a_0U^3+a_1U^2V+a_2UV^2+a_3V^3 \\
&= V^3\left((a_1e_2+a_0e_3+a_2e_1)U^2+a_2e_3V^2\right)
\end{split}\end{equation}
from which \eqref{3compinf} also follows.

\section{Survey}
\label{app:Survey}

This appendix gives a detailed list of properties of the models and an account of how we arrived at the list of models given in the survey in Appendix~\ref{app:Models}. Recall that we consider models with two $U(1)$ symmetries, and thus there are two types of factorizations of the spectral cover, i.e. {\bf 2+2+1} and {\bf 3+1+1}, which we now discuss in turn.


\subsection{2+2+1}

We now discuss in detail how we arrive at the models in Appendix~\ref{app:Models}. We begin with the spectral covers that factor as ${\bf 2+2+1}$. 
A generic solution to the Dudas-Palti relations here takes the form
\begin{equation}\begin{array}{c|c}\text{Matter Curve} & F_Y \\ \hline
\mathbf{10}^{(a)} & -M-N\\
\mathbf{10}^{(d)} & M \\
\mathbf{10}^{(e)} & N \\
\mathbf{\overline{5}}^{(aa)} & -M-N-P \\
\mathbf{\overline{5}}^{(ad)} & P+Q \\
\mathbf{\overline{5}}^{(dd)} & M-Q \\
\mathbf{\overline{5}}^{(ae)} & M+N+P-Q \\
\mathbf{\overline{5}}^{(de)} & -M-P+Q
\end{array}\end{equation}

There are three types of charged singlets that can break one of our $U(1)$'s through an expectation value.  In turn, they have weights $\pm (\lambda_a-\lambda_d)$, $\pm(\lambda_a-\lambda_e)$, and $\pm(\lambda_d-\lambda_e)$.  To make our analysis systematic, we will consider in turn what kind of models we can get by letting one of these singlets pick up a nonzero vev.  The second and third cases are related by symmetry so it will be sufficient to consider only the first two.

\subsubsection{Singlet with weight $\lambda_d-\lambda_a$}

A singlet with weight $\pm (\lambda_a-\lambda_d)$ will be unable to lift any exotics on the $\mathbf{10}^{(e)}$ curve so we must set $N=0$.  Doing this, the spectrum on each matter curve is as follows
\begin{equation}\begin{array}{c|c|c|c}\text{Matter Curve} & (1,1)_{+1} & (3,2)_{+1/6} & (\overline{3},1)_{-2/3} \\ \hline
\mathbf{10}^{(a)} & G_a-M & G_a & G_a+M \\
\mathbf{10}^{(d)} & G_d+M & G_d & G_d-M \\
\mathbf{10}^{(e)} & G_e & G_e & G_e
\end{array}\end{equation}

\begin{equation}\begin{array}{c|c|c}\text{Matter Curve} & (\overline{3},1)_{+1/3} & (1,2)_{-1/2} \\ \hline
\mathbf{\overline{5}}^{(aa)} & G_{aa} & G_{aa}+M+P \\
\mathbf{\overline{5}}^{(ad)} & G_{ad} & G_{ad} -P-Q \\
\mathbf{\overline{5}}^{(dd)} & G_{dd} & G_{dd} +Q-M \\
\mathbf{\overline{5}}^{(ae)} & G_{ae} & G_{ae}+Q-M-P \\
\mathbf{\overline{5}}^{(de)} & G_{de} & G_{de}+M+P-Q
\end{array}\end{equation}

We must put $\mathbf{10}_M$ on one of $\mathbf{10}^{(a)}$ or $\mathbf{10}^{(d)}$.  Without loss of generality, let us put it on $\mathbf{10}^{(a)}$.  This means that $\mathbf{5}_H$ must live on the curve $\mathbf{\overline{5}}^{(aa)}$.

Because all three generations of $\mathbf{10}_M$ live on $\mathbf{10}^{(a)}$, any exotics that localize there must have the same chirality as the $\mathbf{10}_M$ fields.  We therefore have two choices.  If there are exotics on $\mathbf{10}^{(a)}$, then in order to lift them the sign of our singlet weight is fixed to
\begin{equation}\text{Weight of singlet is }\lambda_d-\lambda_a\end{equation}
We shall investigate this possibility for now and return to the case $\lambda_a-\lambda_d$ with no exotics on $\mathbf{10}^{(a)}$ later.

When the singlet weight is $\lambda_d-\lambda_a$, the allowed mass terms are
\begin{equation}X \mathbf{10}^{(a)}\mathbf{\overline{10}}^{(d)} + X\mathbf{\overline{5}}^{(aa)}\mathbf{5}^{(ad)}+X\mathbf{\overline{5}}^{(ad)}\mathbf{5}^{(dd)}+X\mathbf{\overline{5}}^{(ae)}\mathbf{5}^{(de)}\end{equation}

To get a top Yukawa coupling we must put $H_u$ on $\mathbf{\overline{5}}^{(aa)}$.  There can be no excess of up-type doublets because they do not participate in the allowed mass coupling.  We therefore see that
\begin{equation}G_{aa}+M+P=-1\label{Gaacond}\end{equation}

For the assignments of the remaining $\mathbf{\overline{5}}$ matter curves there are several possibilities.  Note, however, that the net chirality of both doublets and triplets on $\mathbf{\overline{5}}^{(ae)}$ and $\mathbf{\overline{5}}^{(de)}$ is vanishing.  This means that $\mathbf{\overline{5}}_H$ must lie on $\mathbf{\overline{5}}^{(ad)}$ or $\mathbf{\overline{5}}^{(dd)}$.  In each case, the assignment of $\mathbf{\overline{5}}_M$ is fixed by the desire to get a down type Yukawa coupling.  We consider each of these possibilities in turn.

\begin{itemize}

\item $\mathbf{\overline{5}}_M\rightarrow \mathbf{\overline{5}}^{(de)}$ and $\mathbf{\overline{5}}_H\rightarrow \mathbf{\overline{5}}^{(ad)}$

In this case we need three $\mathbf{\overline{5}}$'s on $\mathbf{\overline{5}}^{(de)}$.  There can be no excess of $\mathbf{\overline{5}}$'s there because they do not participate in the mass term.  As a result, we have
\begin{equation}G_{de}=3\qquad G_{ae}=0\qquad Q=M+P\label{qmp1}\end{equation}

We need at least one down-type doublet on $\mathbf{\overline{5}}^{(ad)}$ to play the role of $H_d$.  
There can be an excess of these doublets in principle because they can be lifted through the mass term.  The net chirality on $\mathbf{\overline{5}}^{(ad)}$ and $\mathbf{\overline{5}}^{(dd)}$ has to be +1, though so we find
\begin{equation}\left[G_{dd}+Q-M\right]+\left[G_{ad}-P-Q\right]=1\quad\implies\quad G_{dd}+G_{ad}-Q=1\label{doubcond1}\end{equation}

Turning now to the triplets, there are two cases to consider.  We can have down-type triplets on $\mathbf{\overline{5}}^{(ad)}$ that pair with up-type triplets on $\mathbf{5}^{(dd)}$.  In this case, we cannot have any triplets on $\mathbf{\overline{5}}^{(aa)}$.  The other option is to have up-type triplets on $\mathbf{5}^{(ad)}$ that pair up with down-type triplets on $\mathbf{\overline{5}}^{(aa)}$.  In this case, we cannot have any triplets on $\mathbf{\overline{5}}^{(dd)}$.  We summarize the two cases below.

\begin{itemize}
\item Down-type triplets on $\mathbf{\overline{5}}^{(ad)}$

In this case, we must have
\begin{equation}G_{ad}=-G_{dd}>0\qquad G_{aa}=0\end{equation}
The condition \eqref{doubcond1} becomes
\begin{equation}Q=-1\end{equation}
while \eqref{Gaacond} implies
\begin{equation}M+P=-1\end{equation}
which is consistent with \eqref{qmp1}.  Summarizing, we have
\begin{equation}\begin{array}{c|c|c|c|c|c}
\text{Matter Curve} & G & F_Y & (1,1)_{+1} & (3,2)_{+1/6} & (\overline{3},1)_{-2/3} \\ \hline
\mathbf{10}^{(a)} & 3+\tilde{G} & P+1 & 3+\tilde{G}+(P+1) & 3+\tilde{G} & 3+\tilde{G}-(P+1)\\
\mathbf{10}^{(d)} & -\tilde{G} & -(P+1) & -\tilde{G}-(P+1) & -\tilde{G} & -\tilde{G}+(P+1) \\
\mathbf{10}^{(e)} & 0 & 0 & 0 & 0 & 0
\end{array}\end{equation}
\begin{equation}\begin{array}{c|c|c|c|c}
\text{Matter Curve} & G & F_Y & (\overline{3},1)_{+1/3} & (1,2)_{-1/2} \\ \hline
\mathbf{\overline{5}}^{(aa)} & 0 & 1 & 0 & -1\\
\mathbf{\overline{5}}^{(ad)} & G_{ad} & P-1 & G_{ad} & (G_{ad}-P)+1 \\
\mathbf{\overline{5}}^{(dd)} & -G_{ad} & -P & -G_{ad} & -(G_{ad}-P) \\
\mathbf{\overline{5}}^{(ae)} & 0 & 0 & 0 & 0 \\
\mathbf{\overline{5}}^{(de)} & 3 & 0 & 3 & 3
\end{array}\end{equation}
where
\begin{equation}G_{ad}\ge0\qquad (G_{ad}-P)\ge 0\qquad \tilde{G}\ge |P+1|\end{equation}

\item Up-type triplets on $\mathbf{\overline{5}}^{(ad)}$

In this case, we must have
\begin{equation}G_{aa}=-G_{ad}>0\qquad G_{dd}=0\end{equation}
The condition \eqref{doubcond1} becomes
\begin{equation}G_{aa}=-(Q+1)\end{equation}
while \eqref{Gaacond} implies
\begin{equation}G_{aa}=-(M+P+1)\end{equation}
which is consistent with \eqref{qmp1}.  Summarizing, we have
\begin{equation}\begin{array}{c|c|c|c|c|c}
\text{M.C.} & G & F_Y & (1,1)_{+1} & (3,2)_{+1/6} & (\overline{3},1)_{-2/3} \\ \hline
\mathbf{10}^{(a)} & 3+\tilde{G} & P+G_{aa}+1 & 3+\tilde{G}+(P+G_{aa}+1) & 3+\tilde{G} & 3+\tilde{G}-(P+G_{aa}+1) \\
\mathbf{10}^{(d)} & -\tilde{G} & -(P+G_{aa}+1) & -\tilde{G}-(P+G_{aa}+1) & -\tilde{G} & -\tilde{G}+(P+G_{aa}+1) \\
\mathbf{10}^{(e)} & 0 & 0 & 0 & 0 & 0
\end{array}\end{equation}
\begin{equation}\begin{array}{c|c|c|c|c}
\text{Matter Curve} & G & F_Y & (\overline{3},1)_{+1/3} & (1,2)_{-1/2} \\ \hline
\mathbf{\overline{5}}^{(aa)} & G_{aa} & 1+G_{aa} & G_{aa} & -1 \\
\mathbf{\overline{5}}^{(ad)} & -G_{aa} & P-1-G_{aa} & -G_{aa} & 1-P \\
\mathbf{\overline{5}}^{(dd)} & 0 & -P & 0 & P \\
\mathbf{\overline{5}}^{(ae)} & 0 & 0 & 0 & 0 \\
\mathbf{\overline{5}}^{(de)} & 3 & 0 & 3 & 3
\end{array}\end{equation}
where
\begin{equation}G_{aa}\ge 0\qquad (-P)\ge 0\qquad \tilde{G}\ge|P+G_{aa}+1|\end{equation}

\end{itemize}

\item $\mathbf{\overline{5}}_M\rightarrow \mathbf{\overline{5}}^{(ae)}$ and $\mathbf{\overline{5}}_H\rightarrow \mathbf{\overline{5}}^{(dd)}$

With this assignment, we need one down type doublet on $\mathbf{\overline{5}}^{(dd)}$ and there can be no excess of these doublets because they do not participate in the mass term.  This means that
\begin{equation}G_{dd}+Q-M=1\end{equation}
in addition to the condition we had before from up-type doublets on $\mathbf{5}^{(aa)}$
\begin{equation}G_{aa}+P+M=-1.\end{equation}

Our $\mathbf{\overline{5}}_M$'s sit on $\mathbf{\overline{5}}^{(ae)}$.  There can be an excess of down-type doublets and triplets here because they can be paired with up-type doublets and triplets on $\mathbf{5}^{(de)}$.  In general, we must have
\begin{equation}G_{ae}=3+\hat{G}\quad G_{de}=-\hat{G}\quad \hat{G}>0\quad Q\ge M+P=-(1+G_{aa})\end{equation}
where we included the information from \eqref{Gaacond}.

Turning to the triplets, we again have two cases depending on whether $\mathbf{\overline{5}}^{(ad)}$ houses up- or down-type triplets.

\begin{itemize}
\item Down-type triplets on $\mathbf{\overline{5}}^{(ad)}$

In this case, we must have
\begin{equation}G_{ad}=-G_{dd}\ge 0\qquad G_{aa}=0.\end{equation}
The resulting fluxes and spectra are
\begin{equation}\begin{array}{c|c|c|c|c|c}
\text{M.C.} & G & F_Y & (1,1)_{+1} & (3,2)_{+1/6} & (\overline{3},1)_{-2/3} \\ \hline
\mathbf{10}^{(a)} & 3+\tilde{G} & P+1 & 3+\tilde{G} +(P+1) & 3+\tilde{G} & 3+\tilde{G}-(P+1) \\
\mathbf{10}^{(d)} & -\tilde{G} & -(P+1) & -\tilde{G}-(P+1) & -\tilde{G} & -\tilde{G}+(P+1)  \\
\mathbf{10}^{(e)} & 0 & 0 & 0 & 0 & 0
\end{array}\end{equation}
\begin{equation}\begin{array}{c|c|c|c|c}
\text{Matter Curve} & G & F_Y & (\overline{3},1)_{+1/3} & (1,2)_{-1/2} \\ \hline
\mathbf{\overline{5}}^{(aa)} & 0 & 1 & 0 & -1 \\
\mathbf{\overline{5}}^{(ad)} & G_{ad} & G_{ad} & G_{ad} & 0 \\
\mathbf{\overline{5}}^{(dd)} & -G_{ad} & -G_{ad}-1 & -G_{ad} & 1 \\
\mathbf{\overline{5}}^{(ae)} & 3+\hat{G} & P-G_{ad}-1 & 3+\hat{G} & 3+\hat{G}+G_{ad}+1-P \\
\mathbf{\overline{5}}^{(de)} & -\hat{G} & -P+G_{ad}+1 & -\hat{G} & -\hat{G}-G_{ad}-1+P 
\end{array}\end{equation}
where
\begin{equation}G_{ad}\ge 0\qquad \hat{G}\ge 0\qquad \hat{G}+G_{ad}+1-P\ge 0\qquad \tilde{G}>|P+1|.\end{equation}

\item Up-type triplets on $\mathbf{\overline{5}}^{(ad)}$
In this case, we must have
\begin{equation}G_{aa}=-G_{ad}\ge 0\qquad G_{dd}=0\end{equation}
The resulting fluxes and spectra are
\begin{equation}\begin{array}{c|c|c|c|c|c}
\text{M.C.} & G & F_Y & (1,1)_{+1} & (3,2)_{+1/6} & (\overline{3},1)_{-2/3} \\ \hline
\mathbf{10}^{(a)} & 3+\tilde{G} & G_{aa}+P+1 & 3+\tilde{G}+(G_{aa}+P+1) & 3+\tilde{G} & 3+\tilde{G}-(G_{aa}+P+1) \\
\mathbf{10}^{(d)} & -\tilde{G} & -(G_{aa}+P+1)& -\tilde{G}-(G_{aa}+P+1)& -\tilde{G} & -\tilde{G}+(G_{aa}+P+1  \\
\mathbf{10}^{(e)} & 0 & 0 & 0 & 0 & 0
\end{array}\end{equation}
\begin{equation}\begin{array}{c|c|c|c|c}
\text{Matter Curve} & G & F_Y & (\overline{3},1)_{+1/3} & (1,2)_{-1/2} \\ \hline
\mathbf{\overline{5}}^{(aa)} & G_{aa} & G_{aa}+1 & G_{aa} & -1 \\
\mathbf{\overline{5}}^{(ad)} & -G_{aa} & -G_{aa} & -G_{aa} & 0 \\
\mathbf{\overline{5}}^{(dd)} & 0 & -1 & 0 & 1\\
\mathbf{\overline{5}}^{(ae)} & 3+\hat{G} & P-1 & 3+\hat{G} & 3+\hat{G}-(P-1)\\
\mathbf{\overline{5}}^{(de)} & -\hat{G} & -(P-1) & -\hat{G} & -\hat{G}+(P-1) \\
\end{array}\end{equation}
where
\begin{equation}G_{aa}\ge 0\qquad \hat{G}\ge 0 \qquad \hat{G}-(P-1)\ge 0\qquad \tilde{G}\ge |G_{aa}+P+1|.\end{equation}

\end{itemize}

\end{itemize}

\subsubsection{Singlet with weight $\lambda_a-\lambda_d$}

Now that we have exhaustively analyzed the case where the singlet weight is $\lambda_d-\lambda_a$, let us return to the other possibility
\begin{equation}\text{The singlet weight is }\lambda_a-\lambda_d\end{equation}
Because $\mathbf{10}_M$ is on $\mathbf{10}^{(a)}$, any exotics on $\mathbf{10}^{(a)}$ must come from the $\mathbf{10}$ so that their weight is $\lambda_a$ and they cannot be lifted by our singlet.  This means that we must choose
\begin{equation}M=N=0\end{equation}
Further, the allowed mass terms are
\begin{equation}X\mathbf{\overline{5}}^{(ad)}\mathbf{5}^{(aa)} + X\mathbf{\overline{5}}^{(dd)}\mathbf{5}^{(ad)}+\mathbf{\overline{5}}^{(de)}\mathbf{5}^{(ae)}\end{equation}
For the assignments of $\mathbf{\overline{5}}_H$ and $\mathbf{\overline{5}}_M$, there are several possibilities.  Note that the net chirality of both doublets and triplets are identical on $\mathbf{\overline{5}}^{(ae)}$ and $\mathbf{\overline{5}}^{(de)}$, which can only pair up with one another via a mass term.  This means that we cannot put $H_d$ on either of these curves.  We are also unable to put $H_d$ on $\mathbf{\overline{5}}^{(ad)}$ because down type doublets there can pair up with up type doublets on $\mathbf{\overline{5}}^{(aa)}$ via the mass term.  As a result, our only option is
\begin{equation}\mathbf{\overline{5}}_M\rightarrow \mathbf{\overline{5}}^{(ae)}\qquad\mathbf{\overline{5}}_H\rightarrow\mathbf{\overline{5}}^{(dd)}\end{equation}
There are now four choices depending on whether we have down type or up type triplets and doublets on $\mathbf{\overline{5}}^{(ad)}$
\begin{itemize}
\item Down type triplets and down type doublets on $\mathbf{\overline{5}}^{(ad)}$

In this case, we must have exactly 1 down type doublet and zero down type triplets on $\mathbf{\overline{5}}^{(dd)}$.  The net chirality of doublets on $\mathbf{\overline{5}}^{(ad)}$ and $\mathbf{5}^{(aa)}$ must then be -1.  We find
\begin{equation}G_{ad}=-G_{aa}\ge 0 \qquad G_{dd}=0\qquad G_{aa}+P=-1\qquad Q=1.\end{equation}
Finally, we note that exotics on $\mathbf{\overline{5}}^{(ae)}$ must come from the $\mathbf{\overline{5}}$ because they represent an excess beyond the three generations of $\mathbf{\overline{5}}_M$ that we put there.  Among other things, this means that the $U(1)_Y$ flux through $\mathbf{\overline{5}}^{(ae)}$ and $\mathbf{\overline{5}}^{(de)}$ must vanish
\begin{equation}P=1\end{equation}
In the end, we have
\begin{equation}\begin{array}{c|c|c|c|c|c}
\text{M.C.} & G & F_Y & (1,1)_{+1} & (3,2)_{+1/6} & (\overline{3},1)_{-2/3} \\ \hline
\mathbf{10}^{(a)} & 3 & 0 & 3 & 3 & 3 \\
\mathbf{10}^{(d)} & 0 & 0 & 0 & 0 & 0 \\
\mathbf{10}^{(e)} & 0 & 0 & 0 & 0 & 0
\end{array}\end{equation}
\begin{equation}\begin{array}{c|c|c|c|c}
\text{Matter Curve} & G & F_Y & (\overline{3},1)_{+1/3} & (1,2)_{-1/2} \\ \hline
\mathbf{\overline{5}}^{(aa)} & -G_{ad} & -1 & -G_{ad} & -G_{ad}+1 \\
\mathbf{\overline{5}}^{(ad)} & G_{ad} & 2 & G_{ad} & G_{ad}-2 \\
\mathbf{\overline{5}}^{(dd)} & 0 & -1 & 0 & 1 \\
\mathbf{\overline{5}}^{(ae)} & 3 & 0 & 3 & 3 \\
\mathbf{\overline{5}}^{(de)} & 0 & 0 & 0 & 0 \\
\end{array}\end{equation}
where
\begin{equation}G_{ad}\ge 2.\end{equation}

\item Down type triplets and up type doublets on $\mathbf{\overline{5}}^{(ad)}$

In this case, we must have exactly one up type doublet on $\mathbf{5}^{(aa)}$ and zero triplets on $\mathbf{\overline{5}}^{(dd)}$.  Further, the net chirality of doublets on $\mathbf{\overline{5}}^{(ad)}$ and $\mathbf{\overline{5}}^{(dd)}$ must be +1.  We find
\begin{equation}G_{ad}=-G_{aa}\ge 0\qquad G_{dd}=0\qquad -G_{ad}+P=-1.\end{equation}
Finally, we note that exotics on $\mathbf{\overline{5}}^{(ae)}$ must come from the $\mathbf{\overline{5}}$ because they represent an excess beyond the three generations of $\mathbf{\overline{5}}_M$ that we put there.  Among other things, this means that the $U(1)_Y$ flux through $\mathbf{\overline{5}}^{(ae)}$ and $\mathbf{\overline{5}}^{(de)}$ must vanish
\begin{equation}G_{ad}-Q-1.\end{equation}
This leads to
\begin{equation}\begin{array}{c|c|c|c|c|c}
\text{M.C.} & G & F_Y & (1,1)_{+1} & (3,2)_{+1/6} & (\overline{3},1)_{-2/3} \\ \hline
\mathbf{10}^{(a)} & 3 & 0 & 3 & 3 & 3 \\
\mathbf{10}^{(d)} & 0 & 0 & 0 & 0 & 0 \\
\mathbf{10}^{(e)} & 0 & 0 & 0 & 0 & 0
\end{array}\end{equation}
\begin{equation}\begin{array}{c|c|c|c|c}
\text{Matter Curve} & G & F_Y & (\overline{3},1)_{+1/3} & (1,2)_{-1/2} \\ \hline
\mathbf{\overline{5}}^{(aa)} & -G_{ad} & -G_{ad}+1 & -G_{ad} & -1 \\
\mathbf{\overline{5}}^{(ad)} & G_{ad} & 2G_{ad}-2 & G_{ad} & 2-G_{ad} \\
\mathbf{\overline{5}}^{(dd)} & 0 & -G_{ad}+1 & 0 & G_{ad}-1 \\
\mathbf{\overline{5}}^{(ae)} & 3 & 0 & 3 & 3 \\
\mathbf{\overline{5}}^{(de)} & 0 & 0 & 0 & 0 \\
\end{array}\end{equation}
where
\begin{equation}G_{ad}\ge 2.\end{equation}

\item Up type triplets and down type doublets on $\mathbf{\overline{5}}^{(ad)}$

In this case, we must have zero triplets on $\mathbf{\overline{5}}^{(aa)}$ and exactly 1 down type doublet on $\mathbf{\overline{5}}^{(dd)}$.  The net chirality of doublets on $\mathbf{\overline{5}}^{(aa)}$ and $\mathbf{\overline{5}}^{(ad)}$ must be -1 while the net chirality of triplets on $\mathbf{\overline{5}}^{(ad)}$ and $\mathbf{\overline{5}}^{(dd)}$ must be 0.  This all leads to
\begin{equation}G_{dd}=-G_{ad}\ge 0\qquad G_{aa}=0\qquad G_{dd}+Q=1.\end{equation}
Finally, we note that exotics on $\mathbf{\overline{5}}^{(ae)}$ must come from the $\mathbf{\overline{5}}$ because they represent an excess beyond the three generations of $\mathbf{\overline{5}}_M$ that we put there.  Among other things, this means that the $U(1)_Y$ flux through $\mathbf{\overline{5}}^{(ae)}$ and $\mathbf{\overline{5}}^{(de)}$ must vanish
\begin{equation}P-1+G_{dd}=0.\end{equation}
This leads to
\begin{equation}\begin{array}{c|c|c|c|c|c}
\text{M.C.} & G & F_Y & (1,1)_{+1} & (3,2)_{+1/6} & (\overline{3},1)_{-2/3} \\ \hline
\mathbf{10}^{(a)} & 3 & 0 & 3 & 3 & 3 \\
\mathbf{10}^{(d)} & 0 & 0 & 0 & 0 & 0 \\
\mathbf{10}^{(e)} & 0 & 0 & 0 & 0 & 0
\end{array}\end{equation}
\begin{equation}\begin{array}{c|c|c|c|c}
\text{Matter Curve} & G & F_Y & (\overline{3},1)_{+1/3} & (1,2)_{-1/2} \\ \hline
\mathbf{\overline{5}}^{(aa)} & 0 & G_{dd}-1 & 0 & 1-G_{dd} \\
\mathbf{\overline{5}}^{(ad)} & -G_{dd} & 2(1-G_{dd}) & -G_{dd} & -2+G_{dd} \\
\mathbf{\overline{5}}^{(dd)} & G_{dd} & G_{dd}-1 & G_{dd} & 1 \\
\mathbf{\overline{5}}^{(ae)} & 3 & 0 & 3 & 3 \\
\mathbf{\overline{5}}^{(de)} & 0 & 0 & 0 & 0 \\
\end{array}\end{equation}
where
\begin{equation}G_{dd}\ge 2.\end{equation}

\item Up type triplets and up type doublets on $\mathbf{\overline{5}}^{(ad)}$

In this case, we must have 0 triplets and 1 up type doublet on $\mathbf{\overline{5}}^{(aa)}$.  The net chirality of triplets on $\mathbf{\overline{5}}^{(ad)}$ and $\mathbf{\overline{5}}^{(dd)}$ must be 0 while the net chirality of doublets must be 1.  This leads to
\begin{equation}G_{dd}=-G_{ad}\ge 0\qquad G_{aa}=0\qquad P=-1.\end{equation}
Finally, we note that exotics on $\mathbf{\overline{5}}^{(ae)}$ must come from the $\mathbf{\overline{5}}$ because they represent an excess beyond the three generations of $\mathbf{\overline{5}}_M$ that we put there.  Among other things, this means that the $U(1)_Y$ flux through $\mathbf{\overline{5}}^{(ae)}$ and $\mathbf{\overline{5}}^{(de)}$ must vanish
\begin{equation}Q=-1.\end{equation}
This leads to
\begin{equation}\begin{array}{c|c|c|c|c|c}
\text{M.C.} & G & F_Y & (1,1)_{+1} & (3,2)_{+1/6} & (\overline{3},1)_{-2/3} \\ \hline
\mathbf{10}^{(a)} & 3 & 0 & 3 & 3 & 3 \\
\mathbf{10}^{(d)} & 0 & 0 & 0 & 0 & 0 \\
\mathbf{10}^{(e)} & 0 & 0 & 0 & 0 & 0
\end{array}\end{equation}
\begin{equation}\begin{array}{c|c|c|c|c}
\text{Matter Curve} & G & F_Y & (\overline{3},1)_{+1/3} & (1,2)_{-1/2} \\ \hline
\mathbf{\overline{5}}^{(aa)} & 0 & 1 & 0 & -1 \\
\mathbf{\overline{5}}^{(ad)} & -G_{dd} & -2 & -G_{dd} & -G_{dd}+2 \\
\mathbf{\overline{5}}^{(dd)} & G_{dd} & 1 & G_{dd} & G_{dd}-1 \\
\mathbf{\overline{5}}^{(ae)} & 3 & 0 & 3 & 3 \\
\mathbf{\overline{5}}^{(de)} & 0 & 0 & 0 & 0 \\
\end{array}\end{equation}
where
\begin{equation}G_{dd}\ge 2.\end{equation}

\end{itemize}

\subsubsection{Singlet with weight $\pm (\lambda_a-\lambda_e)$ with $\mathbf{10}_M$ on $\mathbf{10}^{(a)}$}

Because our singlet treats states with weights $\lambda_a$ and $\lambda_e$ differently, the choice of where $\mathbf{10}_M$ fields localize becomes important.  In this subsection, we take the $\mathbf{10}_M$ fields to localize on $\mathbf{10}^{(a)}$.  We cannot lift any zero modes that localize on $\mathbf{10}^{(d)}$ so it better be that the flux there vanishes.  This means that $M=0$.  The spectrum on each matter curve, in this case, is as follows
\begin{equation}\begin{array}{c|c|c|c}\text{Matter Curve} & (1,1)_{+1} & (3,2)_{+1/6} & (\overline{3},1)_{-2/3} \\ \hline
\mathbf{10}^{(a)} & G_a-N & G_a & G_a+N \\
\mathbf{10}^{(d)} & 0 & 0 & 0 \\
\mathbf{10}^{(e)} & G_e+N & G_e & G_e-N
\end{array}\end{equation}

\begin{equation}\begin{array}{c|c|c}\text{Matter Curve} & (\overline{3},1)_{+1/3} & (1,2)_{-1/2} \\ \hline
\mathbf{\overline{5}}^{(aa)} & G_{aa} & G_{aa}+N+P \\
\mathbf{\overline{5}}^{(ad)} & G_{ad} & G_{ad} -P-Q\\
\mathbf{\overline{5}}^{(dd)} & G_{dd} & G_{dd} + Q \\
\mathbf{\overline{5}}^{(ae)} & G_{ae} & G_{ae} + Q-N-P \\
\mathbf{\overline{5}}^{(de)} & G_{de} & G_{de} + P-Q
\end{array}\end{equation}

There are now two subcases depending on the sign of the singlet weight.

\begin{itemize}

\item Singlet weight is $\lambda_e-\lambda_a$

In this case, the allowed mass terms are
\begin{equation}X\mathbf{10}^{(a)}\mathbf{\overline{10}}^{(e)} + X\mathbf{\overline{5}}^{(aa)}\mathbf{5}^{(ae)}+X\mathbf{\overline{5}}^{(ad)}\mathbf{5}^{(de)}.\end{equation}
Note that $\mathbf{\overline{5}}^{(dd)}$ does not participate in any mass couplings.

Consider now the assignment of $\mathbf{\overline{5}}$ matter curves.  To get a top Yukawa coupling, we must get $H_u$ from $\mathbf{5}^{(aa)}$.  To get a bottom Yukawa coupling, we can then place $\mathbf{\overline{5}}_M$ and $\mathbf{\overline{5}}_H$ on either the pair $\mathbf{\overline{5}}^{(ad)}/\mathbf{\overline{5}}^{(de)}$ or the pair $\mathbf{\overline{5}}^{(ae)}/\mathbf{\overline{5}}^{(dd)}$.

Suppose first that we put $\mathbf{\overline{5}}_M/\mathbf{\overline{5}}_H$ on the pair $\mathbf{\overline{5}}^{(ad)}/\mathbf{\overline{5}}^{(de)}$.  Because fields on these two curves can only get mass by pairing with one another, we see that the net chirality of triplets on both must be 3 while the net chirality of doublets must be 4.  This amounts to the conditions
\begin{equation}G_{ad}+G_{de}=3,\qquad -2Q=1.\end{equation}
Because $Q$ is an integer, we see that there are no solutions.

We are therefore forced to put $\mathbf{\overline{5}}_M/\mathbf{\overline{5}}_H$ on the pair $\mathbf{\overline{5}}^{(ae)}/\mathbf{\overline{5}}^{(dd)}$.  Regardless of which of the two possibilities for this assignment we have, the net chiralities of both doublets and triplets on $\mathbf{\overline{5}}^{(ad)}$ and $\mathbf{\overline{5}}^{(de)}$ must both vanish.  Further, we know that all doublets and triplets on $\mathbf{\overline{5}}^{(ad)}$ must be down type in order to participate in the mass coupling while those on $\mathbf{\overline{5}}^{(de)}$ must be up-type.  This means that
\begin{equation}G_{ad}-G_{de}\ge 0,\qquad G_{ad}-P\ge 0,\qquad Q=0\end{equation}
Now, because $Q=0$ there can be no $U(1)_Y$ flux threading $\mathbf{\overline{5}}^{(dd)}$.  We are therefore forced to take
\begin{equation}\mathbf{\overline{5}}_M\rightarrow \mathbf{\overline{5}}^{(dd)}\qquad \mathbf{\overline{5}}_H\rightarrow \mathbf{\overline{5}}^{(ae)} .\end{equation}
The down-type doublets on $\mathbf{\overline{5}}^{(ae)}$ do not participate in the mass coupling so there must be exactly one of those.  We can have up-type triplets there that become massive by coupling to down-type triplets on $\mathbf{\overline{5}}^{(aa)}$.  Of course, the up-type doublets on $\mathbf{\overline{5}}^{(aa)}$ do not enter into the mass term so we better have exactly one of those as well.  All of these conditions mean that
\begin{equation}G_{aa}=-G_{ae}\ge 0,\qquad G_{aa}+N+P=1,\qquad G_{ae}-N-P=-1\end{equation}
where these equations exhibit one redundancy.

Taking everything together, the fluxes and spectra are as follows
\begin{equation}G_{aa}=-G_{ad}\ge 0,\qquad G_{dd}=0.\end{equation}
The resulting fluxes and spectra are
\begin{equation}\begin{array}{c|c|c|c|c|c}
\text{M.C.} & G & F_Y & (1,1)_{+1} & (3,2)_{+1/6} & (\overline{3},1)_{-2/3} \\ \hline
\mathbf{10}^{(a)} & 3+\tilde{G} & G_{aa}+P+1 & 3+\tilde{G}+(G_{aa}+P+1) & 3+\tilde{G} & 3+\tilde{G} - (G_{aa}+P+1) \\
\mathbf{10}^{(d)} & 0 & 0 & 0 & 0 & 0 \\
\mathbf{10}^{(e)} & -\tilde{G} & -(G_{aa}+P+1) & -\tilde{G}-(G_{aa}+P+1) & -\tilde{G} & -\tilde{G}+(G_{aa}+P+1) \\
\end{array}\end{equation}
\begin{equation}\begin{array}{c|c|c|c|c}
\text{Matter Curve} & G & F_Y & (\overline{3},1)_{+1/3} & (1,2)_{-1/2} \\ \hline
\mathbf{\overline{5}}^{(aa)} & G_{aa} & G_{aa}+1 & G_{aa} & -1 \\
\mathbf{\overline{5}}^{(ad)} & G_{ad} & P & G_{ad} & G_{ad}-P \\
\mathbf{\overline{5}}^{(dd)} & 3 & 0 & 3 & 3 \\
\mathbf{\overline{5}}^{(ae)} & -G_{aa} & -G_{aa}-1 & -G_{aa} & 1 \\
\mathbf{\overline{5}}^{(de)} & -G_{ad} & -P & -G_{ad} & -(G_{ad}-P) \\
\end{array}\end{equation}
where
\begin{equation}G_{aa}\ge 0,\qquad G_{ad}\ge 0,\qquad (G_{ad}-P)\ge 0,\qquad \tilde{G}\ge |G_{aa}+P+1|.\end{equation}

\item Singlet weight is $\lambda_a-\lambda_e$

In this case, we cannot lift any potential exotics on $\mathbf{10}^{(a)}$ since they must come from a $\mathbf{10}$.  As a result, we have
\begin{equation}N=0.\end{equation}
The mass terms of interest are therefore
\begin{equation}X\mathbf{\overline{5}}^{(ae)}\mathbf{5}^{(aa)}+X\mathbf{\overline{5}}^{(de)}\mathbf{5}^{(ad)}.\end{equation}
As before, the presence of a top Yukawa coupling means that we must get $H_u$ from $\mathbf{5}^{(aa)}$.  To get a bottom Yukawa coupling, we can then place $\mathbf{\overline{5}}_M$ and $\mathbf{\overline{5}}_H$ on either the pair $\mathbf{\overline{5}}^{(ad)}/\mathbf{\overline{5}}^{(de)}$ or the pair $\mathbf{\overline{5}}^{(ae)}/\mathbf{\overline{5}}^{(dd)}$.

Suppose first that we put $\mathbf{\overline{5}}_M/\mathbf{\overline{5}}_H$ on the pair $\mathbf{\overline{5}}^{(ad)}/\mathbf{\overline{5}}^{(de)}$.  Because fields on these two curves can only get mass by pairing with one another, we see that the net chirality of triplets on both must be 3 while the net chirality of doublets must be 3+1=4.  This amounts to the conditions
\begin{equation}G_{ad}+G_{de}=3,\qquad -2Q=1.\end{equation}
Because $Q$ is an integer, we see that there are no solutions.

We are therefore forced to put $\mathbf{\overline{5}}_M/\mathbf{\overline{5}}_H$ on the pair $\mathbf{\overline{5}}^{(ae)}/\mathbf{\overline{5}}^{(dd)}$.  Regardless of which of the two possibilities for this assignment we have, the net chiralities of both doublets and triplets on $\mathbf{\overline{5}}^{(ad)}$ and $\mathbf{\overline{5}}^{(de)}$ must vanish.  Further, we know that all doublets and triplets on $\mathbf{\overline{5}}^{(ad)}$ must be up type in order to participate in the mass coupling while those on $\mathbf{\overline{5}}^{(de)}$ must be down type.  This means that
\begin{equation}G_{de}+G_{ad}\ge 0,\qquad G_{de}+P\ge 0,\qquad G=0.\end{equation}
Now, because $Q=0$ there can be no $U(1)_Y$ flux threading $\mathbf{\overline{5}}^{(dd)}$.  We are therefore forced to take
\begin{equation}\mathbf{\overline{5}}_M\rightarrow\mathbf{\overline{5}}^{(dd)}\qquad \mathbf{\overline{5}}_H\rightarrow\mathbf{\overline{5}}^{(ae)}.\end{equation}
Any down type doublets on $\mathbf{\overline{5}}^{(ae)}$ will pair with up type doublets on $\mathbf{5}^{(aa)}$, though, making it impossible to obtain massless $H_u$ and $H_d$.  We therefore conclude that
\begin{equation}\text{There are no viable models with singlet weight }\lambda_a-\lambda_e.\end{equation}

\end{itemize}

\subsubsection{Singlet with weight $\pm (\lambda_a-\lambda_e)$ with $\mathbf{10}_M$ on $\mathbf{10}^{(d)}$}

As in the last case, we cannot have any exotics on $\mathbf{10}^{(d)}$ so we must have $M=0$.  The spectrum on each matter curve with this choice is given by
\begin{equation}\begin{array}{c|c|c|c}\text{Matter Curve} & (1,1)_{+1} & (3,2)_{+1/6} & (\overline{3},1)_{-2/3} \\ \hline
\mathbf{10}^{(a)} & G_a-N & G_a & G_a+N \\
\mathbf{10}^{(d)} & 0 & 0 & 0 \\
\mathbf{10}^{(e)} & G_e+N & G_e & G_e-N
\end{array}\end{equation}

\begin{equation}\begin{array}{c|c|c}\text{Matter Curve} & (\overline{3},1)_{+1/3} & (1,2)_{-1/2} \\ \hline
\mathbf{\overline{5}}^{(aa)} & G_{aa} & G_{aa}+N+P \\
\mathbf{\overline{5}}^{(ad)} & G_{ad} & G_{ad} -P-Q\\
\mathbf{\overline{5}}^{(dd)} & G_{dd} & G_{dd} + Q \\
\mathbf{\overline{5}}^{(ae)} & G_{ae} & G_{ae} + Q-N-P \\
\mathbf{\overline{5}}^{(de)} & G_{de} & G_{de} + P-Q
\end{array}\end{equation}

The chiralities of exotics on $\mathbf{10}^{(a)}$ can in principle be arbitrary so we cannot conclude anything about the sign of the singlet weight at this point. 

Because $\mathbf{10}_M$ is on $\mathbf{10}^{(a)}$, we must get $H_u$ from $\mathbf{\overline{5}}^{(dd)}$.  Further, we know that $\mathbf{\overline{5}}^{(dd)}$ does not participate in any mass terms for either sign of the singlet weight.  This means that we must have
\begin{equation}G_{dd}=0\qquad Q=-1.\end{equation}

Let us turn now to the $\mathbf{\overline{5}}$'s.  In order to get a down type Yukawa coupling we must put $\mathbf{\overline{5}}_M$ and $\mathbf{\overline{5}}_H$ on either $\mathbf{\overline{5}}^{(aa)}/\mathbf{\overline{5}}^{(de)}$ or $\mathbf{\overline{5}}^{(ae)}/\mathbf{\overline{5}}^{(ad)}$.  Note that in neither case can fields on the $\mathbf{\overline{5}}_M$ matter curve couple to $\mathbf{\overline{5}}_H$ matter curve through the mass term.  Wherever we put $\mathbf{\overline{5}}_M$, it must be that the net chirality of doublets and triplets on this curve and the one that it couples to via the mass term is 3.  Take a closer look at the curves that can be connected by singlet vevs, though
\begin{equation}\mathbf{\overline{5}}^{(aa)}\leftrightarrow\mathbf{\overline{5}}^{(ae)}\qquad \mathbf{\overline{5}}^{(ad)}\leftrightarrow \mathbf{\overline{5}}^{(de)}.\end{equation}
The net chiralities of triplets and doublets on $\mathbf{\overline{5}}^{(aa)}$ and $\mathbf{\overline{5}}^{(ae)}$ are $G_{aa}+G_{ae}$ and $G_{aa}+G_{ae}+Q$, respectively.  Since $Q=-1$, these can never be equivalent.  Similarly, the net chiralities of triplets and doublets on $\mathbf{\overline{5}}^{(ad)}$ and $\mathbf{\overline{5}}^{(de)}$ are $G_{ad}+G_{de}$ and $G_{ad}+G_{de}-2Q$, respectively.  Again, since $Q=-1$ these can never be equivalent.  We therefore see that
\begin{equation}\text{There are no viable models with singlet weight }\pm (\lambda_a-\lambda_e)\text{ and }\mathbf{10}_M\text{ on }\mathbf{10}^{(d)}.\end{equation}

\subsection{3+1+1}

A generic solution to the Dudas-Palti relations here takes the form
\begin{equation}\begin{array}{c|c}\text{Matter Curve} & F_Y \\ \hline
\mathbf{10}^{(a)} & -M-N \\
\mathbf{10}^{(d)} & M \\
\mathbf{10}^{(e)} & N \\
\mathbf{\overline{5}}^{(aa)} & -2(M+N)-P \\
\mathbf{\overline{5}}^{(ad)} & 2M+N+P \\
\mathbf{\overline{5}}^{(ae)} & M+2N+P \\
\mathbf{\overline{5}}^{(de)} & -M-N-P
\end{array}\end{equation}

In order to get a top Yukawa coupling we must always realize $\mathbf{10}_M$ on $\mathbf{10}^{(a)}$ and $H_u$ on $\mathbf{\overline{5}}^{(aa)}$.  Without loss of generality, there are two choices of singlet weights that we can take as $\pm (\lambda_a-\lambda_d)$ and $\pm (\lambda_d-\lambda_e)$.  We treat each of these in turn.

\subsubsection{Singlet with weight $\pm (\lambda_a-\lambda_d)$}

A singlet with this weight will be unable to lift any exotics on the $\mathbf{10}^{(e)}$ curve so we must set
\begin{equation}N=0.\end{equation}
Doing this, the spectrum we get on each matter curve is as follows

\begin{equation}\begin{array}{c|c|c|c}\text{Matter Curve} & (1,1)_{+1} & (3,2)_{+1/6} & (\overline{3},1)_{-2/3} \\ \hline
\mathbf{10}^{(a)} & G_a-M & G_a & G_a+M \\
\mathbf{10}^{(d)} & G_d+M & G_d & G_d-M \\
\mathbf{10}^{(e)} & G_e & G_e & G_e
\end{array}\end{equation}

\begin{equation}\begin{array}{c|c|c}\text{Matter Curve} & (\overline{3},1)_{+1/3} & (1,2)_{-1/2} \\ \hline
\mathbf{\overline{5}}^{(aa)} & G_{aa} & G_{aa}+2M+P \\
\mathbf{\overline{5}}^{(ad)} & G_{ad} & G_{ad} -(2M+P) \\
\mathbf{\overline{5}}^{(ae)} & G_{ae} & G_{ae} -(M+P) \\
\mathbf{\overline{5}}^{(de)} & G_{de} & G_{de} +(M+P)
\end{array}\end{equation}

In addition, any exotics on $\mathbf{10}^{(a)}$ must come from the $\mathbf{10}$ since they arise as excesses of zero modes in addition to the three generations of $\mathbf{10}_M$'s that localize there.  If we have any such exotics, the singlet weight must be
\begin{equation}\text{Singlet weight is }\lambda_d-\lambda_a.\end{equation}
We focus on this case for now, which gives rise to the mass terms
\begin{equation}X\mathbf{10}^{(a)}\mathbf{\overline{10}}^{(d)}+X\mathbf{\overline{5}}^{(aa)}\mathbf{5}^{(ad)}+X\mathbf{\overline{5}}^{(ae)}\mathbf{5}^{(de)}.\end{equation}
Because $H_u$ localizes on $\mathbf{\overline{5}}^{(aa)}$ and up type doublets do not participate in the mass term, the number of doublet zero modes there must be exactly -1
\begin{equation}G_{aa}+2M+P=-1.\end{equation}

To get a down type Yukawa coupling we must put $\mathbf{\overline{5}}_H/\mathbf{\overline{5}}_M$ on $\mathbf{\overline{5}}^{(ad)}/\mathbf{\overline{5}}^{(ae)}$.  If we put $\mathbf{\overline{5}}_M$ on $\mathbf{\overline{5}}^{(ad)}$, though, we encounter a problem.  Fields on $\mathbf{\overline{5}}^{(ad)}$ become massive by pairing with fields from $\mathbf{\overline{5}}^{(aa)}$, where our Higgs doublet $H_u$ lives.  The net chirality of doublets on $\mathbf{\overline{5}}^{(aa)}$ and $\mathbf{\overline{5}}^{(ad)}$ is equivalent to the net chirality of triplets, though, so it is impossible to get the desired spectrum, which has a net chirality of 3 triplets and 3-1=2 doublets.

Suppose we instead put $\mathbf{\overline{5}}_M$ on $\mathbf{\overline{5}}^{(ae)}$.  In this case, $H_d$ goes on $\mathbf{\overline{5}}^{(ad)}$.  Down type doublets on $\mathbf{\overline{5}}^{(ad)}$ do not participate in the mass term, though, so we must have exactly 1 such zero mode there.  This implies that
\begin{equation}G_{ad}-(2M+P)=1.\end{equation}
We can have up type triplets on $\mathbf{\overline{5}}^{(ad)}$ that pair with down type triplets on $\mathbf{\overline{5}}^{(aa)}$.  This leads to
\begin{equation}G_{aa} = -G_{ad} \ge 0.\end{equation}

We turn finally to the $\mathbf{\overline{5}}_M$ fields on $\mathbf{\overline{5}}^{(ae)}$.  The net chirality of both doublets and triplets on $\mathbf{\overline{5}}^{(ae)}$ and $\mathbf{5}^{(de)}$ must be 3 which leads to
\begin{equation}G_{ae}=3+\hat{G},\qquad G_{de}=-\hat{G},\qquad M+P=0.\end{equation}
where
\begin{equation}\hat{G}\ge 0.\end{equation}
In the end, we therefore find the following lone possibility for the spectrum and fluxes
\begin{equation}G_{aa}=-G_{ad}\ge 0,\qquad G_{dd}=0.\end{equation}
The resulting fluxes and spectra are
\begin{equation}\begin{array}{c|c|c|c|c|c}
\text{M.C.} & G & F_Y & (1,1)_{+1} & (3,2)_{+1/6} & (\overline{3},1)_{-2/3} \\ \hline
\mathbf{10}^{(a)} & 3+\tilde{G} & G_{aa}+1 & 3+\tilde{G}+(G_{aa}+1) & 3+\tilde{G} & 3+\tilde{G}-(G_{aa}+1)\\
\mathbf{10}^{(d)} & -\tilde{G} &-(G_{aa}+1) & -\tilde{G}-(G_{aa}+1) & -\tilde{G} & -\tilde{G}+(G_{aa}+1) \\
\mathbf{10}^{(e)} & 0 & 0 & 0 & 0 & 0 \\
\end{array}\end{equation}
\begin{equation}\begin{array}{c|c|c|c|c}
\text{Matter Curve} & G & F_Y & (\overline{3},1)_{+1/3} & (1,2)_{-1/2} \\ \hline
\mathbf{\overline{5}}^{(aa)} & G_{aa} & G_{aa}+1 & G_{aa} & -1 \\
\mathbf{\overline{5}}^{(ad)} & -G_{aa} & -G_{aa}-1 & -G_{aa} & 1 \\
\mathbf{\overline{5}}^{(ae)} & 3+\hat{G} & 0 & 3+\hat{G} & 3+\hat{G} \\
\mathbf{\overline{5}}^{(de)} & -\hat{G} & 0 & -\hat{G} & -\hat{G} 
\end{array}\end{equation}
where
\begin{equation}G_{aa}\ge 0,\qquad \hat{G}\ge 0,\qquad \tilde{G}\ge |G_{aa}+1|.\end{equation}

Let us now return to the possibility of singlet weight $\lambda_a-\lambda_d$.  In this case, we cannot lift any exotics from $\mathbf{10}$ curves so both $M$ and $N$ must vanish.  The mass terms take the form
\begin{equation}X\mathbf{\overline{5}}^{(ad)}\mathbf{5}^{(aa)}+X\mathbf{\overline{5}}^{(de)}\mathbf{5}^{(ae)}.\end{equation}
A top Yukawa coupling forces us to put $H_u$ on $\mathbf{5}^{(aa)}$.  A bottom Yukawa coupling then requires that $\mathbf{\overline{5}}_H/\mathbf{\overline{5}}_M$ be placed on $\mathbf{\overline{5}}^{(ad)}/\mathbf{\overline{5}}^{(ae)}$.  Either way, we must have at least one non-exotic massless down type doublet on $\mathbf{\overline{5}}^{(ad)}$.  Because of the mass term, we cannot simultaneously keep a massless up type doublet on $\mathbf{5}^{(aa)}$ and a massless down type doublet on $\mathbf{\overline{5}}^{(ad)}$ so we conclude that
\begin{equation}\text{There are no viable models with singlet weight }\lambda_a-\lambda_d.\end{equation}

\subsubsection{Singlet with weight $\pm (\lambda_d-\lambda_e)$}

A singlet with this weight will be unable to lift any exotics on the $\mathbf{10}^{(a)}$ curve so we must set
\begin{equation}M+N=0.\end{equation}
Doing this, the spectrum we get on each matter curve is as follows
\begin{equation}\begin{array}{c|c|c|c}\text{Matter Curve} & (1,1)_{+1} & (3,2)_{+1/6} & (\overline{3},1)_{-2/3} \\ \hline
\mathbf{10}^{(a)} & G_a & G_a & G_a \\
\mathbf{10}^{(d)} & G_d+M & G_d & G_d-M \\
\mathbf{10}^{(e)} & G_e-M & G_e & G_e+M
\end{array}\end{equation}

\begin{equation}\begin{array}{c|c|c}\text{Matter Curve} & (\overline{3},1)_{+1/3} & (1,2)_{-1/2} \\ \hline
\mathbf{\overline{5}}^{(aa)} & G_{aa} & G_{aa} +P \\
\mathbf{\overline{5}}^{(ad)} & G_{ad} & G_{ad} -M-P \\
\mathbf{\overline{5}}^{(ae)} & G_{ae} & G_{ae} +M-P \\
\mathbf{\overline{5}}^{(de)} & G_{de} & G_{de} +P
\end{array}\end{equation}

When the singlet weight is $\pm (\lambda_d-\lambda_e)$, none of the fields on $\mathbf{\overline{5}}^{(aa)}$ or $\mathbf{\overline{5}}^{(de)}$ can participate in any mass terms.  Since we must have $H_u$ on $\mathbf{\overline{5}}^{(aa)}$ in order to get a top Yukawa coupling we must take
\begin{equation}G_{aa}=0\qquad P=-1.\end{equation}
When we do this, however, the zero modes on $\mathbf{\overline{5}}^{(de)}$ cannot comprise a complete GUT multiplet.  We are therefore forced to put $\mathbf{\overline{5}}_H$ on $\mathbf{\overline{5}}^{(de)}$.  This is impossible, though, because the absence of triplets on $\mathbf{\overline{5}}^{(de)}$ forces
\begin{equation}G_{de}=0\end{equation}
which, along with $P=-1$, leads to an up type doublet on $\mathbf{\overline{5}}^{(de)}$.  We therefore see that
\begin{equation}\text{There are no viable models with singlet weights }\pm (\lambda_d-\lambda_e).\end{equation}



\bibliographystyle{JHEP}

\providecommand{\href}[2]{#2}\begingroup\raggedright\endgroup

\end{document}